\documentclass[11pt,a4]{article}
\usepackage{geometry}
\usepackage{graphicx}
\usepackage{amssymb}
\usepackage{booktabs}
\usepackage{epstopdf}
\usepackage[x11names,table]{xcolor}
\usepackage[font=footnotesize,labelfont=bf]{caption}
\usepackage{titlesec}
\usepackage[OT1]{fontenc}
\usepackage{subfigure}
\usepackage{hyperref}
\usepackage{float}
\usepackage{amsthm,amsmath,dcolumn,graphicx,multicol,fancyhdr}
\usepackage{rotating,amsfonts, color,amssymb,amsmath,amscd,array,enumerate,afterpage}
\usepackage{bm}

\usepackage{varwidth,lipsum}
\newcommand\anon[1]{#1}
\newcommand\hide[1]{#1}

\theoremstyle{lemma}

\theoremstyle{proposition}

\theoremstyle{corollary}

\theoremstyle{remark}

\theoremstyle{definition}

\numberwithin{equation}{section}
\theoremstyle{plain}

\begin{document}
\bibliographystyle{plain}

\title{Spectral Synchronicity in Brain Signals}

\author{\anon{Carolina Eu\'an \footnote{\anon{Centro de Investigaci\'on en Matem\'aticas, Guanajuato, Gto, M\'exico.}} \footnotemark,
Hernando Ombao \footnote{\anon{Department of Statistics, University of California, Irvine, USA.}} \addtocounter{footnote}{-2}\addtocounter{Hfootnote}{-2}\footnote{\anon{UC Irvine Space-Time Modeling Group.}}, Joaqu\'in Ortega \addtocounter{footnote}{-2}\addtocounter{Hfootnote}{-2}\footnotemark }}

\maketitle

\begin{abstract}
Brain activity following stimulus presentation and during resting state are often
the result of highly coordinated responses of large numbers of neurons both locally
(within each region) and globally (across different brain regions). Coordinated
activity of neurons can give rise to oscillations which are captured by
electroencephalograms (EEG). Most studies on resting state are based on
functional magnetic resonance imaging data. In this paper, we examine
EEGs as this is the primary data being used by our collaborators who are studying
coordination of neuronal response
during the execution of tasks such as learning,
and memory formation, retention and retrieval. Developing new methods that
address this research question is important because disruptions in coordination
have been observed in a number of neurological and mental diseases such as schizophrenia,
epilepsy and Alzheimer's disease. In this paper, we develop the spectral merger
clustering (SMC) method that identifies synchronized brain
regions during resting state in a sense that these regions
share similar oscillations or waveforms. The SMC method, produces clusters of EEGs
which serve as a proxy for segmenting the brain cortical surface since the EEGs capture
neuronal activity over a locally distributed region on the cortical surface. The extent of
desynchronicity between a pair of EEGs is measured using the total variation distance
(TVD) which gives the largest possible difference between the spectral densities of
the pair of EEGs. We considered the spectral merger algorithm for clustering EEGs,
which updates the spectral estimate of the cluster from a weighted average of the spectral
estimate obtained from each EEG in the cluster. Numerical experiments suggest
that the SMC method performs very well in producing the correct clusters.
When applied to resting state EEG data, the SMC method gave results that partly
confirms the segmentation based on the anatomy of the cortical surface. In
addition, the method showed how some regions, though not contiguously connected
on the cortical surface, are spectrally synchronized during resting state. Moreover,
the method demonstrates that brain organization, as expressed in cluster formation,
evolves over resting state.
\end{abstract}

\vspace{1in}


\section{Introduction.}
Most of research in neuroscience is currently focused on how populations of neurons respond
to external stimuli or how they behave during resting state. Brain activity following
stimulus presentation and even during resting state are often the result of highly coordinated
responses of large numbers of neurons both locally (within each region) and globally (across
different brain regions) [\cite{Fingel05}]. Most studies on resting state are based on
functional magnetic resonance imaging data where most analyses are centered on functional
and effective connectivity. In this paper, we study brain activity during resting state
using electroencephalograms (EEG). Coordinated activity of neurons can give rise to
macroscopic oscillations which are projected on the scalp. These oscillations are captured by
EEGs which are brain signals that indirectly measure brain neuronal
activity. While EEG signals have excellent temporal resolution, they suffer from poor spatial
resolution. Hence, activity measured at point on the scalp (i.e., a channel) is believed to be
mostly due to distributed neuronal activity on a cortical patch directly perpendicular to the EEG
sensor.

There is a strong interest in studying the coordinated response of a
population of neurons because, as noted, it is often required to execute tasks such as
learning, and memory formation, retention and retrieval. A population of neurons is said to act in ``unison"
if their responses are synchronized (they are highly cross-correlated in time) and/or spectrally synchronized
(they share similar oscillations or waveforms). Moreover, disruptions in coordination have been
observed in a number of neurological and mental diseases such as schizophrenia, obsessive compulsive disorder,
epilepsy and Alzheimer's disease. The main goal in this paper is to identify brain regions
that are spectrally synchronized during resting state.

Time and frequency domain analyses of scalp EEG recordings are used to track changes in brain states. An excellent overview for spectral analysis of general neural signals is given in
\cite{Pesaran}.  \cite{Emery2011}  developed methods that track loss and recovery of consciousness under general anesthesia
using Bayesian state-space algorithm and \cite{Purdon2013} proposed the use of global coherence, the ratio
of the largest eigenvalue to the sum of the eigenvalues of the cross-spectral matrix at a given frequency
and time, to analyze the spatiotemporal dynamics of multivariate time-series. \cite{Krafty2015} studied
time-varying relationship between delta EEG sleep and high frequency heart rate variability (HF-HRV)
in women and examined the effects of sleep apnea and
insomnia symptoms. This paper is one of the few studies that analyze the resting state
brain using EEGs.

To characterize the impact of an external stimulus on coordinated neuronal activity, one approach
is to contrast brain response following stimulus presentation against brain activity during resting
state (or some control condition) which is commonly employed in most functional magnetic resonance
imaging studies. The aspects of ``brain activity" most commonly investigated in
many imaging modalities are changes in mean level of activation and connectivity (mostly
as measured by synchronicity). In this paper, our goal is slightly different from fMRI studies
since we are studying brain activity using a different modality. We seek to cluster EEG channels
that are spectrally synchronized or that exhibit similar spectral densities. We shall adopt the notion of
similarity in EEGs as follows. A pair of EEGs is considered to be highly spectrally synchronized if
both are dominated by similar frequency oscillations. However, a pair of EEGs are not
in spectrally synchronized if one EEG is dominated by low frequency oscillations but they
show very different frequency content

We developed the spectral merger clustering (SMC) method that produces clusters of EEG channels according to the similarity
of their spectra. The resulting clusters serve as a proxy for segmenting the
brain cortical surface since the EEGs capture neuronal activity over a locally distributed
region on the cortical surface. To measure the extent of desynchronicity between a pair of EEGs,
we shall use the total variation distance (TVD) which gives the largest possible difference
between the spectral densities of the pair of EEGs under consideration. TVD has been shown to
perform well as a similarity measure in oceanography [\cite{EOyA2}]. Due to its generality,
we demonstrate that it also works very well in clustering EEG data.

Our procedure for clustering EEGs uses the following hierarchical spectral merger algorithm,
which updates the spectral estimate of the cluster from a weighted average of the spectral
estimate obtained from each EEG in the cluster. While other approaches perform spectral
clustering separately over different frequency bands, our proposed
approach actually takes the spectra in its entirety (i.e., all the frequency bands simultaneously).
This is one advantage of the SMC method: the clusters are formed objectively by taking the distance
across all frequencies rather than by cherry-picking specific frequency bands which could produce
different clustering for different frequency bands. Our proposed approach is more sensible because
each channel is considered to be a unit and hence it is either fully included in a cluster or fully
excluded from a cluster -- based on the entire spectral curve. Hierarchical clustering algorithms
with linkage functions (such as complete, average, ward, and so on) are based on geometrical ideas
where the distances between new clusters and old ones are computed by a linear combination of the distance of their members, which
are not meaningful for clustering time series since these linear combinations do not have a meaning
in terms of the spectral densities. Clustering brain data based on linear Gaussian models have been
used, \cite{Maitra2011, Neumann2008}. However, our main interest is spectral synchronicity and small
changes in coefficients of the linear Gaussian models can produce big changes in the spectral density.
Our method, on the other hand, is rigorously based on spectral theory. It is intuitive and optimal in the following respects.
First, from the hierarchical merger algorithm, concatenating time series gives rise to updated
spectral estimates over finer frequency resolution. Second, in the hierarchical spectral merger algorithm, the updated spectral estimates obtained by
weighted average of the spectral estimates in the cluster, are smoother, less noisy and hence gives
better estimates of the TVD.

Clustering procedures have been used for classification purpose in earthquake and explosion data.
\cite{kakizawa1998} proposed a clustering procedure based on the Kullback-Leibler and Chernoff information measures between the spectral density matrices.
However, when the signal is non-stationary, \cite{Shumway2003} proposed to use the time varying spectrum and a locallized version of the Kullback-Leibler measure.
Also, \cite{Ombao2004} considered the non-stationary case using the time-varying spectrum.
They develop a discriminant scheme based on the Kullback-Leibler divergence between the
SLEX spectra of the different classes. Evolutionary clustering algorithms are also discussed by \cite{Chakra2006}, they
proposed a generic framework for this problem and evolutionary versions of $K$-means and agglomerative hierarchical clustering.

In this paper we consider that EEGs were recorded over the 3 minutes of resting state and then
segmented into 1-second non-overlapping epochs. One of the key scientific questions of
interest is to determine whether the clustering of the channels is stationary over the
the entire resting state. We will address this problem by comparing the cluster over
the early phase (first 60 epochs) and the latter phase (last 60 epochs) of resting state.
Our clustering procedure is applied for each epoch and a ``averaged result'' using the clusterings
obtained for each epoch is used to compare phases (Section \ref{S5.3}).

The remainder of the paper is organized as follows. In Section 2, we provide the background on
spectral theory that is essential for clustering time series. The spectral merger clustering
(SMC) method is developed in Section 3. In Section 4, we compare the SMC method with an
agglomerative hierarchical clustering procedure using the complete linkage function and
several similarity measures of frequent use in time series clustering. Finally, in Section 5,
we apply the SMC method to the resting state EEG data from two participants in the study.
We conclude our paper with a discussion about
limitations of the current work and future directions.

\section{Background on Analysis of EEGs}
Spectral analysis of time series is a natural approach for studying EEG data because it
identifies frequency oscillations that dominate the signal. In this section we present some basic
concepts of spectral analysis; details are discussed in \cite{ShumStof}, \cite{Brockwell}, \cite{Brillinger} and \cite{Priestley}.
\vspace{0.20in}

\noindent \textit{The Spectral Density.}
Define the autocovariance function of a zero mean weakly stationary time series $X_t$ to be
$\gamma(h)$ which is assumed to be absolutely summable, i.e., $\displaystyle \sum_{h=-\infty}^{\infty} |\gamma(h)|<\infty$. The auto-covariance function admits the representation
\begin{equation}
 \gamma(h)=\int_{-1/2}^{1/2}f(\omega)\exp(2\pi i \omega h)\mbox{d}\omega \qquad h=0,\pm 1,\pm 2,\ldots
\end{equation}
as the inverse transform of the spectral density $f(\omega)$, which is given by
\begin{equation}\label{ECSD}
 f(\omega)=\sum_{h=-\infty}^{\infty} \gamma(h)\exp(-2\pi i \omega h).
\end{equation}

\vspace{0.20in}
\noindent \textit{Remark:} The above representation yields the variance decomposition of the time series.
Note that setting the lag $h=0$ gives $\gamma(0)=\mbox{Var}(X_t)=\int_{-1/2}^{1/2} f(\omega)\mbox{d}\omega$. Thus, the total variance
in the time series is decomposed over the frequency domain, where the spectrum at frequency $\omega$
can be roughly interpreted as the variance contributed by the oscillation in a narrow frequency
band around $\omega$ which we further expound below.

Recall that $X_t$ is a mean-zero stationary process. Define its spectral distribution to be
$F(\omega)$. Then there exists a complex-valued stochastic process $Z(\omega)$, on the interval
$\omega \in [−1/2, 1/2]$, having zero mean stationary uncorrelated increments, such that $X_t$ can be
written as the stochastic integral
\begin{equation}\label{SD2}
X_t=\int_{-1/2}^{1/2} \exp(2\pi i \omega t)\mbox{d}Z(\omega),
\end{equation}
where, for $−1/2 \leq \omega_1 \leq \omega_2 \leq 1/2$, $\mbox{var}(Z(\omega_2)-Z(\omega_1))=F (\omega_2)-F(\omega_1)$. Equation (\ref{SD2}) can be interpreted as follows ``any stationary time series
can be looked at as a sum of infinitely many cosine and sine waveforms with random coefficients''.
This is the key idea in spectral analysis. It has many application in neuroscience because EEG
signals can be looked as a superposition of components oscillating at different frequencies.
The range of frequency that can be observed in a signal depends on the sampling frequency, usually measured in Hertz (number of cycles per second). Moreover, the convention for the different frequency bands are as follows: delta (0-4 Hz), theta (4-8 Hz), alpha (8-12 Hz), beta (12-30 Hz) and gamma (30-50 Hz).

\subsection{An example for modeling band oscillations in EEGs}

One way to study time series data is to decompose it into different frequency
oscillations. To motivate these oscillations at specific frequency bands, we shall
consider the second order auto-regressive (AR(2)) model which is defined to be
\begin{equation}
 Z_t=\phi_1 Z_{t-1}+ \phi_2 Z_{t-2}+ \epsilon_t,
\end{equation}
where $\epsilon_t$ is a white noise process. The characteristic polynomial for this model is $\phi(z)= 1-\phi_1 z-\phi_2 z^2$. The roots of the polynomial equation indicate the properties of the
oscillations. If the roots, denoted $z_0^{1}$ and $z_0^{2}$ are complex-valued then they
have to be complex-conjugates, i.e., $z_0^{1}=\overline{z_0^{2}}$. These roots have a polar
representation
\begin{equation}\label{AR2}
 |z_0^{1}|=|z_0^{2}|=M, \qquad \qquad \arg(z_0)= \frac{2 \pi \eta}{F_s},
\end{equation}
where $F_s$ denotes the sampling frequency; $M$ is the amplitude or magnitude of the root
($M >1$ for causality); and $\eta$ is the frequency index. The spectrum of the AR$(2)$ process
with polynomial roots above will have peak frequency at $\eta$. The peak becomes broader as $M \to \infty$; it becomes narrower as $M \to 1^{+}$.

\begin{figure}
\centering
\includegraphics[scale=.5]{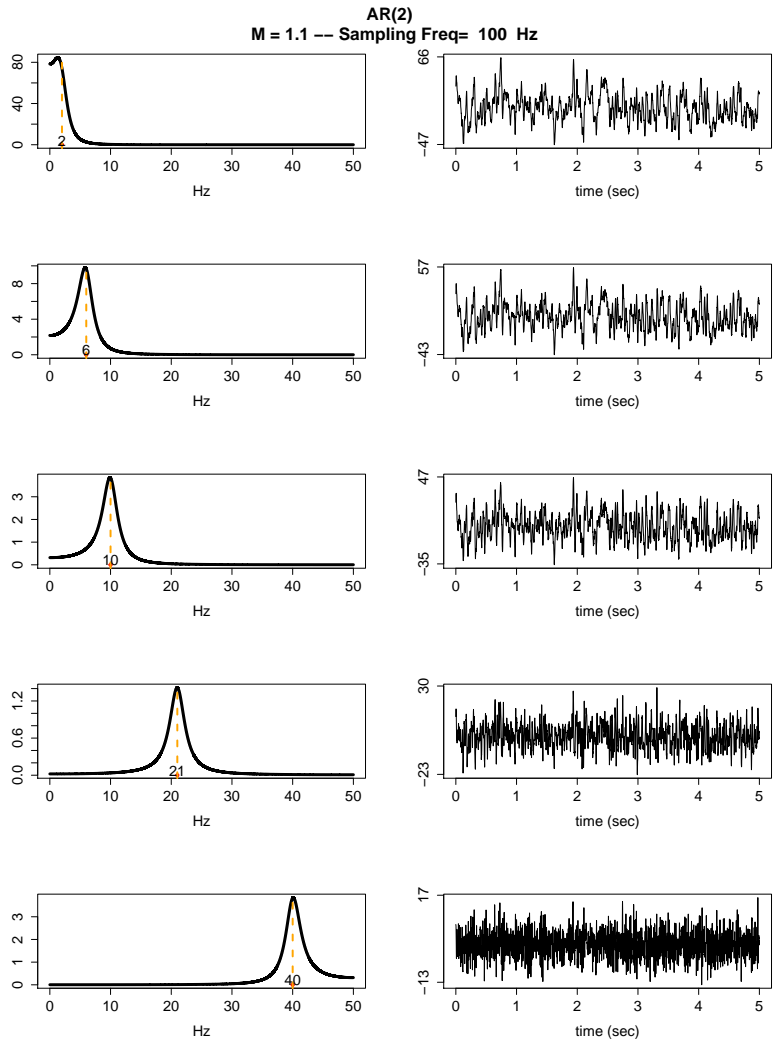}
\caption{Left: Spectra for the AR(2) process with different peak frequency; $\eta=2,6,10,21,40$ which correspond to delta, theta, alpha, beta and gamma frequencies.
Right: Realization from the corresponding AR(2) process.}\label{F2}
\end{figure}

To illustrate the type of oscillatory patterns that can be observed in time series
from processes with corresponding spectra,  we plot in Figure \ref{F2} the spectra (left)
for different values of $\eta$, $M=1.1$ and $F_s=100 Hz$; and the generated time series (right).
Larger values of $\eta$ gives rise to faster oscillation of the signal.

\subsection{Nonparametric Estimation of the Spectrum}
\begin{figure}
\centering
\subfigure[ \label{F3a}]{\includegraphics[scale=.35]{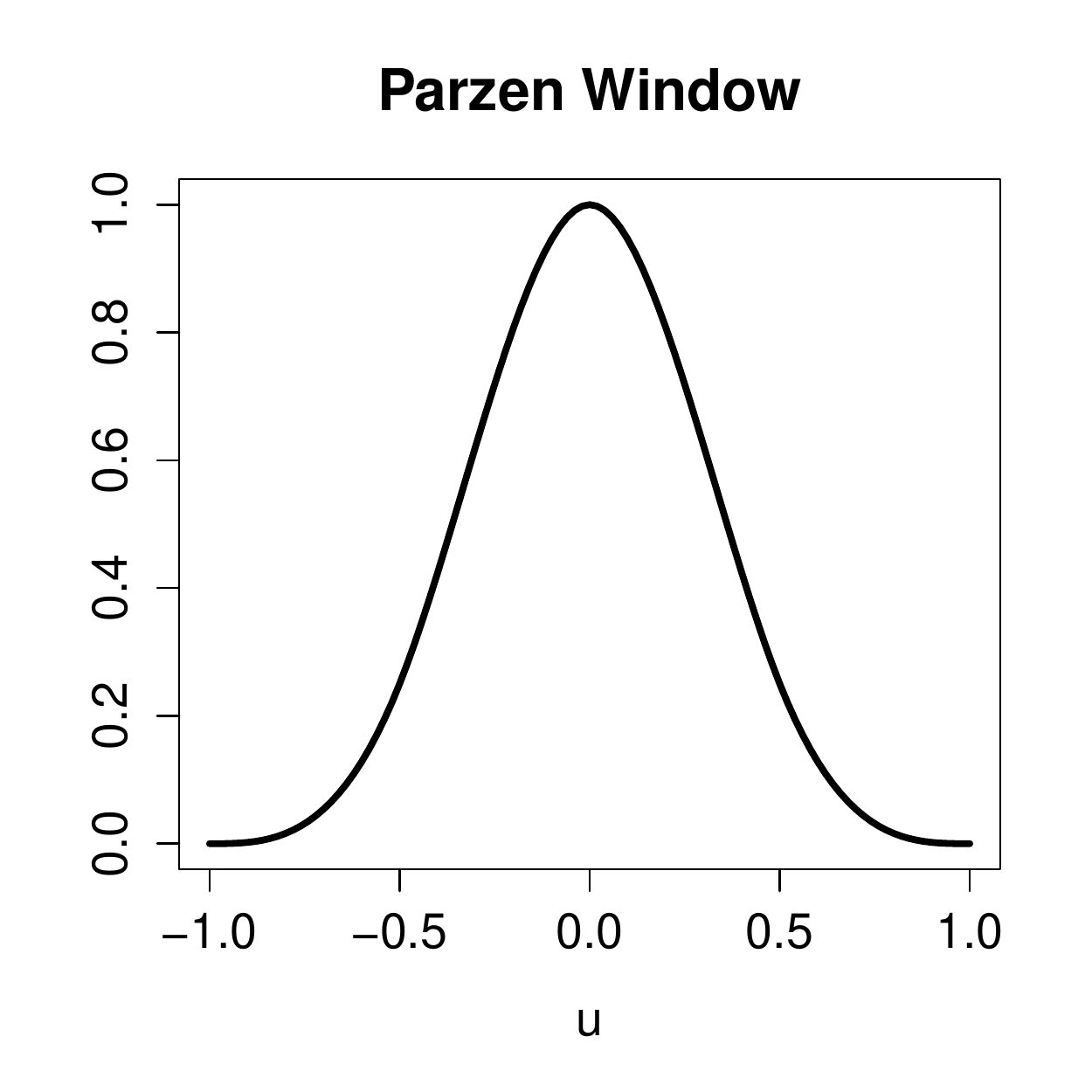}}
\subfigure[ \label{F3b}]{\includegraphics[scale=.35]{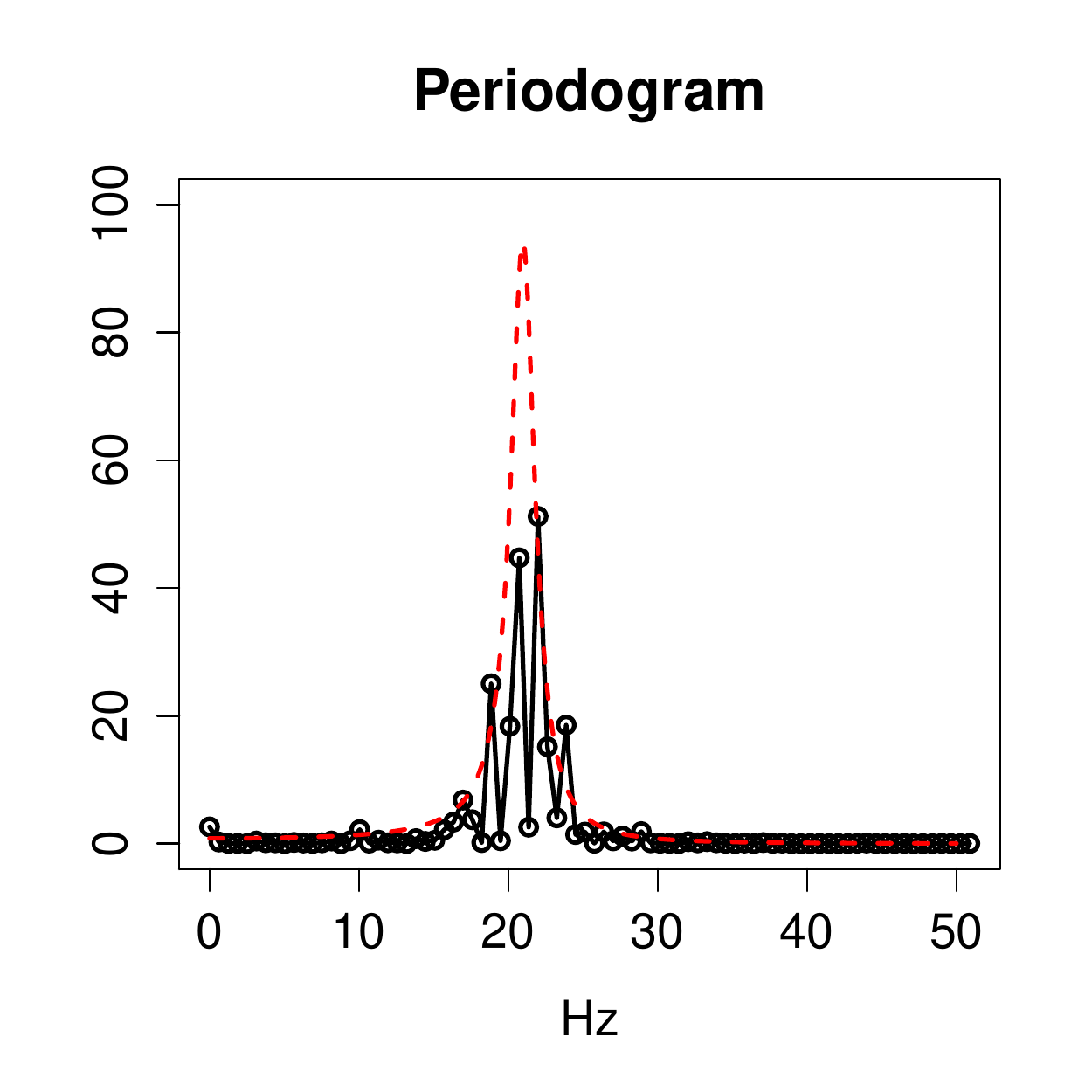}}
\subfigure[ \label{F3c}]{\includegraphics[scale=.35]{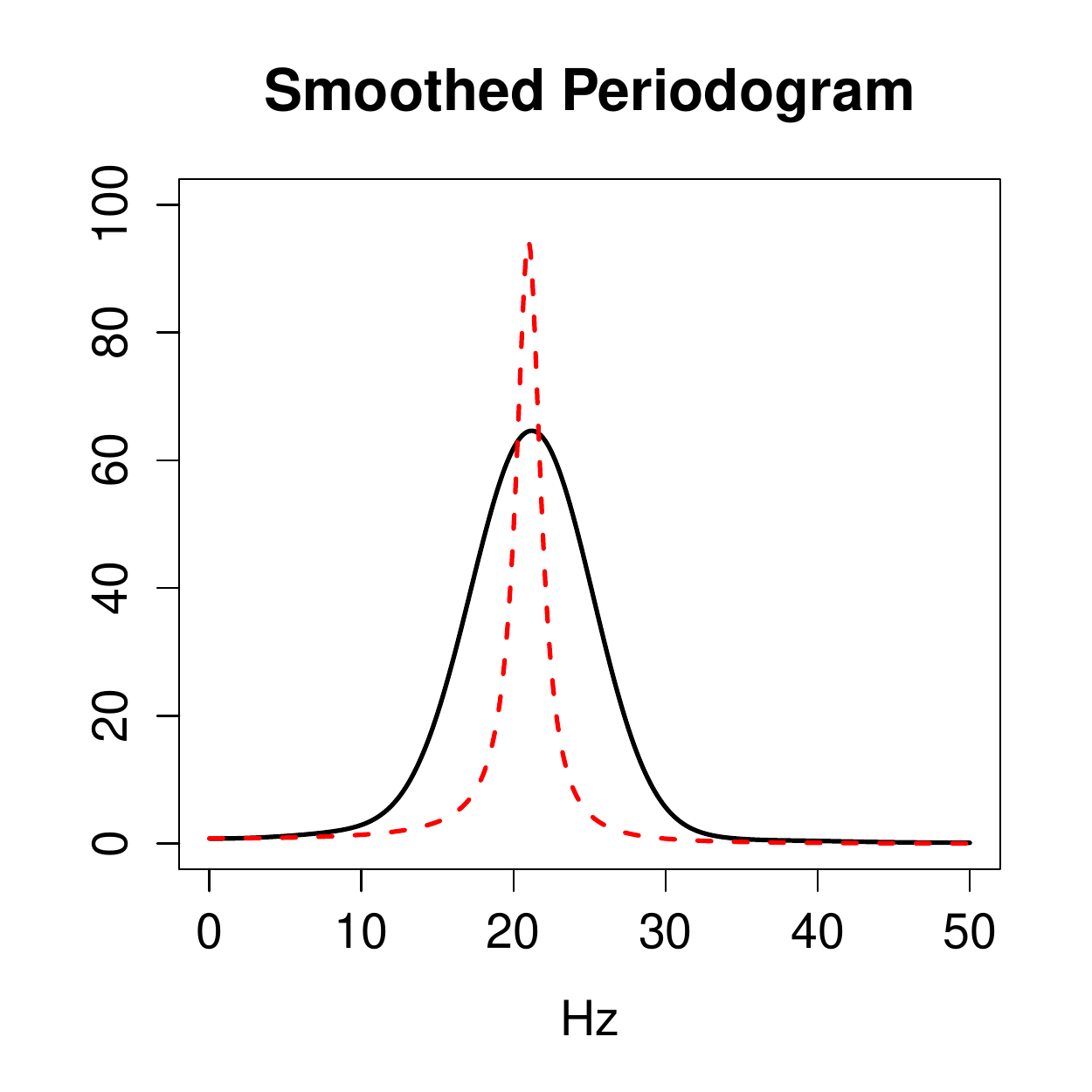}}
\caption{Nonparametric estimation of the spectral density: (a) Parzen window, (b) Periodogram and (c) Smoothed Periodogram with a bandwidth of 100. The red dashed curve is the real spectra }
\end{figure}

Based on Equation (\ref{ECSD}), a natural nonparametric estimator for the spectral density is the
periodogram which is defined as
\begin{equation}
I_X(\omega_k) = T^{-1} \Big|\sum_{t=1}^T X_t e^{-i 2\pi\omega_k t}\Big|^2=\sum_{h=-(n-1)}^{n-1} \widehat{\gamma}(h) e^{- i 2\pi \omega_k h}
\end{equation}
for a stationary and centered time series $X_t$, $t=1, \ldots, T$, at the fundamental Fourier
frequencies $\omega_k= k/T, \ k=1, \dots, n$ with $n=\lfloor(T-1)/2\rfloor$. This estimator
is asymptotically unbiased but the variance does not go to $0$ as $T$ increases. For this reason and
to provide an estimator that is a continuous function on the interval $(-0.5, 0.5)$, we work with the
smoothed periodogram. Here, we use the Parzen window with bandwidth $a$ for smoothing (which is a type of a lag window estimator), so our estimator only uses the real part and is defined as
\begin{equation}\label{Parzen}
\tilde{f}(\omega)= (2 \pi)^{-1} \sum_{|h|\leq a} w(h/a)\widehat{\gamma}(h)e^{-i 2 \pi \omega h},
\end{equation}
where $\widehat{\gamma}$ is the sample covariance and the Parzen window is defined as
\begin{equation*}
 w(u)=\begin{cases}
       1-6|u|^2 +6|u|^3, & \mbox{if  } |u|<\frac{1}{2}\\
       2(1-|u|)^3, & \mbox{if  } \frac{1}{2}\leq |u| \leq 1, \\
       0, & \mbox{otherwise.}
      \end{cases}
\end{equation*}
This window is plotted in Figure \ref{F3a}; the estimation process is represented in Figure \ref{F3b};
the smoothed periodogram obtained with a Parzen window with bandwidth $a=100$ is in Figure \ref{F3c}. We estimated the spectral density using a time series of length $T=1000$ with sampling frequency of $100$ Hz, the real spectral density is the red dashed curve. The smoothed periodogram has less noise and recovers the true peak frequency.

\section{SMC Method for Clustering Multi-channel Electroencephalograms}

Our goal is to develop a method that finds groups or clusters that represent spectrally synchronized channels. We developed a new time series clustering algorithm.  \cite{Liao05} classified time series clustering algorithms in three different types: methods
based on comparison of raw data, feature-based methods and methods based on models adjusted to the data. Our proposal, the spectral merger clustering (SMC) method, is feature-based and the spectral density of the time series is considered the central feature for classification purposes. Clustering methods seek to find groups in data so that members of the same group are similar, while members of different groups are as dissimilar as possible. The signal recorded at each channel will be characterized by its spectral density and our proposed method for clustering EEGs will produce groups with channels having the similar spectral densities, i.e channels that are spectrally synchronized.

Other clustering approaches have been applied in brain signals. \cite{Lachiche2005} developed a method for fMRI where the area between the variations of the signals around their means was used as similarity measure together with a Growing Neural Gas (GNG) clustering algorithm. For  detecting phase synchronization of analog signals of the cerebral cortex, the shortest distance on a torus between time series of phase differences was used with a modified K-means cluster algorithm [\cite{Hutt09}]. Our proposal is similar to the above in the following sense; 1) the SMC method can be applied to signals with different lengths and 2) the SMC method does not restrict the clusters to be spatially close. However, the SMC method integrates the similarity measure across all frequencies which is more general than the approaches mentioned before.

To build a clustering  algorithm the first question is how to measure the similarity between spectral densities. We propose the use of the total variation distance (TVD) as a measure of similarity. This measure has shown good results in simulation studies carried out to compare it with other similarity measures based on the spectrum. These results will be presented in section \ref{CS}.

\subsection{Total Variation Distance}
In general, the TVD can be defined for any pair of probability measures that are defined on the same $\sigma$-algebra of sets. We will be focus here in the case when these probability measures have density functions with respect to the Lebesgue measure.
The TVD between two densities, $f$ and $g$, is defined
\begin{eqnarray}\label{TVd_1}
d_{TV}(f,g)=1-\int\min\{f(\omega),g(\omega)\}\mbox{d}\omega.
\end{eqnarray}
In comparison with other similarity measures, the TVD has some desirable properties.
\begin{itemize}
 \item The TVD is a metric. It satisfies symmetry and the triangle inequality which are two reasonable properties expected from a similarity measure. In this sense the TVD may be a better choice than the Kullback-Leibler divergence.
 \item The TVD is bounded, $0 \leq d_{TV}(f,g)\leq 1$ and can be interpreted in terms of the common area between two densities. Having a bounded range ($[0,1]$) is a desirable feature, since this gives a very intuitive sense to the values attained by the similarity measure. A value near $0$ means that the densities are similar while a value near $1$ indicates they are highly dissimilar. In contrast, both the $L^2$ distance and the Kullback-Leibler divergence are not bounded above and thus lack this intuitive interpretation.
\end{itemize}

Since spectral densities are not probability densities, they have to be normalized by dividing the estimated spectral density by the sample variance $\widehat\gamma(0)$; we are going to denote $\tilde{f}(\omega)=\widehat{g}(\omega)/\widehat\gamma(0)$ the normalized estimated spectral density. Since neuroscientists are more interested in the range of frequencies where the energy is transferring than in the quantity of energy transfered, normalizing the spectral densities does not affect the questions of interest.

\subsection{Proposal: Hierarchical Merger Algorithm}
There are two general families of clustering algorithms: partitioning and hierarchical. Among partitioning algorithms, K-means and K-medoids are the two more representative, and for the hierarchical clustering algorithms, the main examples are agglomerative with single-linkage or complete-linkage [\cite{Xu05}]. The clustering algorithms proposed in this work, hierarchical merger and the hierarchical spectral merger, are a modification of the usual agglomerative hierarchical procedure, taking advantage of the spectral point of view for the analysis of the time series.

Let $X_c=(X_c(1),\ldots,X_c(T))$ be EEG at channel $c$, $c=1,\ldots,N$.
The procedure starts with $N$ clusters, each cluster being a unique channel.
\begin{enumerate}
 \item[ Step 1] Estimate the spectral density for each cluster using the smoothed periodogram
 \item[ Step 2] Compute the TVD between their spectra.
 \item[ Step 3\label{St2}] Find the two clusters that have lower TVD, save this value as a characteristic.
 \item[ Step 4\label{St3}] Merge the signals in the two closest clusters and replace the two clusters by this new one.
 \item[ Step 5] Repeat steps 1-4 until there is only one cluster left.
\end{enumerate}

The characteristic saved in Step \ref{St2} represents the ``cost'' of joining two clusters, i. e., having $k-1$ clusters vs $k$ clusters, and the decision on the final number of clusters is based
on these costs. If a significantly large value is observed, then it may be reasonable to choose
$k$ clusters instead of $k-1$.

\begin{table}
\begin{center}
\small
 \begin{tabular}{l}
 \hline
 \textbf{Algorithm:}\\
 \hline \hline \\
\begin{minipage}{5in}
 \begin{enumerate}
 \item  Initial clusters: $\mathbf{C}=\{C_i\}$,  $C_i=X_i$, $i=1,\ldots,N$ \\
 Dissimilarity matrix between clusters $i$ and $j$, \\
 $$D_{ij}=d(C_i,C_j):=d_{TV}(\widehat{f}_i,\widehat{f}_j)$$ $\#$ TVD between the estimated spectra using the signals in each cluster.
 \item \textbf{for} $k$ $in$ $1: N-1$
 \item \hspace{.5cm}$\displaystyle (i_m,j_m)=\arg\!\min_{ij} D_{ij}$\hspace{2.3cm}       $\quad \#$Find the closest clusters
 \item \hspace{.5cm}$\displaystyle min_k=\min_{ij} D_{ij}$
 \item \hspace{.5cm}$C_{new}=C_{i_m} \cup C_{j_m}$\hspace{2.8cm}                         $\quad \#$Join the closest clusters
 \item \hspace{.5cm}$D^{new}=D\setminus \{D_{i_m .} \cup D_{j_m .} \cup D_{. i_m} \cup D_{. j_m}\}$ $\#$Delete rows and columns $i_m,j_m$
 \item \hspace{.5cm}$D_{(N-k)j}^{new}=d_{TV}(C_{new},C_j)$\hspace{1.7cm}    $\quad \#$Compute the TVD
 \item \hspace{.5cm}$D_{i(N-k)}^{new}=d_{TV}(C_i,C_{new})$\hspace{1.7cm}    \hspace{.25cm}$ \#$between new cluster and old ones
 \item \hspace{.5cm}$D = D^{new}$\hspace{4.5cm}    $\#$New matrix $D$
 \item \hspace{.5cm}$\displaystyle \mathbf{C}= \left(\mathbf{C}\setminus \{C_{i_m},C_{i_m}\} \right) \cup C_{new}$ \hspace{1.5cm} $\#$New clusters
 \item \textbf{end}
\end{enumerate}
\end{minipage}
\\
\\
\hline
 \end{tabular}
\caption{Hierarchical Merger Algorithm proposed using the total variation distance and the spectrum.}\label{Algo}
\end{center}
 \end{table}

When two clusters merge, there are two options for estimating the spectral density for the new cluster, either (1.) for the hierarchical merger algorithm, we concatenate both signals and compute the smoothed periodogram with the concatenated signal; or (2.) for the hierarchical spectral merger algorithm, we take the weighted average over all the estimated spectra for each signal in the cluster as the new estimated spectra. It is intuitive and optimal in the following respects.
First, from the hierarchical merger algorithm, concatenating time series gives rise to updated
spectral estimates over finer frequency resolution. Second, in the hierarchical spectral
merger algorithm, the updated spectral estimates obtained by a weighted average of the
spectral estimates in the cluster, are smoother, less noisy and hence give better estimates of the TVD. Both algorithms compute the TVD between the new cluster and the old clusters based on an updated
estimated spectra. We applied both algorithms to simulations and some of the epochs in real data and the results were very similar. However, the procedure may be chosen based on which is more reasonable for the application.  Table \ref{Algo} gives a summary of the hierarchical merger algorithm.

\vspace{0.20in}
\noindent \textit{Remark:} The hierarchical merger algorithm can be useful for applications where the data is observational
and does not come from an experimental design , i.e., they are not responses from some intervention or a stimulus presentation.
Recordings from ocean waves are an example of observational data. However, EEG traces have a well-defined
starting time point which is anchored to the presentation of a stimulus or the start of the trial,
so the hierarchical spectral merger algorithm is more reasonable in this case.
\vspace{0.20in}

Our proposal is model-free and based on the spectrum. While a hierarchical algorithm has a dissimilarity matrix of size $N\times N$ during the whole algorithm, our proposed method reduces this size to $(N-k) \times (N-k)$ at the $k$-th iteration. In a hierarchical algorithm the distance between old and new clusters is computed using linear combinations of the values in the initial dissimilarity matrix, while in our proposal all the series in a cluster are used for estimating the spectral density, and then a new value of the TVD, based on an updated estimated spectra, is computed. The TVD is used as a dissimilarity measure and also as a criterion for selecting the possible number of clusters.

\begin{figure}[ht]
\centering
\subfigure[ \label{F36a}]{\includegraphics[scale=.35]{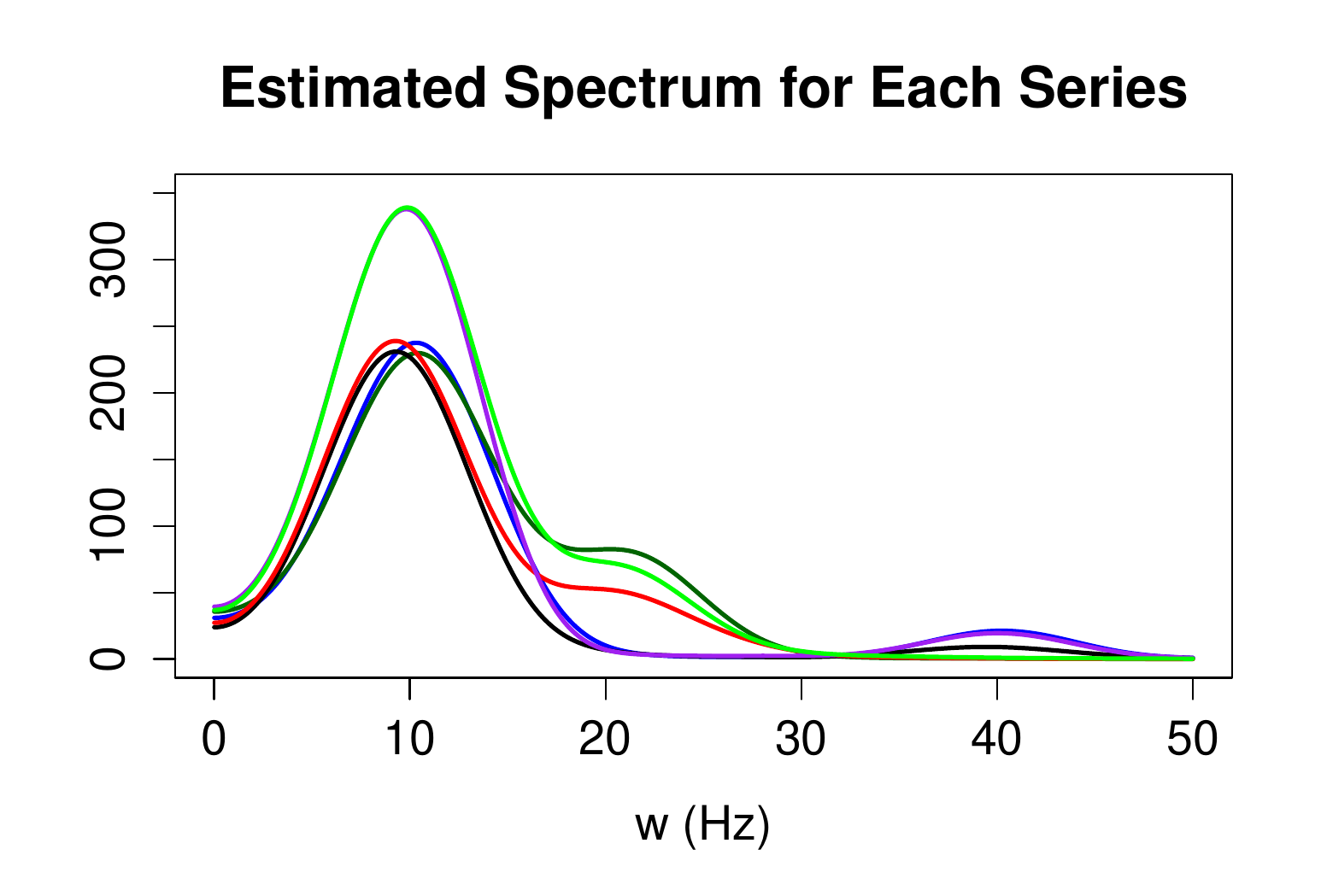}}
\subfigure[ \label{F36b}]{\includegraphics[scale=.35]{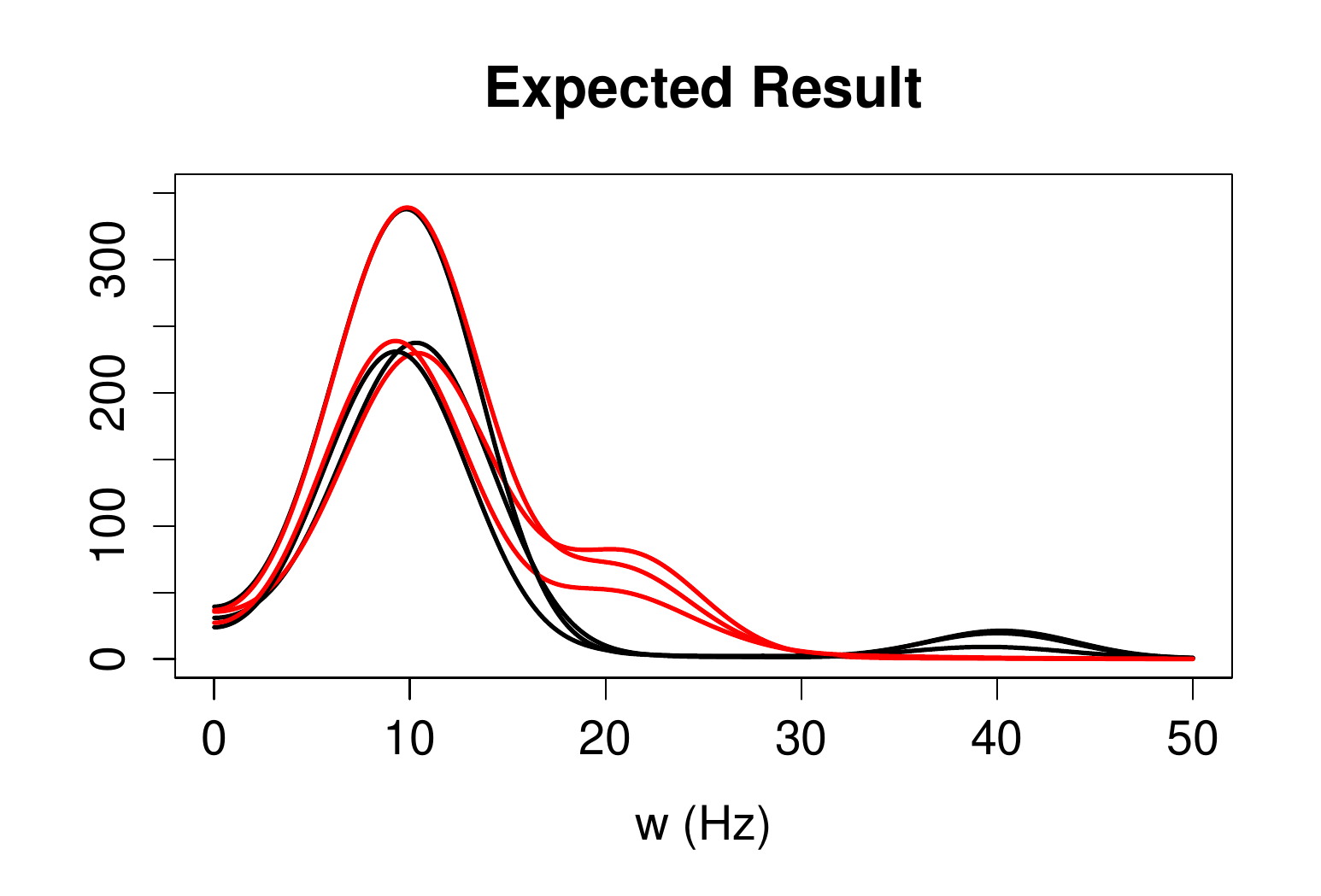}}
\caption{Estimated spectra. (a.) Different colors correspond to different time series, (b.) Red spectra are from the AR(2) model with activity at alpha and beta bands and black spectra are from the AR(2) model with activity at alpha and gamma bands.}
\end{figure}

\begin{figure}
\centering
\subfigure[ \label{F37c}]{\includegraphics[scale=.42]{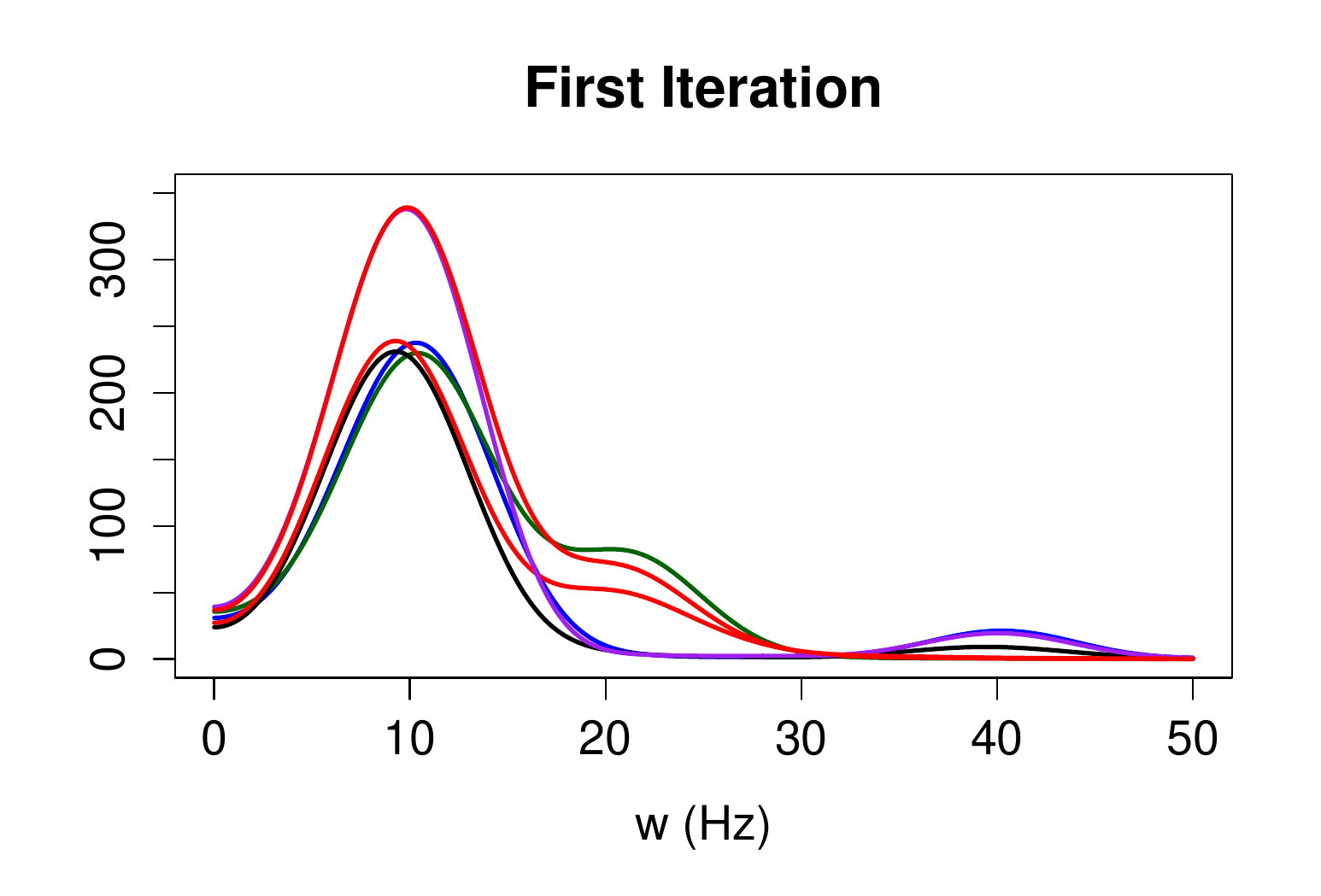}}
\subfigure[ \label{F37d}]{\includegraphics[scale=.42]{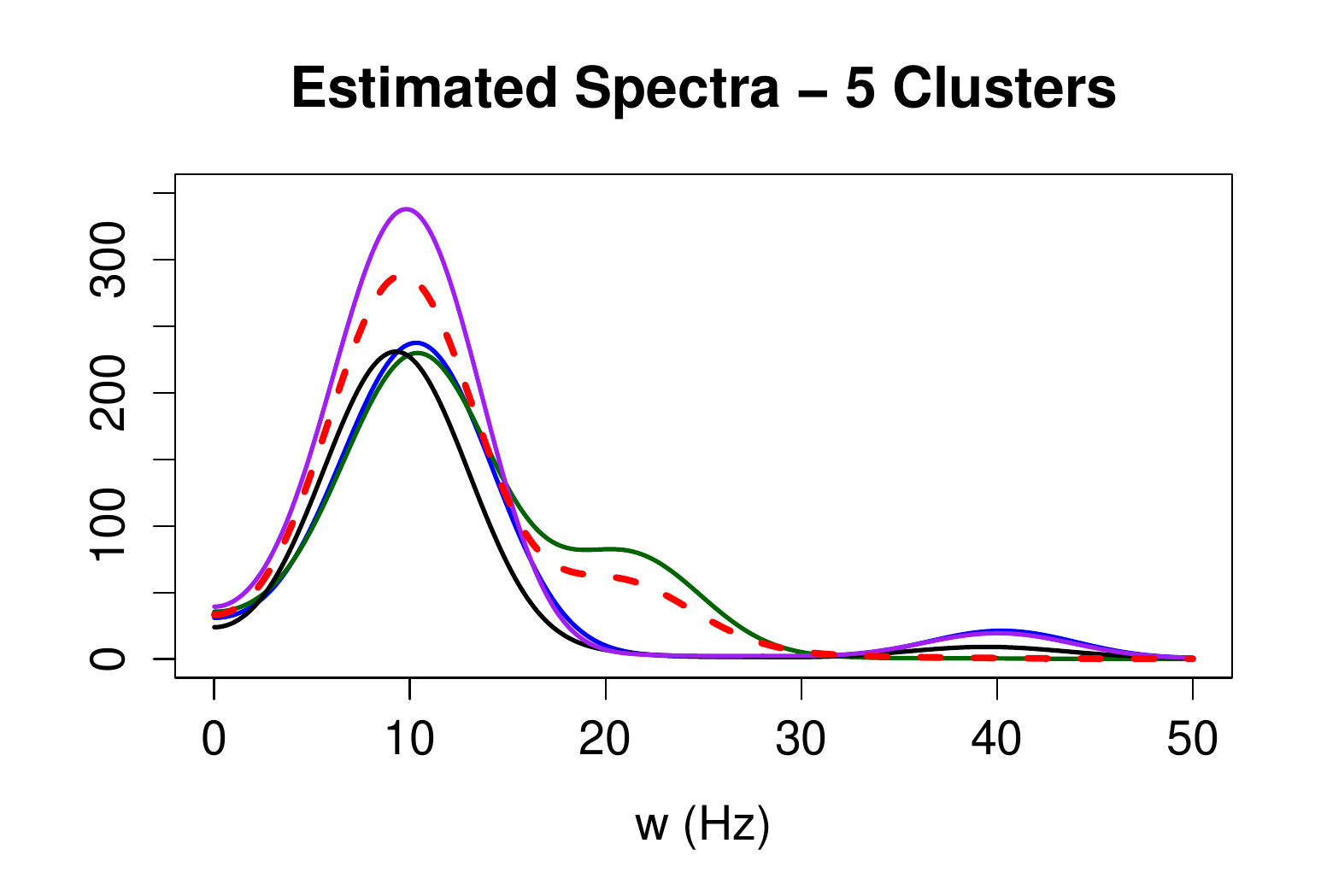}}
\subfigure[ \label{F37e}]{\includegraphics[scale=.42]{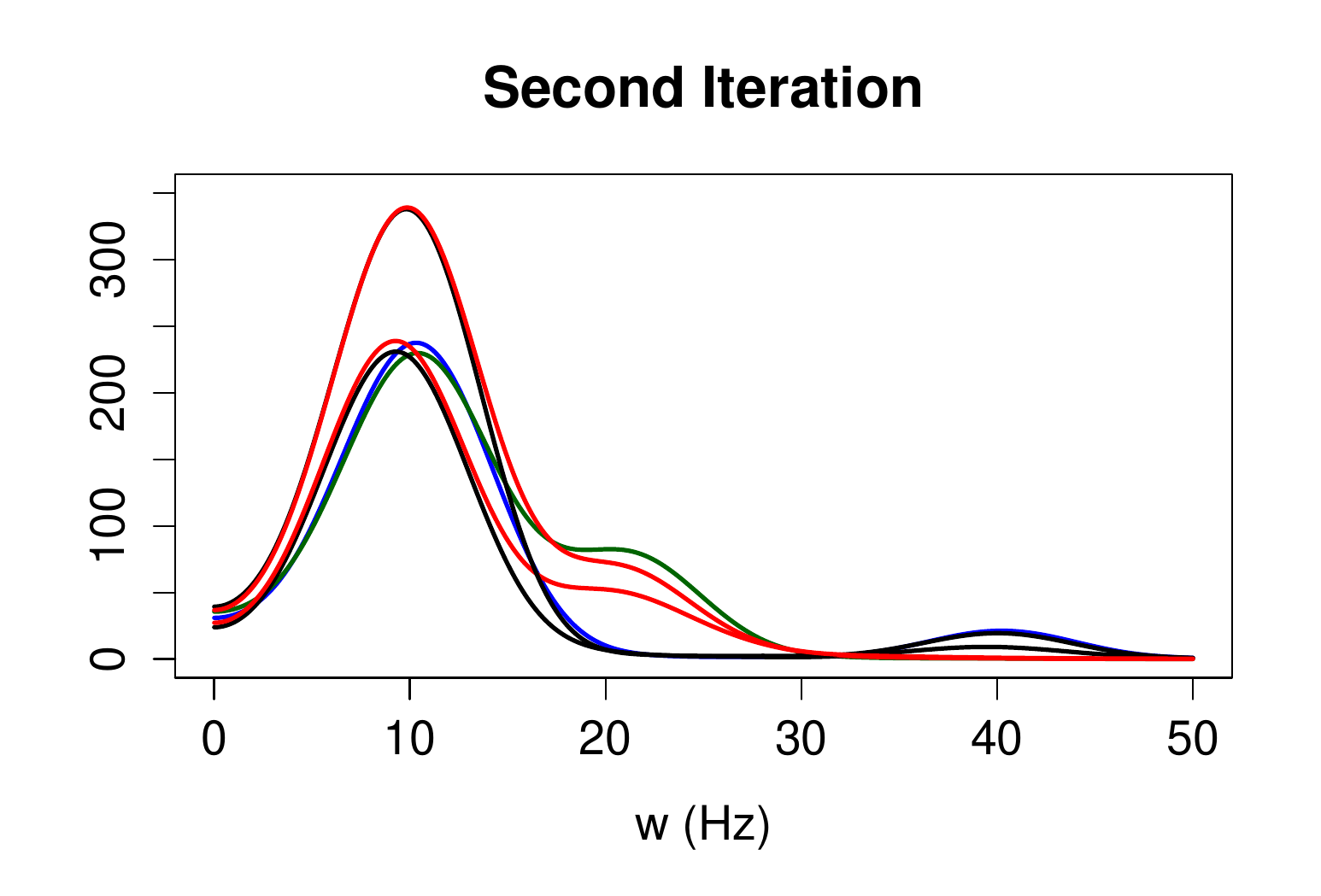}}
\subfigure[ \label{F37f}]{\includegraphics[scale=.42]{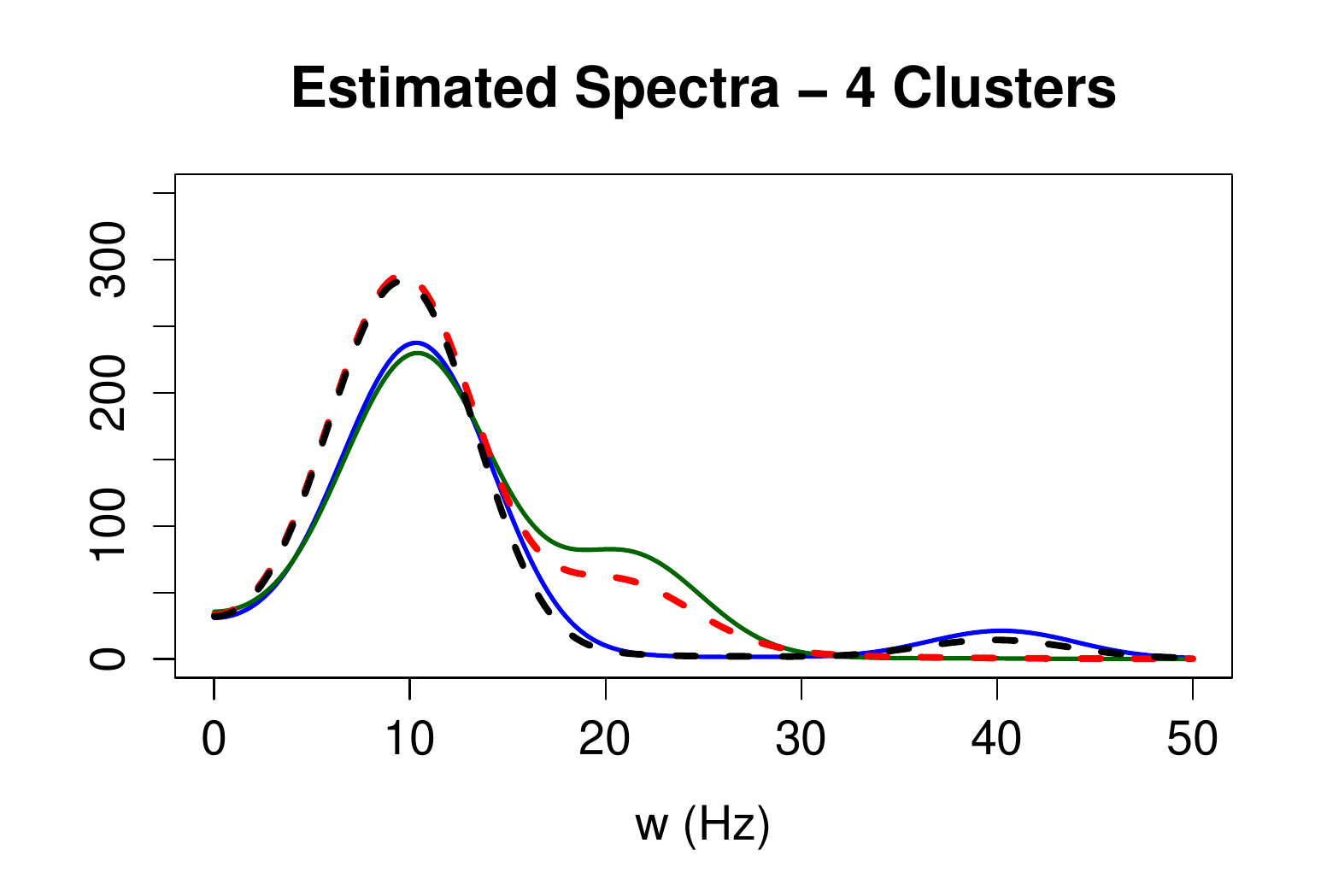}}
\subfigure[ \label{F37g}]{\includegraphics[scale=.42]{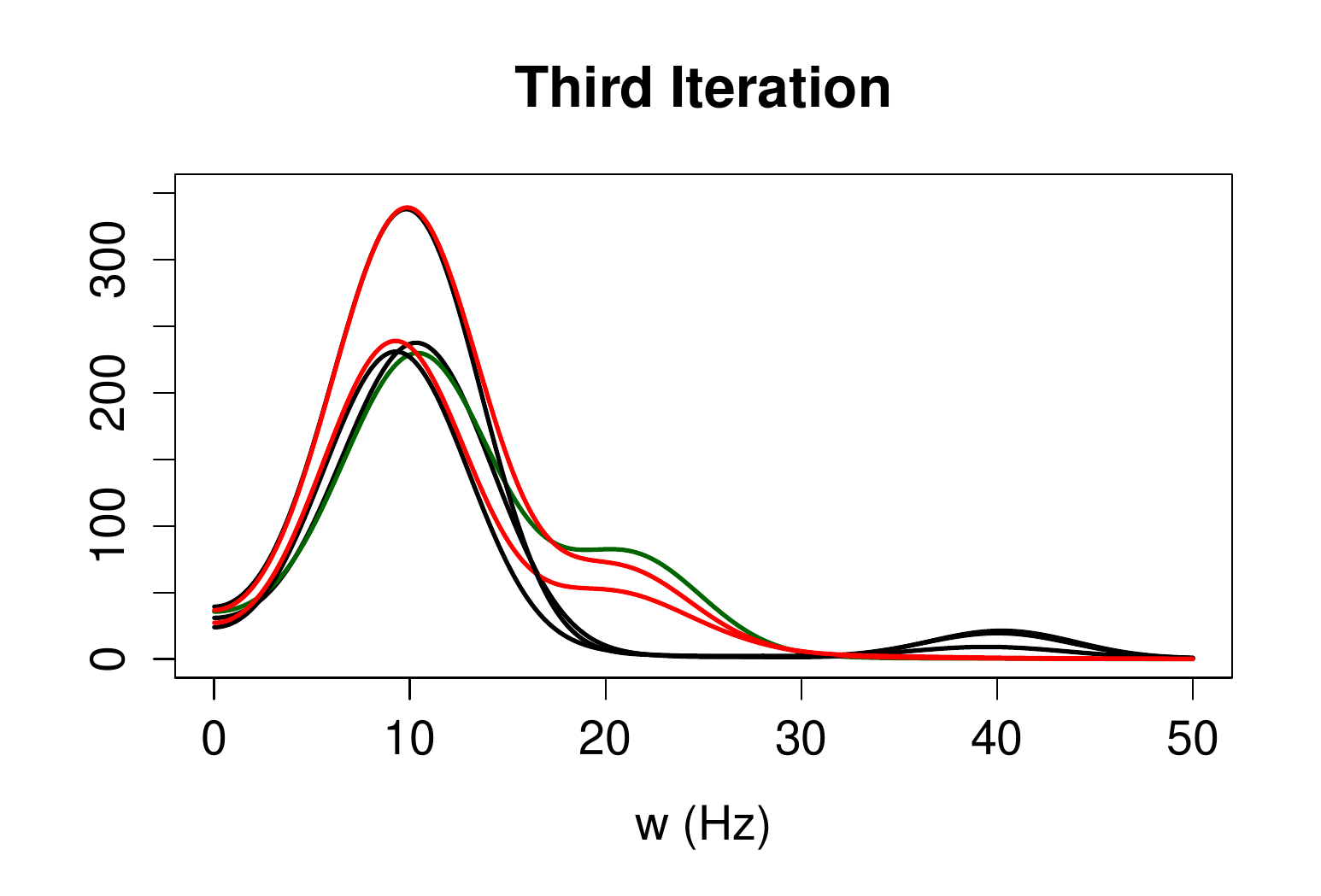}}
\subfigure[ \label{F37h}]{\includegraphics[scale=.42]{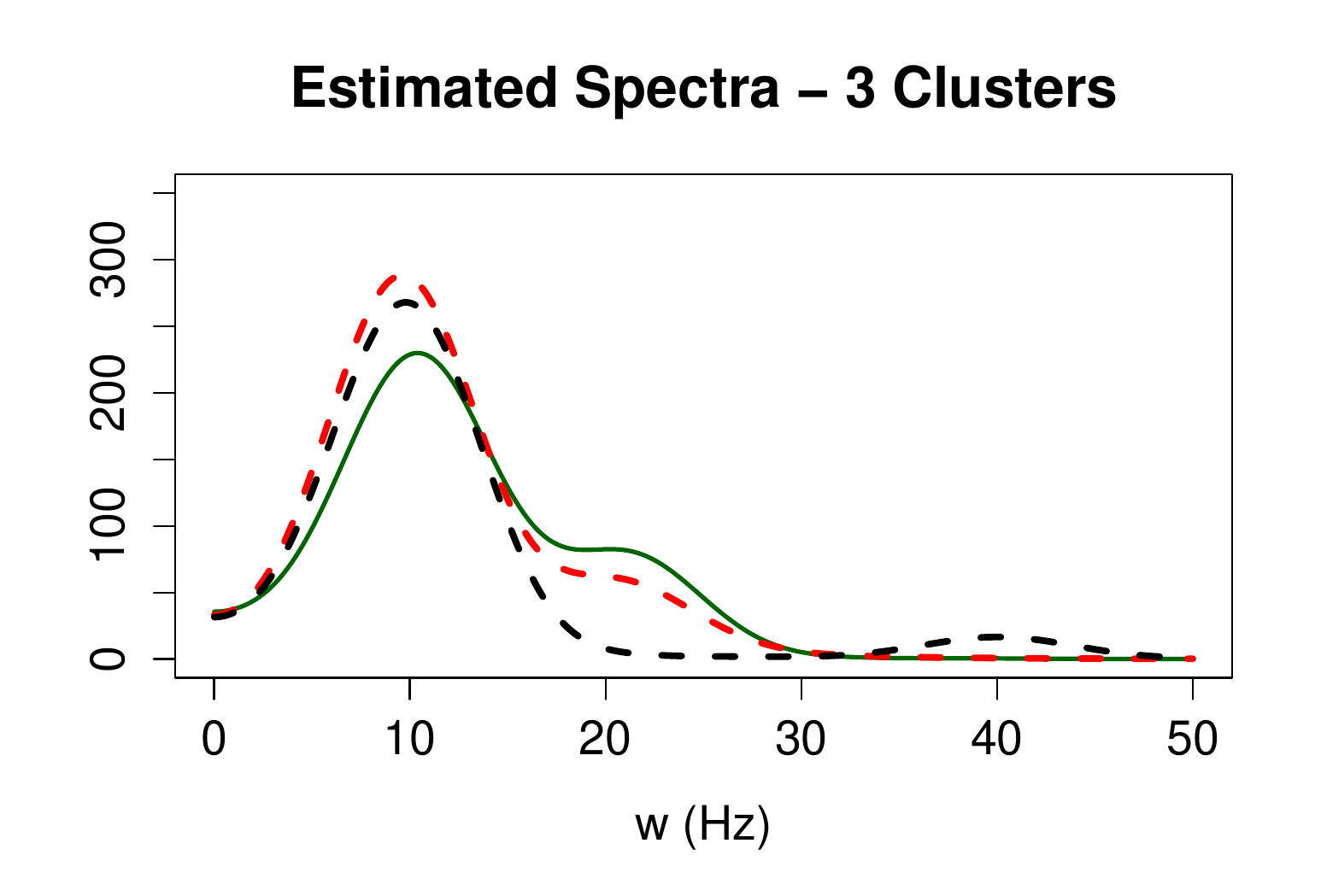}}
\subfigure[ \label{F37i}]{\includegraphics[scale=.42]{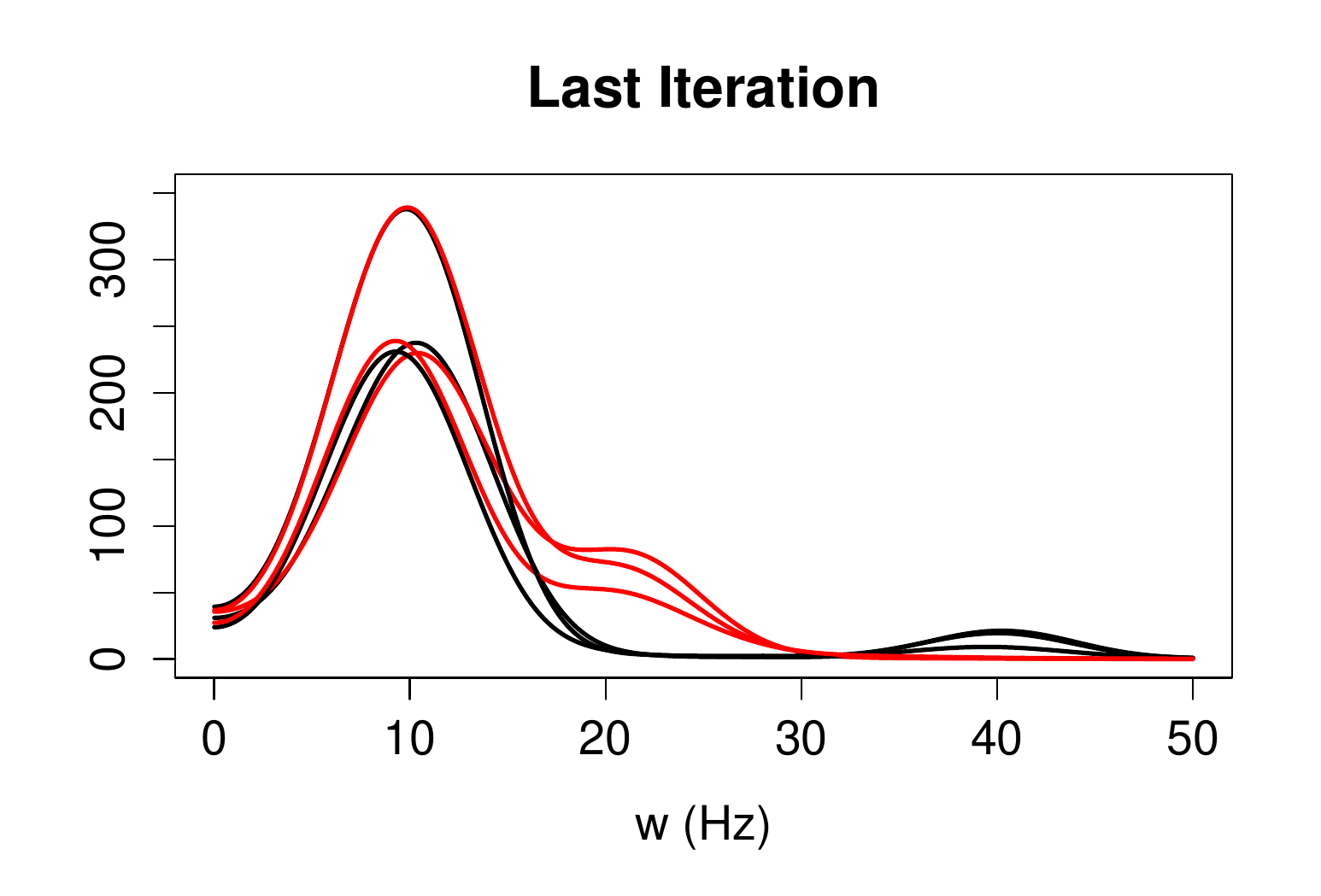}}
\subfigure[ \label{F37j}]{\includegraphics[scale=.42]{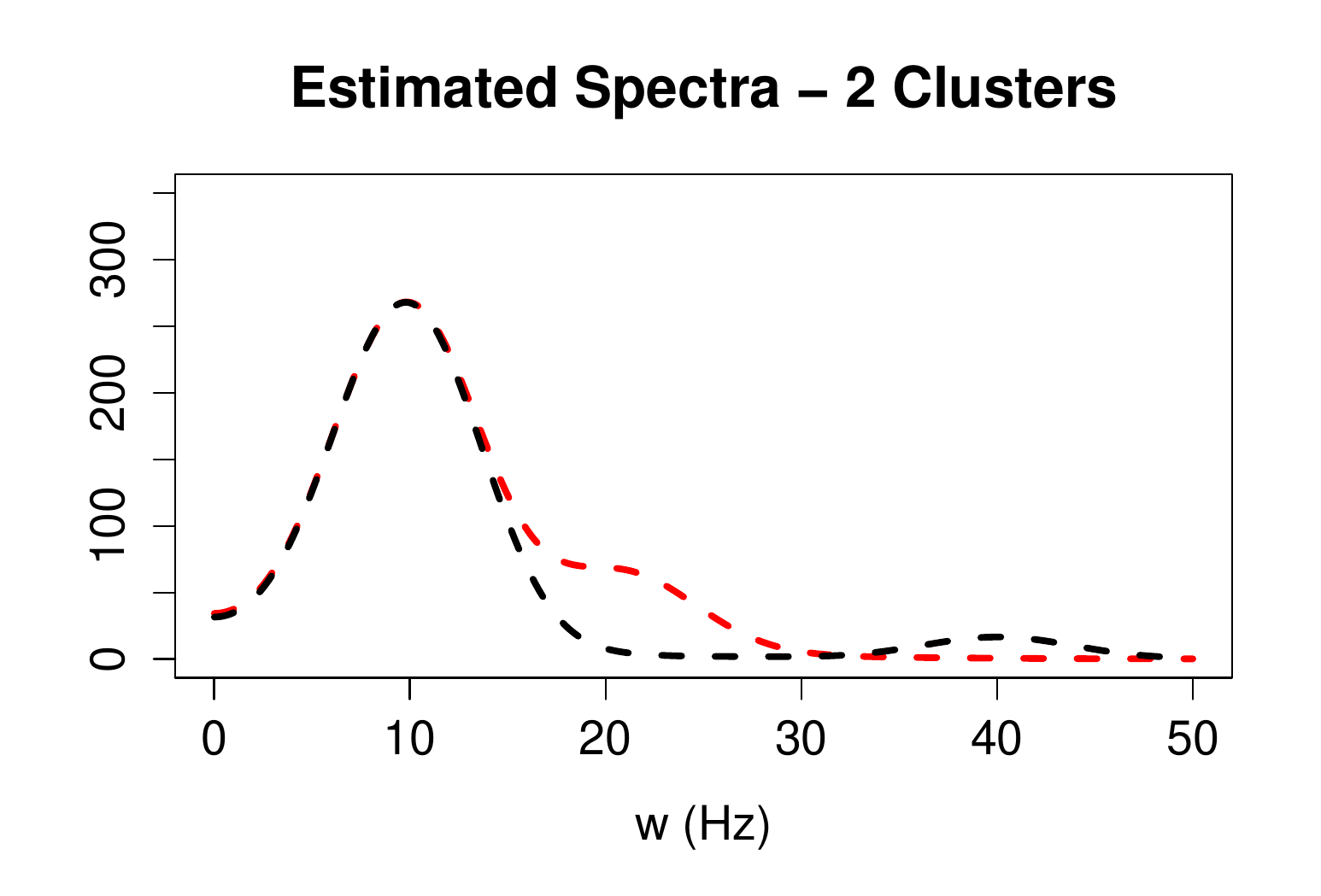}}
\caption{Dynamic of the hierarchical merger algorithm. (a), (c), (e) and (g) show the clustering process for the spectra. (b), (d), (f) and (h) show the evolution of the estimated spectra which improves when we merge the series on the same cluster.}\label{F37}
\end{figure}

To illustrate the SMC method, consider two different AR(2) models with their spectra concentrated on the alpha band (at 10 Hz), however one also contains power in the beta band (at 21 Hz) while the other has power in the gamma band (at 40 Hz). We sample three time series for each process, 10 seconds of each one with a sampling frequency of $100$ Hz ($t=1,\ldots,1000$). Figure \ref{F36a} shows the estimated spectra for each series and Figure \ref{F36b} shows by different color (red and black) which one belongs to the first or second process. If we only look at the spectrum, it could be hard to recognize the number of clusters and their memberships. We probably could not identify some cases, like the red and purple spectrum.

The dynamic of the SMC method is shown in Figure \ref{F37}. We start with six clusters; at the first iteration we find the closest spectra, represented in Figure \ref{F37c} with the same color (red). After the first iteration we merge these series and get $5$ estimated spectra, one per cluster, Figure \ref{F37d} shows the estimated spectra where the new one is represented by the dashed red curve. We can follow the dynamic in Figures \ref{F37e}, \ref{F37f}, \ref{F37g} and \ref{F37h}. By the end of our clustering algorithm the result, Figure \ref{F37i}, is the same as the expected, Figure \ref{F36b}. Also, the estimated spectra for the two clusters, Figure \ref{F37j}, is better than any of the initial spectra and we can identify the dominant frequency bands for each cluster.

\subsection{Further intuition: spectral synchronicity vs. coherence}

\begin{figure}
\centering
\subfigure[ \label{F0a}]{\includegraphics[scale=.15]{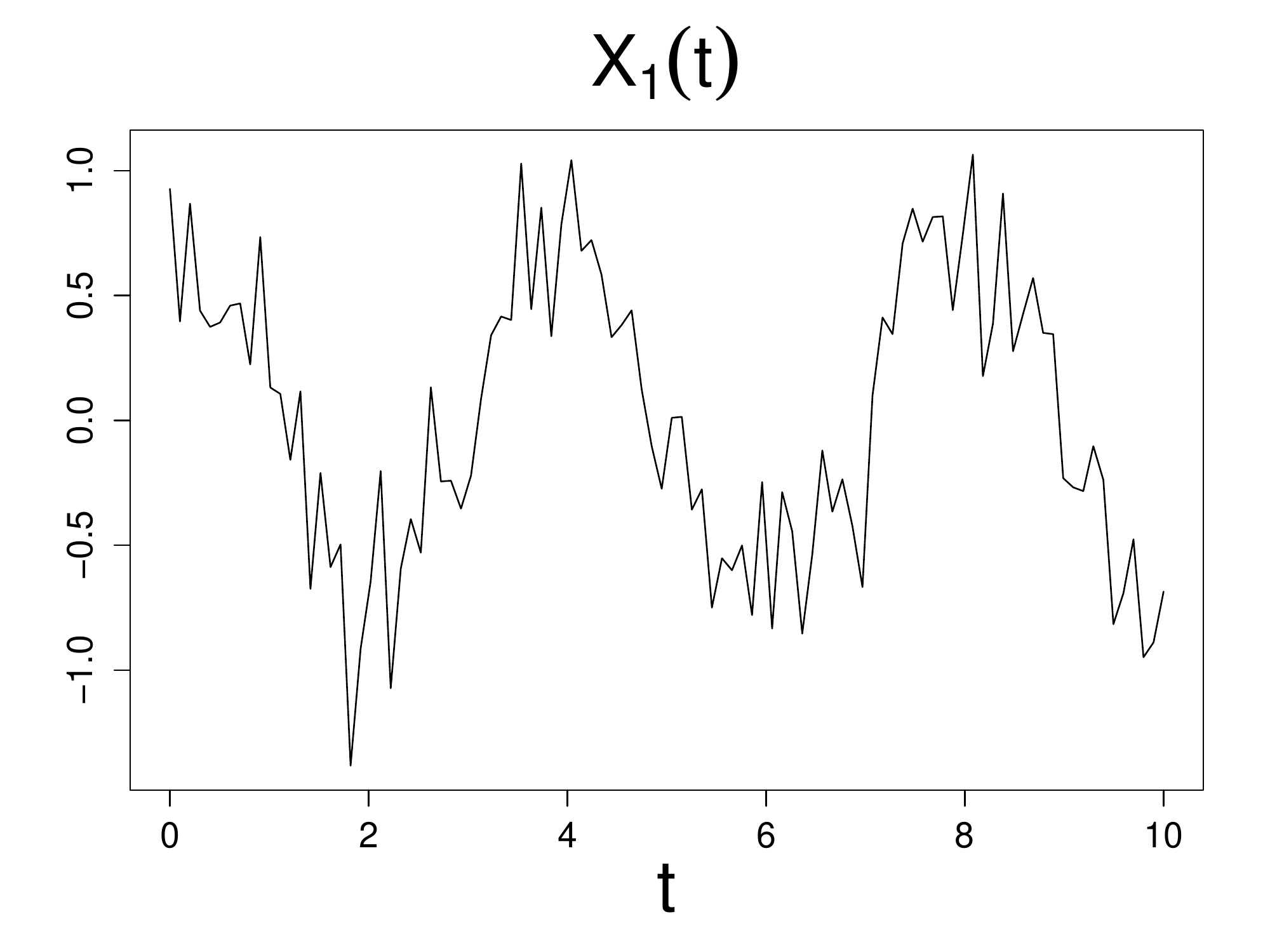}}
\subfigure[ \label{F0b}]{\includegraphics[scale=.15]{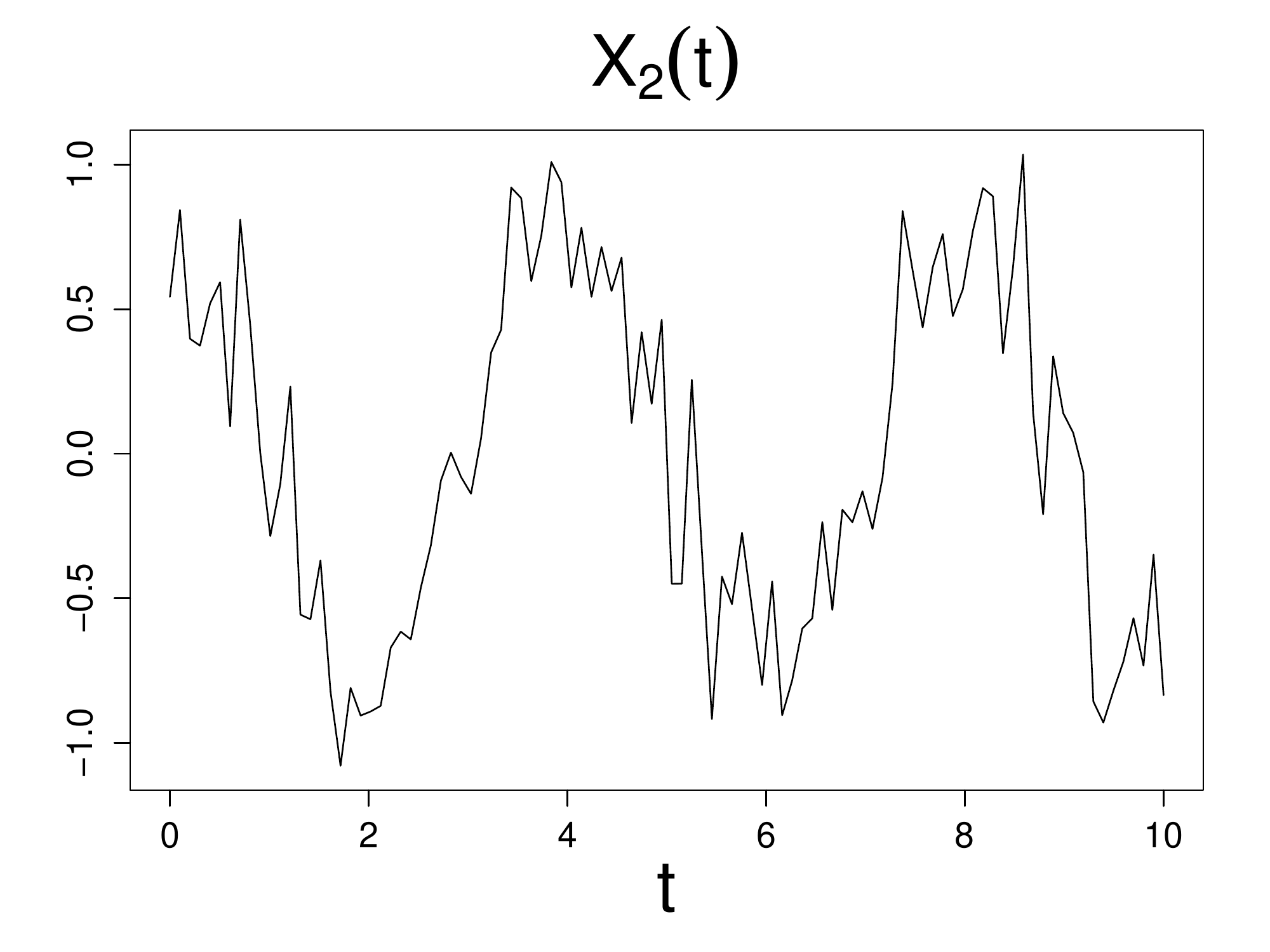}}
\subfigure[ \label{F0d}]{\includegraphics[scale=.15]{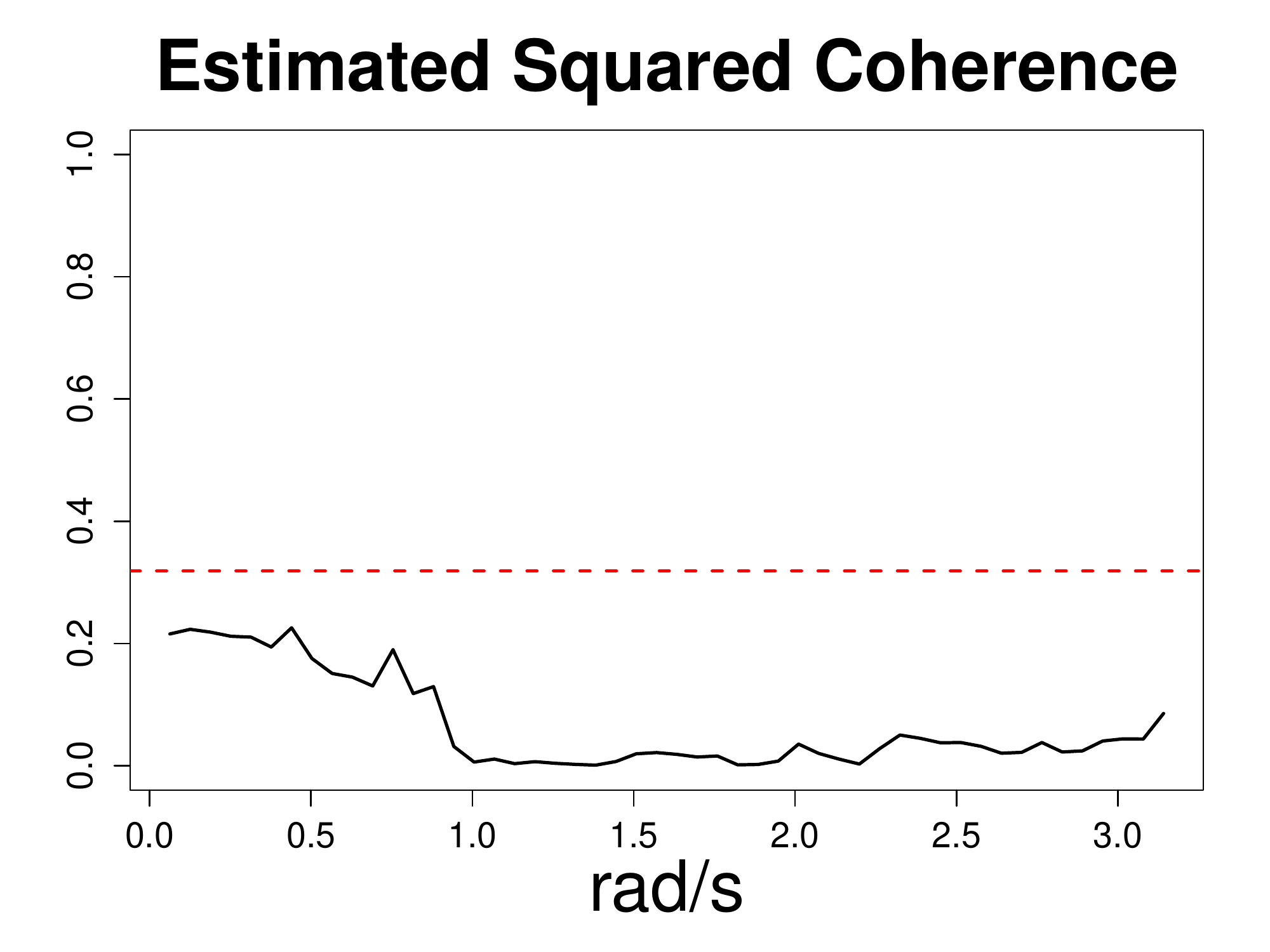}}
\subfigure[ \label{F0c}]{\includegraphics[scale=.15]{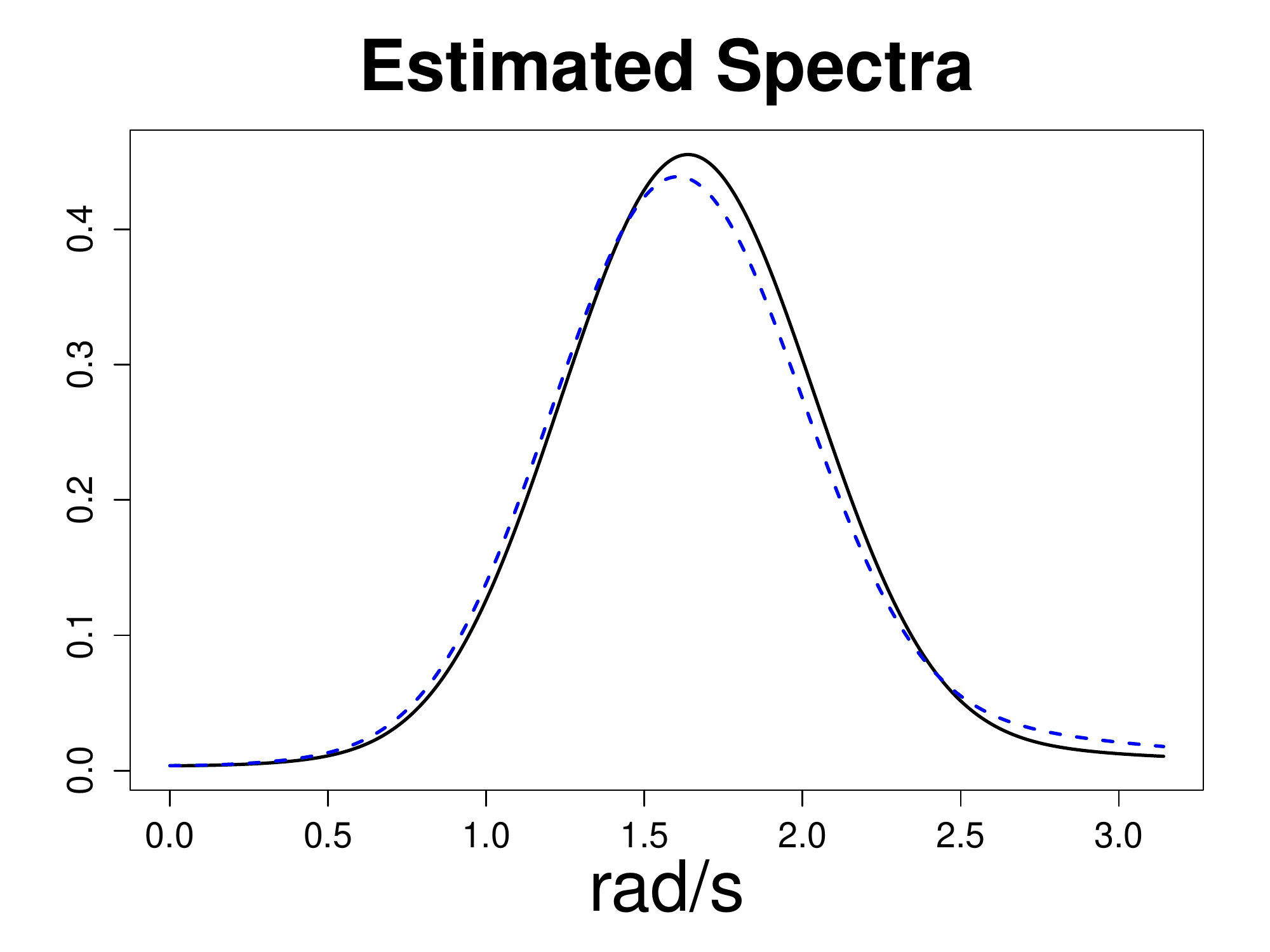}}
\caption{(a.) Plot of time series $X_1(t)$. (b.) Plot of time series $X_2(t)$.
(c.) Estimated squared coherence, red dashed line corresponds to the threshold for 
rejecting zero coherence. (d.) Estimated spectra for each signal, black continuous curve corresponds to $X_1(t)$
and blue dashed curve corresponds to $X_2(t)$.}\label{F0}
\end{figure}

\begin{figure}
\centering
\subfigure[ \label{F01a}]{\includegraphics[scale=.15]{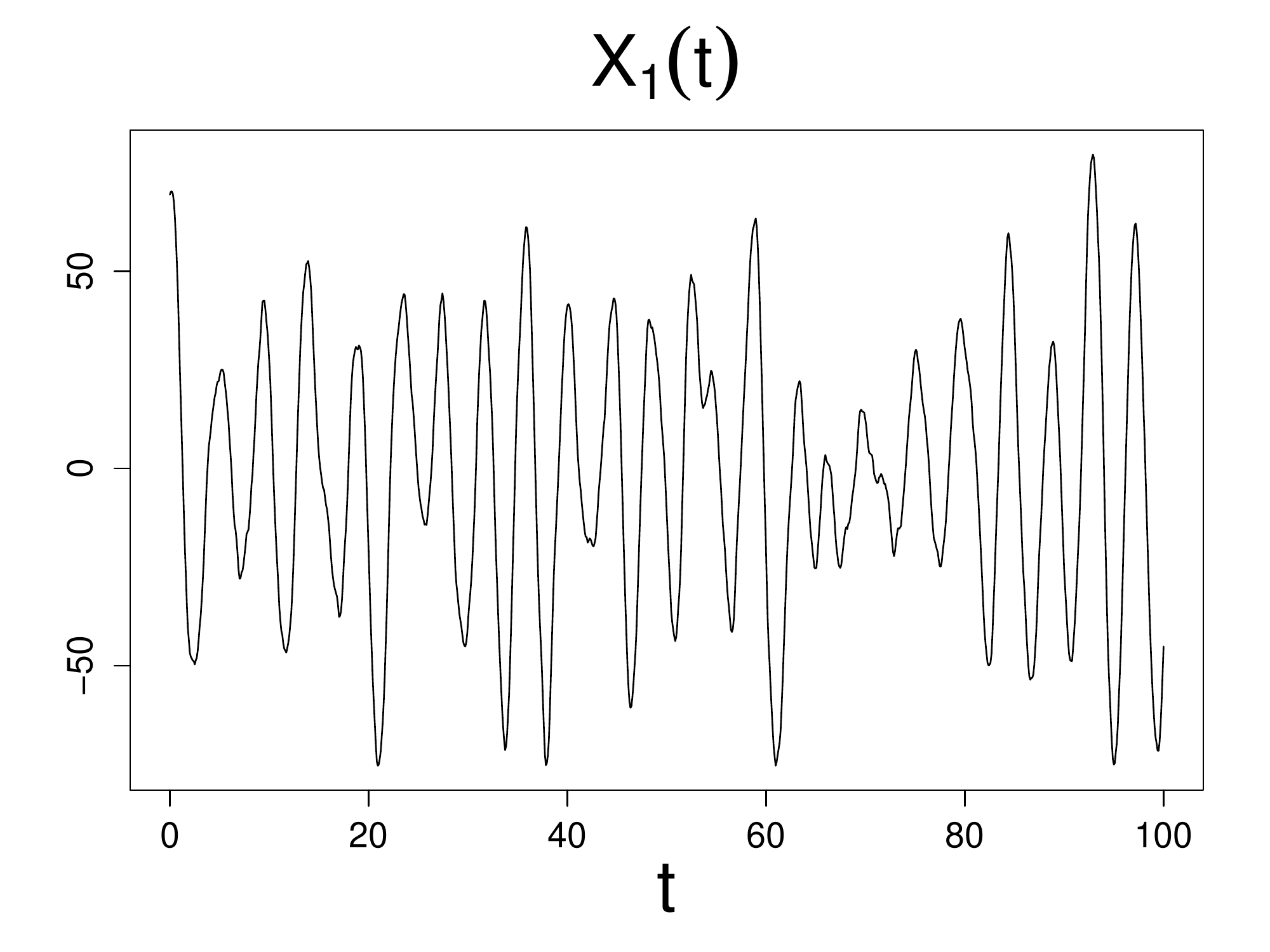}}
\subfigure[ \label{F01b}]{\includegraphics[scale=.15]{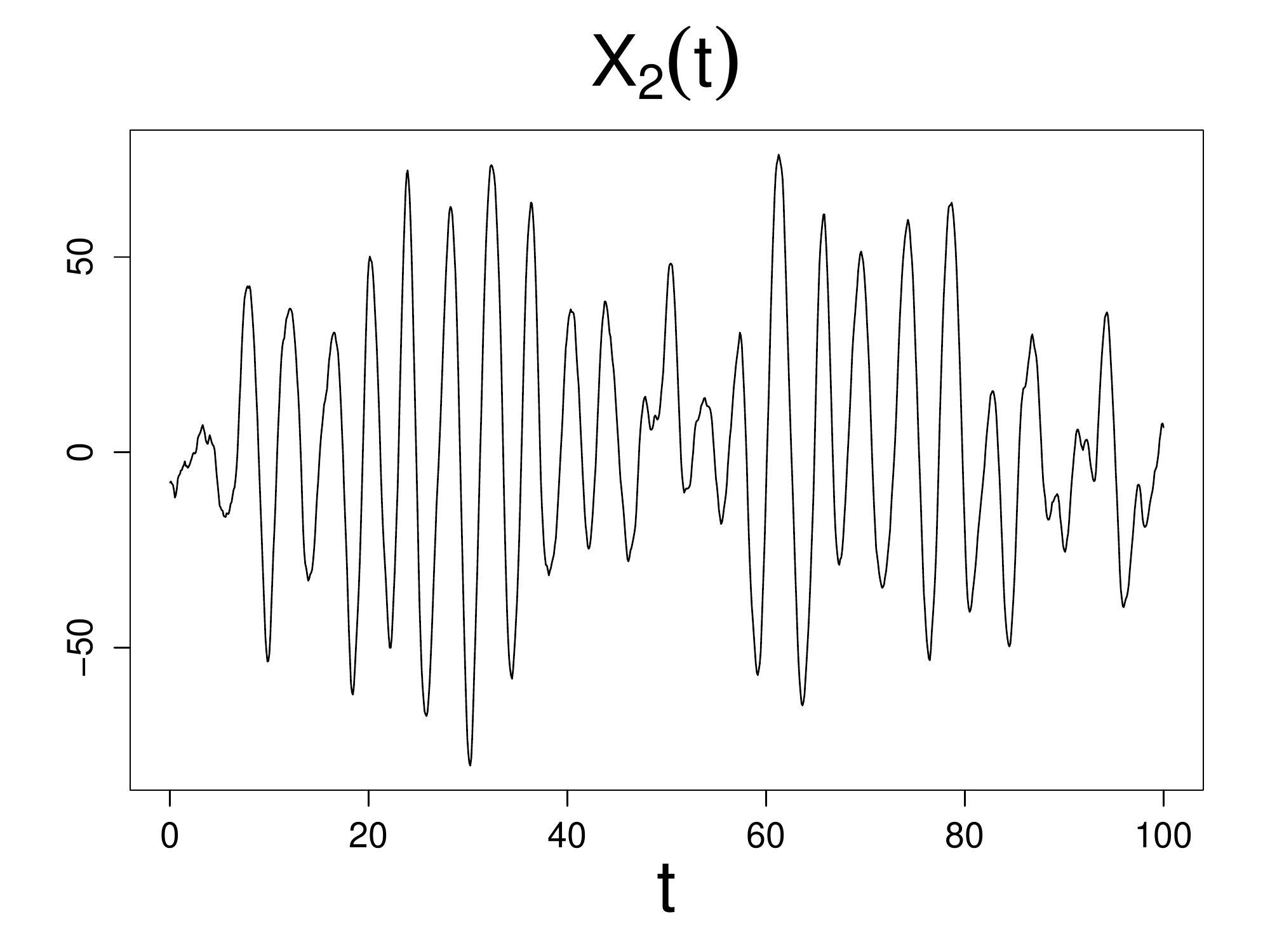}}
\subfigure[ \label{F01d}]{\includegraphics[scale=.15]{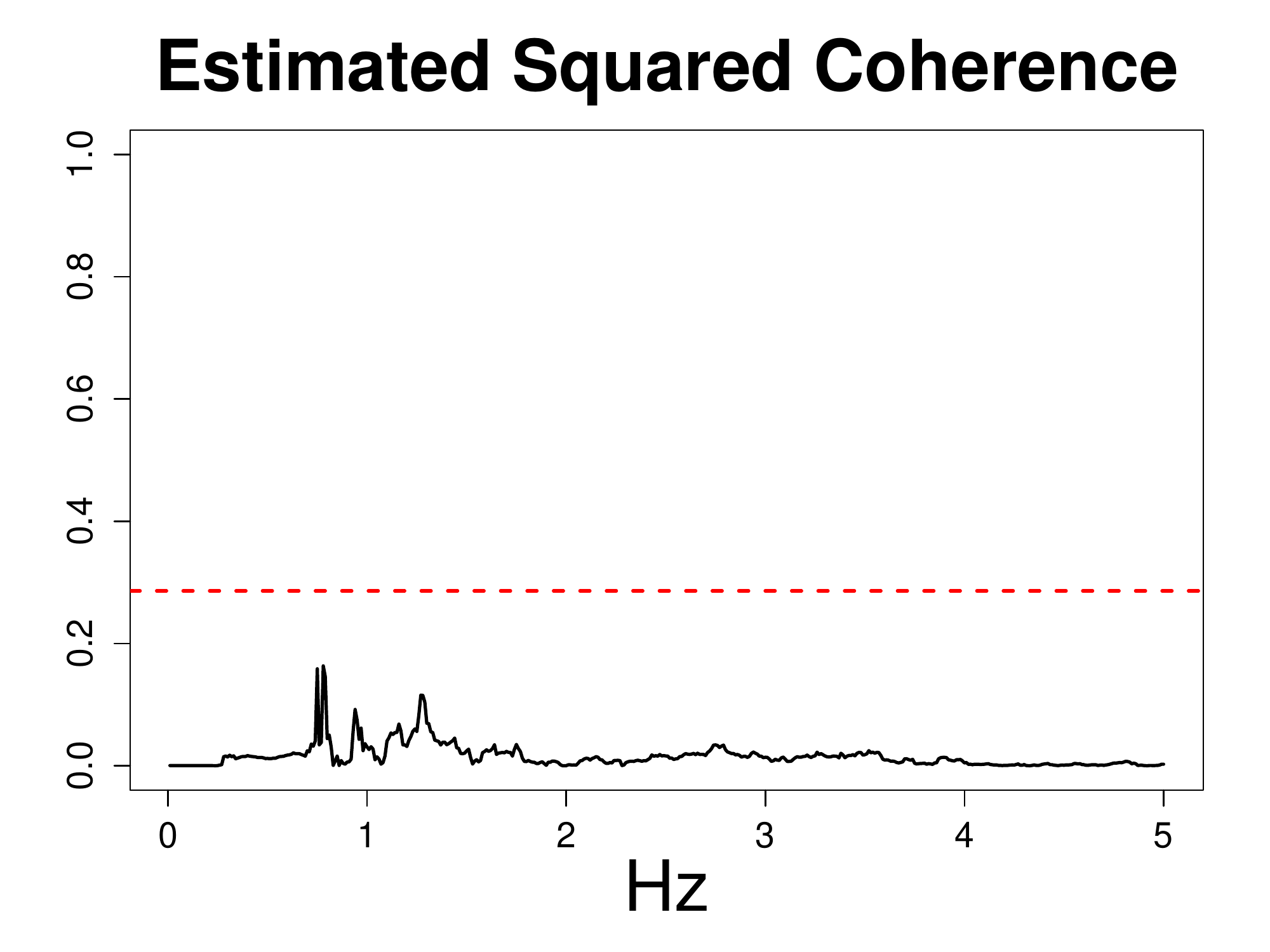}}
\subfigure[ \label{F01c}]{\includegraphics[scale=.15]{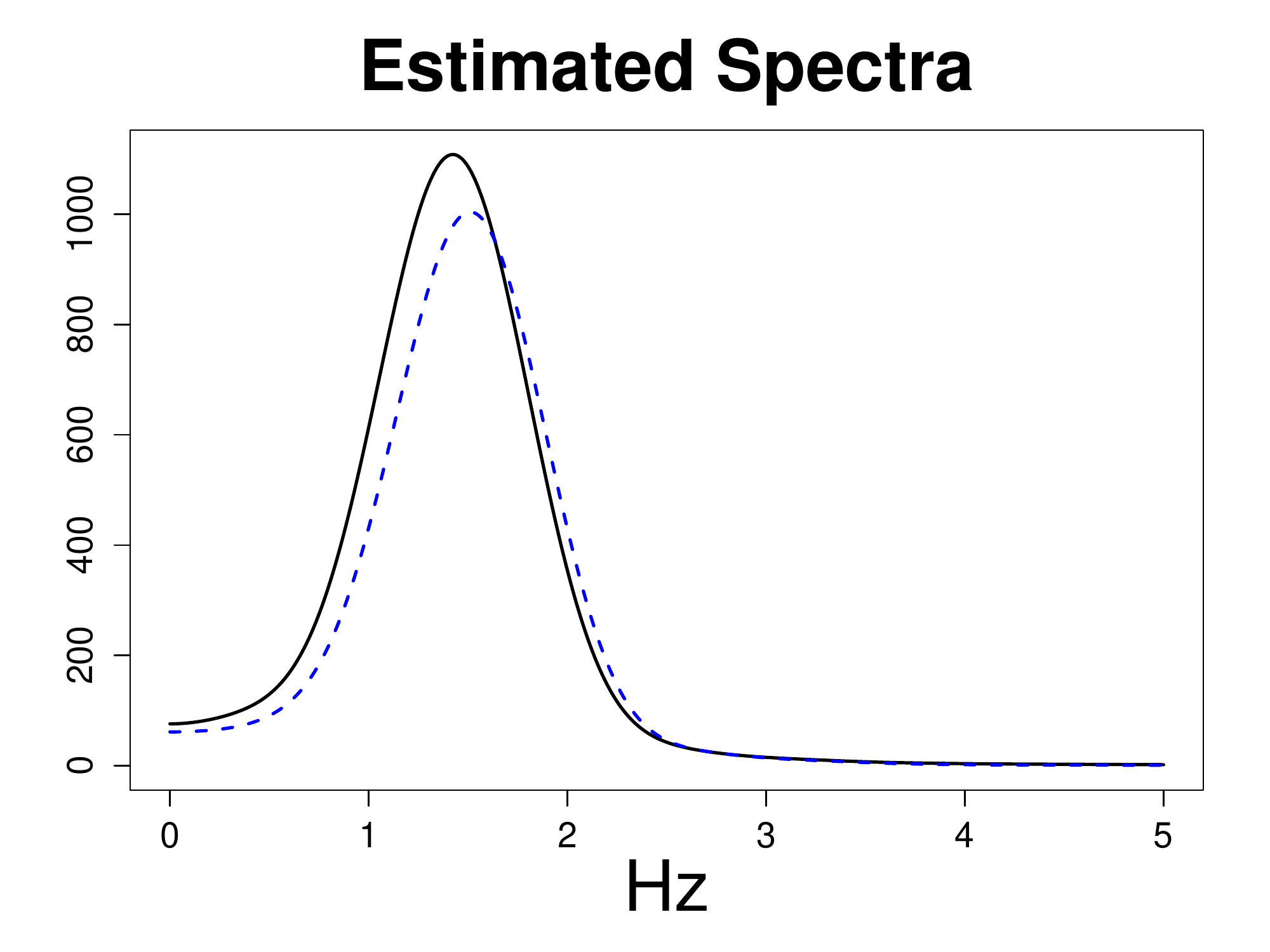}}
\caption{(a.) Plot of $X_1(t)$, AR(2) process. (b.) Plot of $X_2(t)$, AR(2) process.
(c.) Estimated squared coherence, red dashed line corresponds to the threshold for 
rejecting zero coherence. (d.) Estimated spectra for each signal, black continuous curve corresponds to $X_1(t)$ 
and blue dashed curve corresponds to $X_2(t)$.}\label{F01}
\end{figure}

Spectral synchronicity is another notion of functional connectivity that is motivated by
the fact that EEG signals are projections of highly synchronized activity of the neurons.
Due to the rhythmic nature of the EEG signals, spectral synchronicity is natural measure
(see \cite{Nun00}). From the surface, spectral sychronicity might appear to be synonymous
with coherence as defined in \cite{ShumStof} (Section $4.7$, page 217). As we demonstrate here, coherence does not
always sufficiently measure synchrony. Consider two signals $X_1(t) = A_1(\omega) \cos(\omega t)
+ N_1(t)$ and $X_2(t) = A_2(\omega) \cos(\omega t) + N_2(t)$, where $A_1(\omega)$ and $A_2(\omega)$
are zero mean random coefficients and $N_1(t)$ and $N_2(t)$ are independent white noise processes. In this case the coherence between $X_1$ and $X_2$ is zero since they are completely uncorrelated. However, these two process are completely spectrally synchronized  since they contain oscillatory activity at the same frequency $\omega$. Figures \ref{F0a} and \ref{F0b} show a draw for these signals, Figure \ref{F0d} shows the estimated squared coherence and Figure \ref{F0c} shows the estimated spectra for each one. As we mentioned before these two signals have coherence equal zero but they display similar spectral profile. Consider the next example, which is more challenging, $X_1(t)$ and $X_2(t)$ two uncorrelated time series from the AR(2) process with $\eta=1.5$ Hz, $M=1.01$ and $F_s=10$ Hz (see (\ref{AR2})). These signals also have zero coherence (Figure \ref{F01d}) but their estimated spectra look similar, Figure \ref{F01c}. In this case it is more difficult to say that $X_1$ and $X_2$ have the same underlying process only by observing the time series, Figures \ref{F01a} and \ref{F01b}; however their spectral densities show synchrony.

\section{Numerical Experiments}
We now investigate the finite sample performance of our proposed clustering algorithm.
Our experimental design for the simulation study is based on the AR$(2)$ model and the
parametrization defined in (\ref{AR2}).
\subsection{Experimental Design}
Let $Z_t^j$ be the $j$-th component ($j=1, \ldots, 5$) having AR(2) process with $M_j=1.01$
for all $j$ and peak frequency $\eta_j = 2, 6, 10, 21$ and $40$ for $j=1, \ldots, 5$, respectively. $Z_t^{j}$
represents a latent signal oscillating at pre-defined band namely (delta, theta, alpha, beta, gamma).
Define the observed time series to be a mixture of these latent AR$(2)$ processes.
\begin{equation}\label{Sim1}
\begin{pmatrix}
 X_t^1\\
 X_t^2\\
 \vdots\\
 X_t^K\\
\end{pmatrix}_{K \times 1}
=
\begin{pmatrix}
 \boldsymbol{e}_1^T\\
 \boldsymbol{e}_2^T\\
 \vdots\\
 \boldsymbol{e}_K^T\\
\end{pmatrix}_{K \times 5}
\begin{pmatrix}
 Z_t^1 \\
 Z_t^2 \\
 \vdots \\
 Z_t^5\\
\end{pmatrix}_{5 \times 1}
+
\begin{pmatrix}
 \varepsilon_t^1\\
 \varepsilon_t^2\\
 \vdots\\
 \varepsilon_t^K\\
\end{pmatrix}_{K \times 1}
\end{equation}
where $\varepsilon_t^j$ is Gaussian white noise, $X_t^j$ is a signal with oscillatory behavior generated by the linear combination $\boldsymbol{e}_i^T Z_t^{j}$ and $K$ is the number of spectrally synchronized groups or clusters. For each experiment, $K=5$ (number of clusters) and $n$ denotes the
number of replicates of each signal $X_t^{i}$.

\subsection{Comparative Study}\label{CS}
To compare clustering results, we must take into account the ``quality of the clustering" produced
which depends on both the similarity measure and the clustering algorithm used. The SMC method has two main features: The use of the TVD as a similarity measure and the hierarchical merger (spectral merger) algorithm. Since the results are similar, we will only consider the hierarchical merger algorithm. The SMC method will be compared with the usual hierarchical agglomerative clustering algorithm using the complete linkage function, which is one of the standard clustering procedures used in the literature, and the following three similarity measures:
\begin{itemize}
\item The Euclidean distance between the normalized estimated spectra
$$
d_{NP}(X,Y) = \frac{1}{n}\Big( \sum_k \big( \widehat{f}_{X}(\omega_k)- \widehat{f}_{Y}(\omega_k)\big)^2\Big)^{1/2}.
$$
\item The Euclidean distance between the logarithm of the normalized estimated spectra
$$
d_{LNP}(X,Y) = \frac{1}{n}\Big( \sum_k \big( \log \widehat{f}_{X}(\omega_k)- \log \widehat{f}_{Y}(\omega_k)\big)^2\Big)^{1/2}.
$$
\item The symmetric Kullback-Leibler distance between the normalized estimated spectra
$$
d_{KL}(X,Y)= \int \widehat{f}_{X} (\omega) \log\left(\frac{\widehat{f}_X(\omega)}{\widehat{f}_Y(\omega)}\right) \mbox{d}\omega
$$
$$
d_{SKL}(X,Y)=d_{KL}(X,Y)+d_{KL}(Y,X).
$$

\end{itemize}
All similarity measures will use normalized estimated spectra.
The first two are frequently used in time series clustering and have been compared to other measures in different simulations [\cite{Vilar1}, \cite{Mont13}]. The symmetric Kullback-Leibler divergence is used in a hierarchical clustering algorithm in \cite{ShumStof}.

To evaluate the rate of success, we consider the following index which has been already used for comparing
different clusterings [\cite{Vilar1}, \cite{Mont13}, \cite{Gav00}]. Let $\{C_1,\ldots,C_g\}$ and $\{G_1,\ldots,G_k\}$, be the set of the $g$ real groups and a $k$-cluster solution, respectively.
Then,
\begin{equation}\label{PS}
\mbox{Sim}(G,C)=\frac{1}{g}\sum_{i=1}^{g} \max_{1\leq j\leq k} \mbox{Sim}(G_j,C_i),
\end{equation}
where $\quad \displaystyle \mbox{Sim}(G_j,C_i)=\frac{2|G_j \cup C_i|}{|G_j|+|C_i|}$. Note that this similarity measure will return 0 if the two clusterings are completely dissimilar and 1 if they are the same.

\begin{figure}
\begin{minipage}{.45\linewidth}
\centering
\includegraphics[scale=.4]{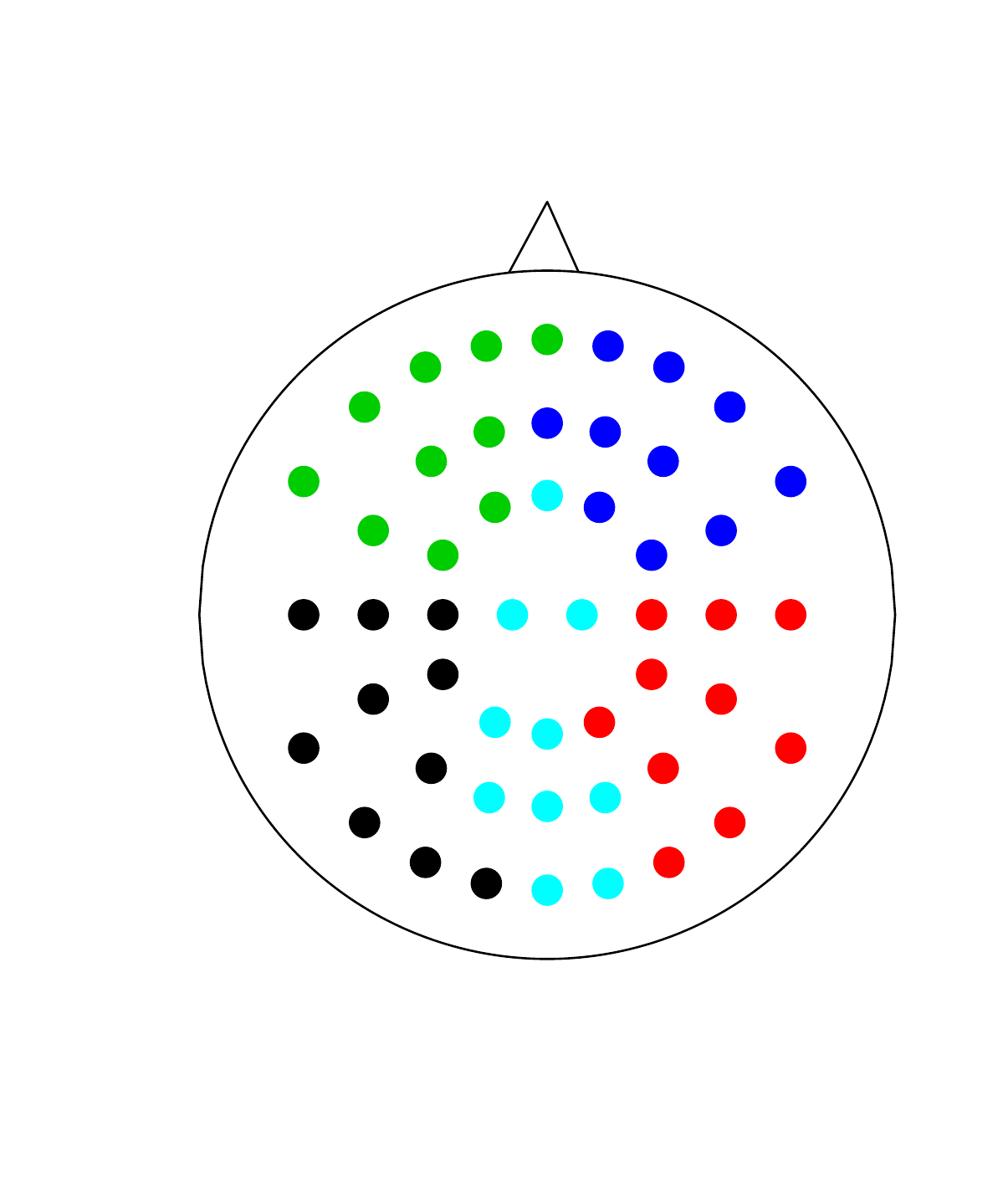}
\caption{Location of clusters in the cortical surface; black channels, red channels, green channels, blue channels and light blue channels belong to underlying process generating by the coefficients $\boldsymbol{e}_1$, $\boldsymbol{e}_2$, $\boldsymbol{e}_3$, $\boldsymbol{e}_4$ and $\boldsymbol{e}_5$ respectively.}\label{F35}
\end{minipage}
\hspace{1cm}
\begin{minipage}{.45\linewidth}
\centering
\includegraphics[scale=.39]{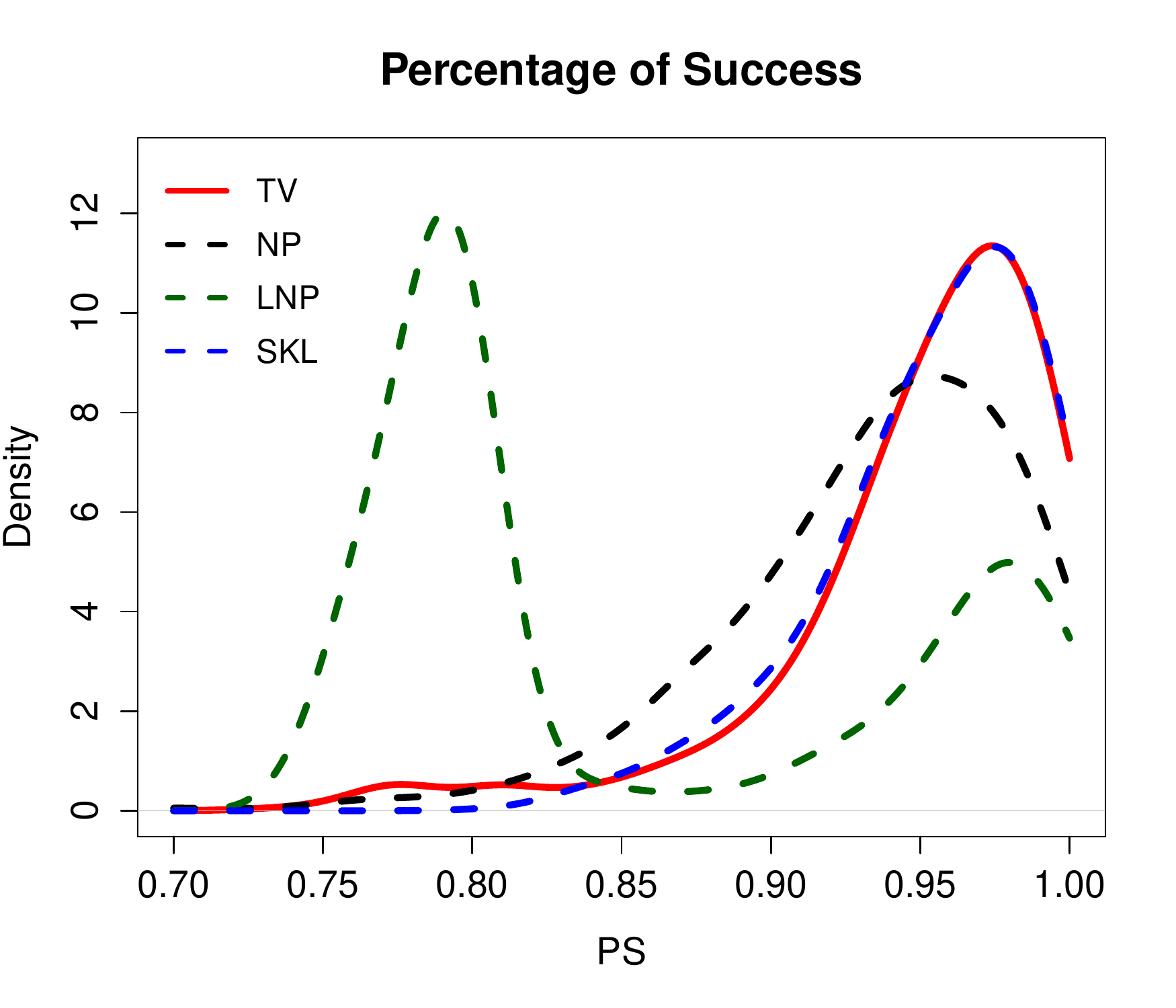}
\caption{Density of the percentage of success (see \ref{PS}) of the different clustering algorithms considered in the comparative study. The continuous red line corresponds to the algorithm proposed in this work.}\label{F4}
\end{minipage}
\end{figure}

In the first experiment, there were $K=5$ clusters with $n=10$ members per cluster and
\begin{equation*}
 \boldsymbol{e}_1=\begin{pmatrix}
                  1 \\
                  2 \\
                  0 \\
                  0 \\
                  0 \\
                  \end{pmatrix},\quad
  \boldsymbol{e}_2=\begin{pmatrix}
                  0 \\
                  1 \\
                  2 \\
                  0 \\
                  0 \\
                  \end{pmatrix},\quad
 \boldsymbol{e}_3=\begin{pmatrix}
                  0 \\
                  0 \\
                  1 \\
                  1 \\
                  0 \\
                  \end{pmatrix},\quad
 \boldsymbol{e}_4=\begin{pmatrix}
                  0 \\
                  0 \\
                  0 \\
                  1 \\
                  1 \\
                  \end{pmatrix},\quad
 \boldsymbol{e}_5=\begin{pmatrix}
                  0 \\
                  0 \\
                  1 \\
                  2 \\
                  0 \\
                  \end{pmatrix}.
\end{equation*}
We generated $1000$ time series $X_t$ simulated form (\ref{Sim1}); $K=5$ is the number
of clusters specified for each of the four clustering procedures previously described.
The location for each signal in the cortical surface is shown at Figure \ref{F35}.

Figure \ref{F4} shows the density of the percentage of success estimated for each algorithm.
One notes that our proposed algorithm outperforms those based on the Euclidean distance. Compared to
 the symmetric Kullback-Leibler distance, there is no clear winner but our proposed method still has the advantage of
being easily interpretable because the KL (or symmetric KL) cannot indicate if the dissimilarity value
is large since it belongs to the range $[0, \infty)$.

\begin{figure}
\centering
\subfigure[]{\includegraphics[scale=.4]{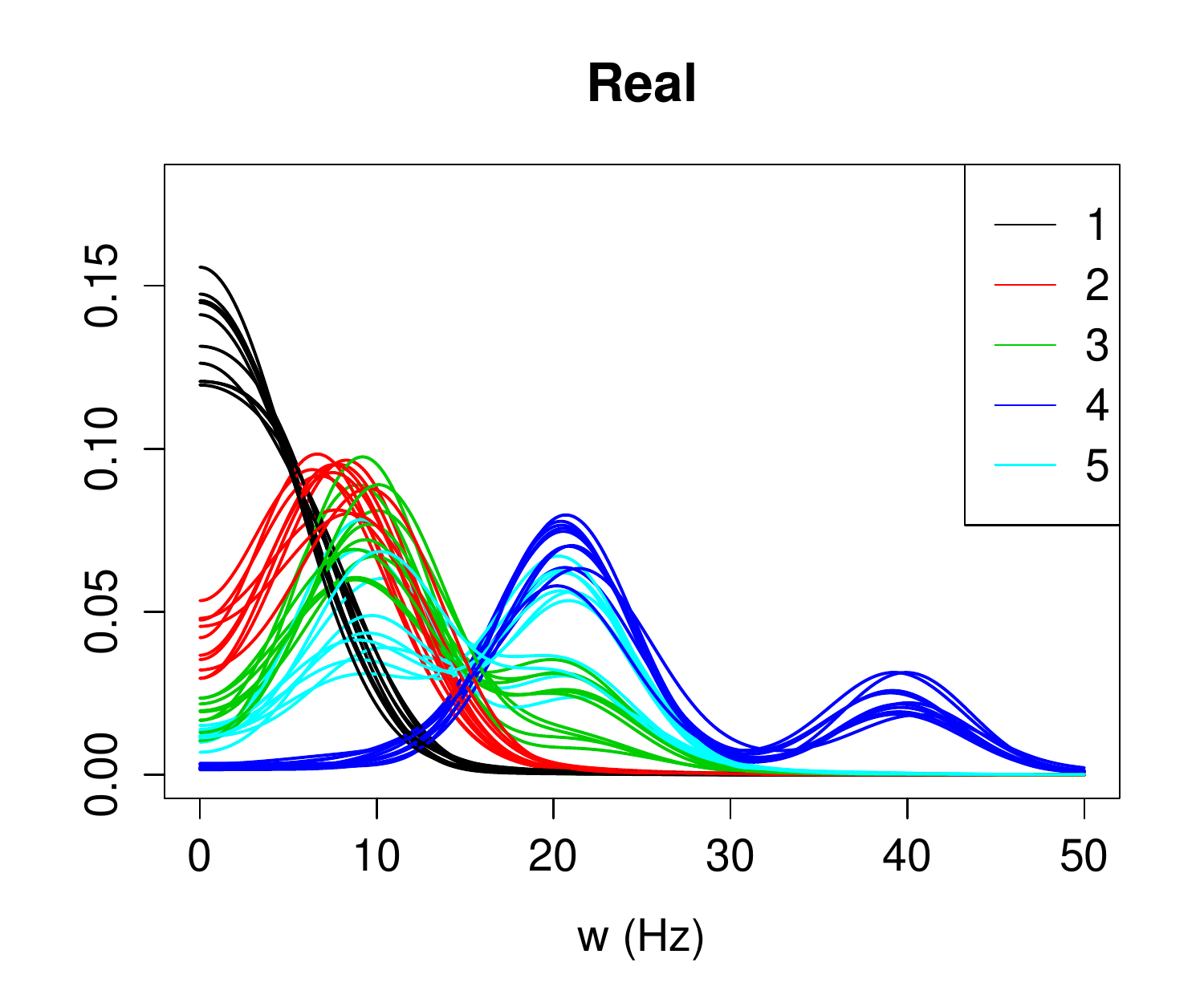}}
\subfigure[]{\includegraphics[scale=.4]{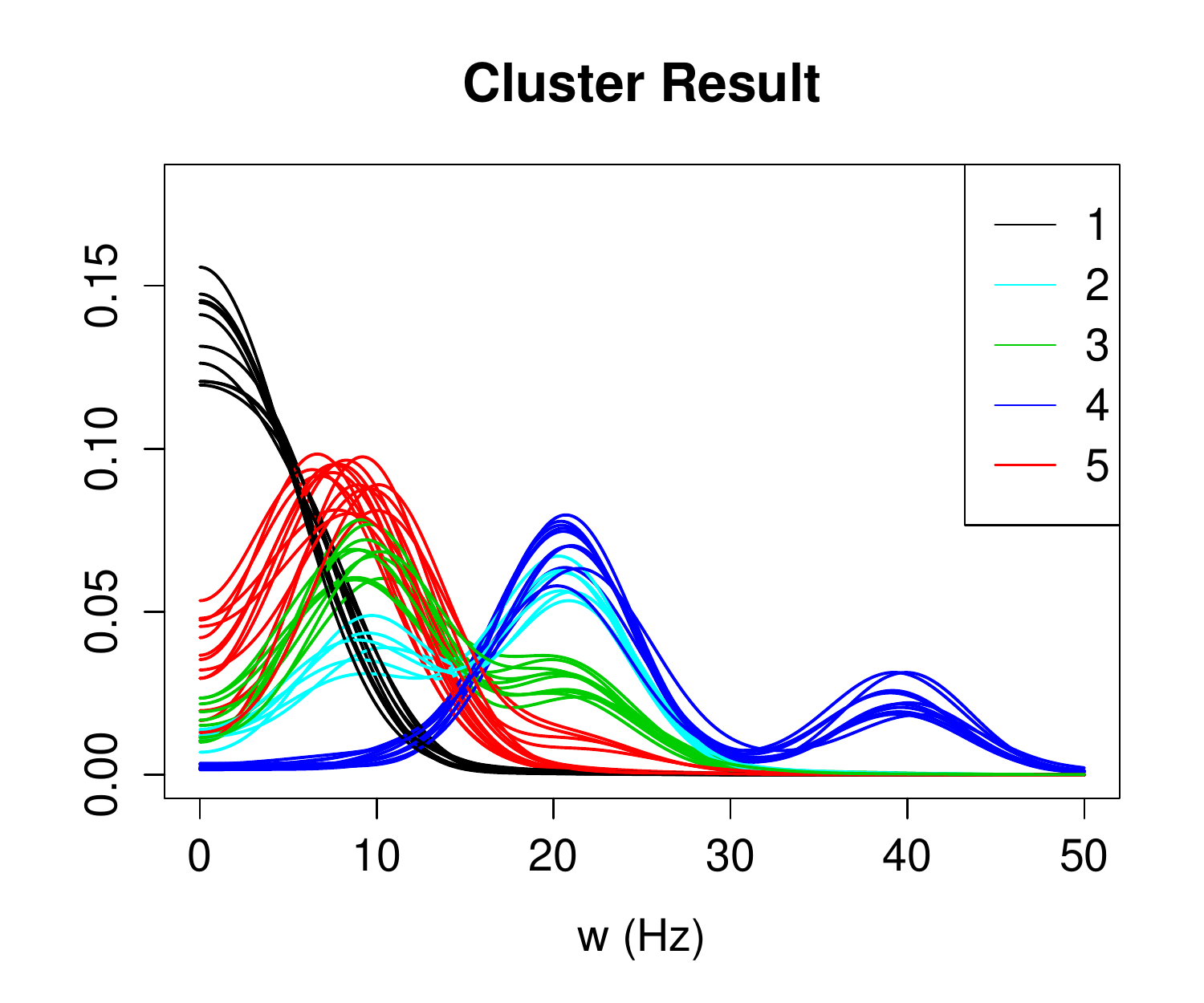}}
\caption{One draw of the first simulation study: (a.) True clusters and (b.) Clustering obtained using the proposed algorithm. }\label{F5}
\end{figure}
One draw from this experiment is shown in Figure \ref{F5}. Note that when the spectral densities are highly similar, it is difficult to distinguish between them because of the estimation error. However the results are good and given that
the TVD and our algorithm have nice established properties, our method has the ability to identify spectrally synchronized signals.

\subsection{When the number of clusters is unknown}

For real data the number of clusters is usually unknown. Thus, an objective criterion is
needed to determine the optimal number of clusters. As mentioned in Step \ref{St2} of our
algorithm, the TVD computed before joining two clusters can be used as a criterion.

\begin{figure}[ht]
\centering
\subfigure[\label{F6a}]{\includegraphics[scale=.25]{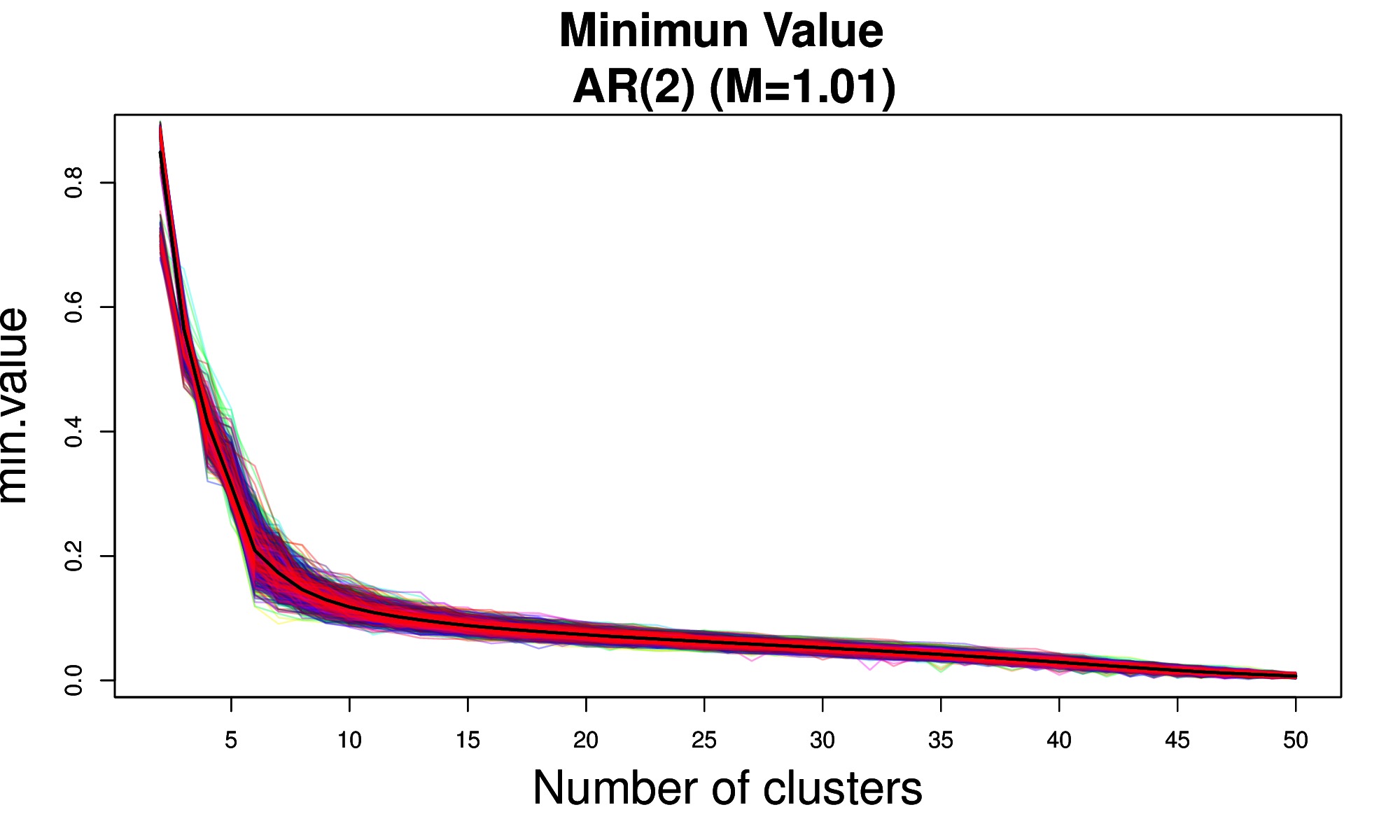}}
\subfigure[\label{F6b}]{\includegraphics[scale=.25]{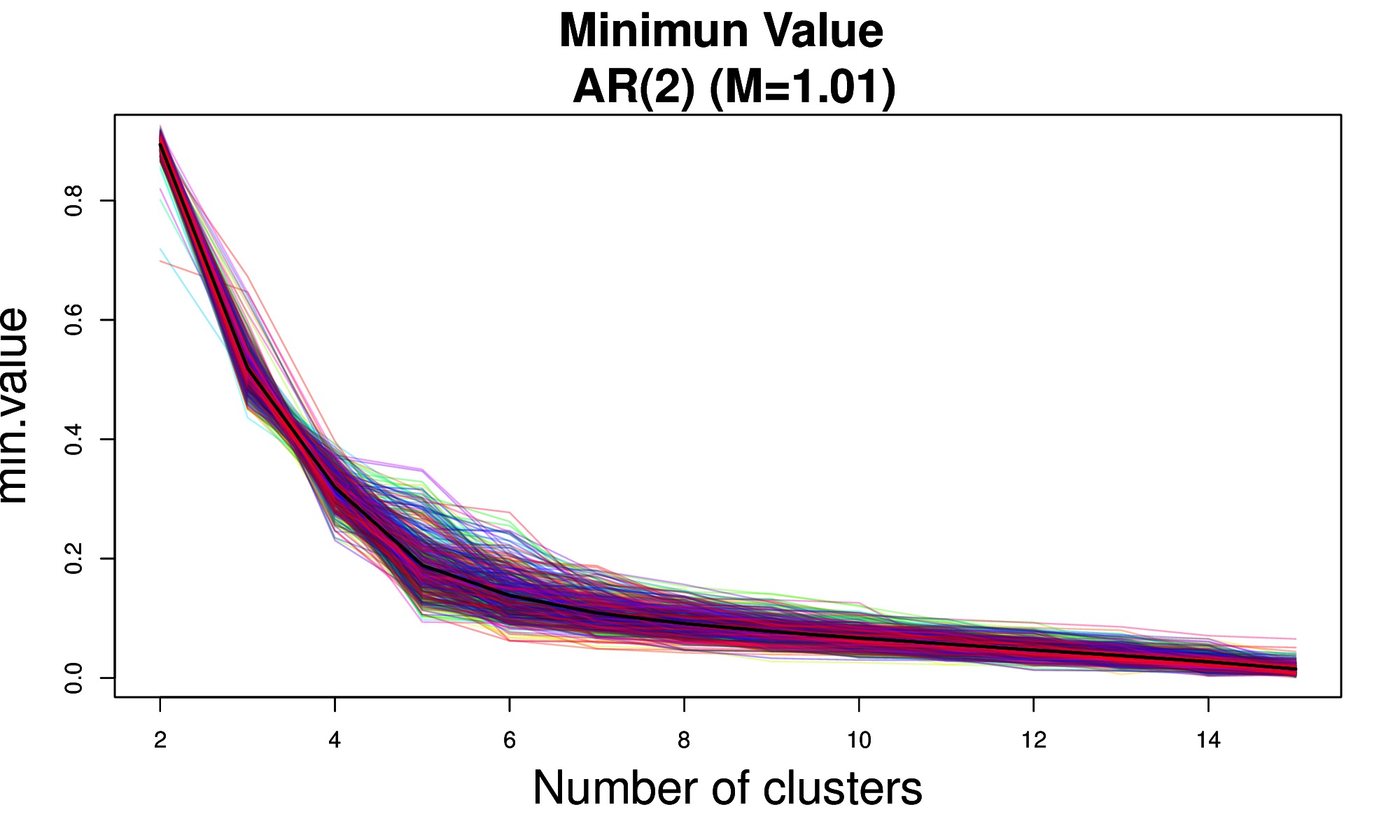}}
\subfigure[\label{F6c}]{\includegraphics[scale=.25]{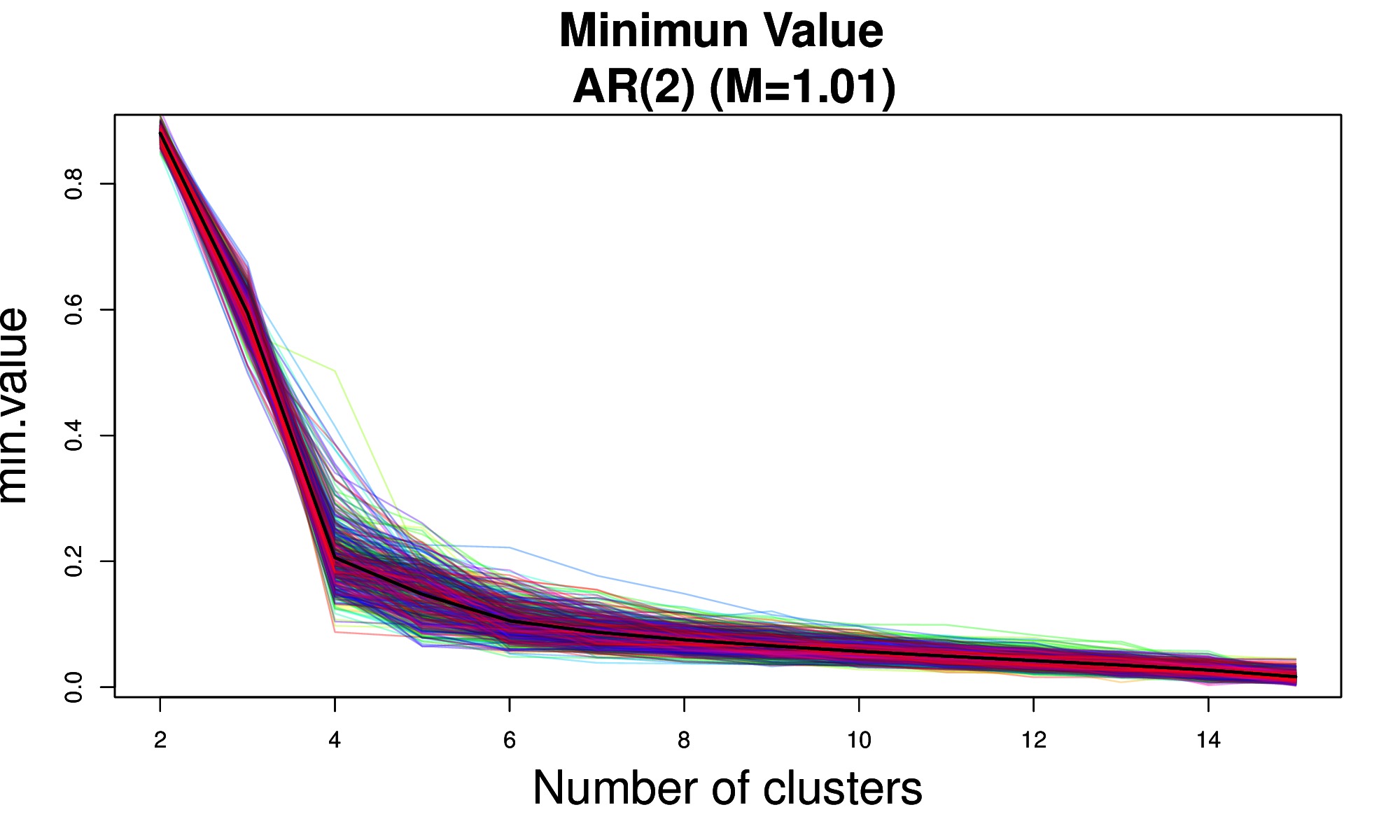}}
\subfigure[\label{F6d}]{\includegraphics[scale=.25]{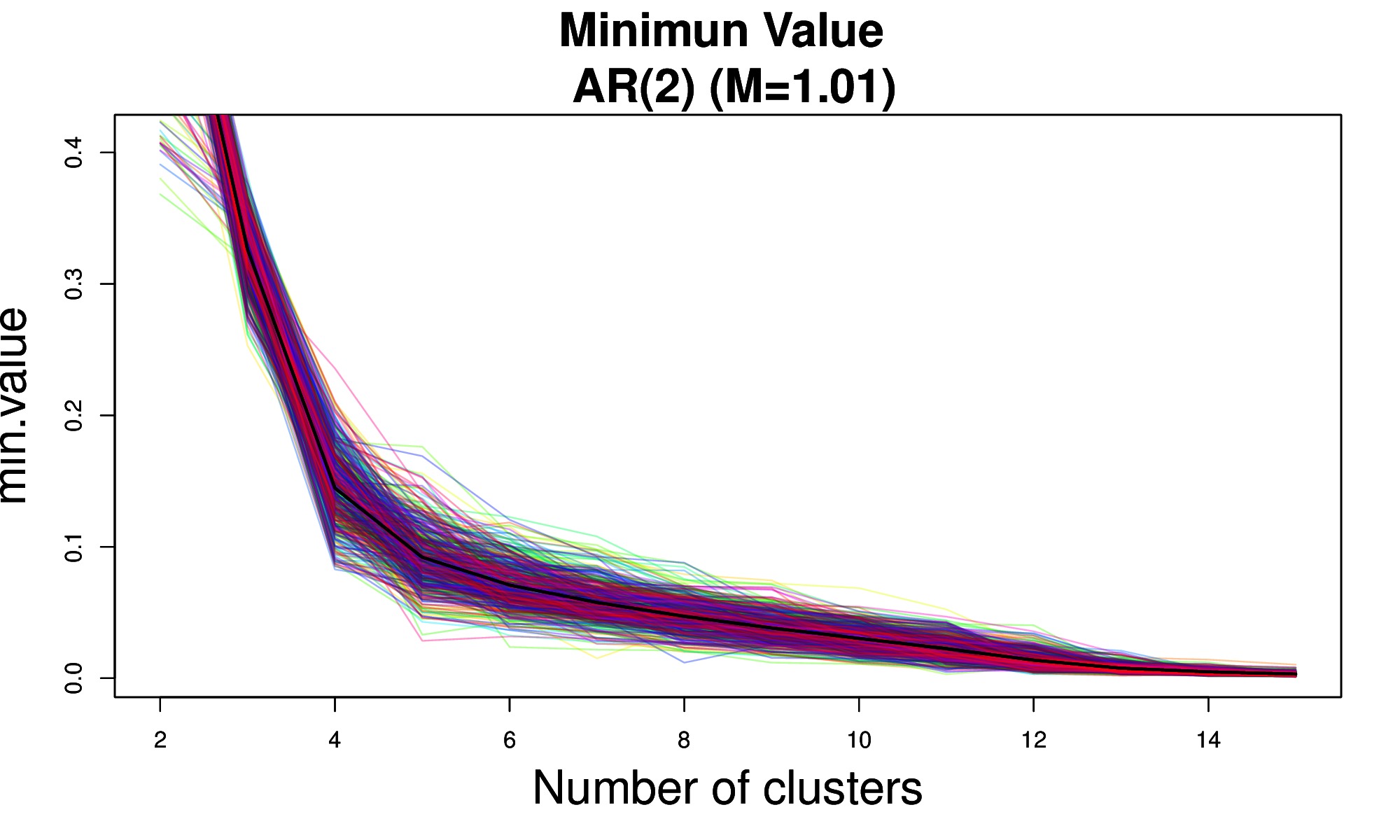}}
\caption{Trajectory of the value of the TVD that is achieved by the algorithm.}\label{F6}
\end{figure}

We propose an empirical criterion. In Figure \ref{F6}, we display
the graphs corresponding to the minimum values of the TVD between clusters as a function of the
number of clusters. Figure \ref{F6a} corresponds to the experimental design presented in the last section. Figure \ref{F6b} corresponds to the experimental design with the same coefficients as the first one but $K=5$, $n=3$ and $500$ draws. Figure \ref{F6c} corresponds to a design with $K=5$, $n=3$ and $500$ draws, with coefficients

\begin{equation*}
 \boldsymbol{e}_1=\begin{pmatrix}
                  \frac{1}{2} \\
                  1 \\
                  0 \\
                  0 \\
                  0 \\
                  \end{pmatrix},\quad
  \boldsymbol{e}_2=\begin{pmatrix}
                  0 \\
                  1 \\
                   \frac{1}{2} \\
                  0 \\
                  0 \\
                  \end{pmatrix},\quad
 \boldsymbol{e}_3=\begin{pmatrix}
                  0 \\
                  0 \\
                  \frac{1}{2} \\
                  1 \\
                  0 \\
                  \end{pmatrix},\quad
 \boldsymbol{e}_4=\begin{pmatrix}
                  0 \\
                  0 \\
                  0 \\
                  1 \\
                   \frac{1}{2} \\
                  \end{pmatrix},\quad
 \boldsymbol{e}_5=\begin{pmatrix}
                  0 \\
                  1 \\
                  0 \\
                  1\\
                  0 \\
                  \end{pmatrix}.
\end{equation*}
Finally, in Figure \ref{F6d} we set $K=5$, $n=3$ and $500$ draws but the coefficients $\boldsymbol{e}_i$ are drawn from a Dirichlet distribution with parameters $(.2,.2,.2,.2,.2)$. All these curves are decreasing and the speed of decrease slows down after the number of clusters is 5, even when the signals involved in each experiment are different. This ``elbow'' that seems to appear with 5 clusters can be used as a empirical criteria to decide the number of clusters. Similar criteria are frequently used in cross validation methods.
In this sense we are going to use this empirical criteria to get the number of possible clusters in real data analysis.


\section{Clustering EEG channels according to spectral synchronicity}

Our goal here is to cluster resting-state EEG signals that are spectrally synchronized, i.e.,
that show similar spectral profiles from subjects in this study.
The participants here are healthy subjects whose EEG clustering will serve as a ``standard" to
which the clustering of stroke patients (with severe motor impairment) will be compared.
In this paper, we will address the following questions of interest to our collaborator:
(1.) How many spectrally synchronized clusters are there during resting-state?
(2.) Does the number of clusters remain fixed across epochs during the entire resting-state?
(3.) Does cluster membership of the channels evolve across the entire resting-state?
(4.) Is there any evidence of difference in EEG clustering between subjects who display
marked improvement in their motor skill performance vs those with poor improvement?
To address these questions, we developed the \textit{HMClust} toolbox written
in R [\cite{RR}] that implements our proposed clustering method. The toolbox can be downloaded from
\url{http://ucispacetime.wix.com/spacetime#!project-a/cxl2}.

\subsection{EEG Data Description}
\begin{figure}
\centering
\includegraphics[scale=.3]{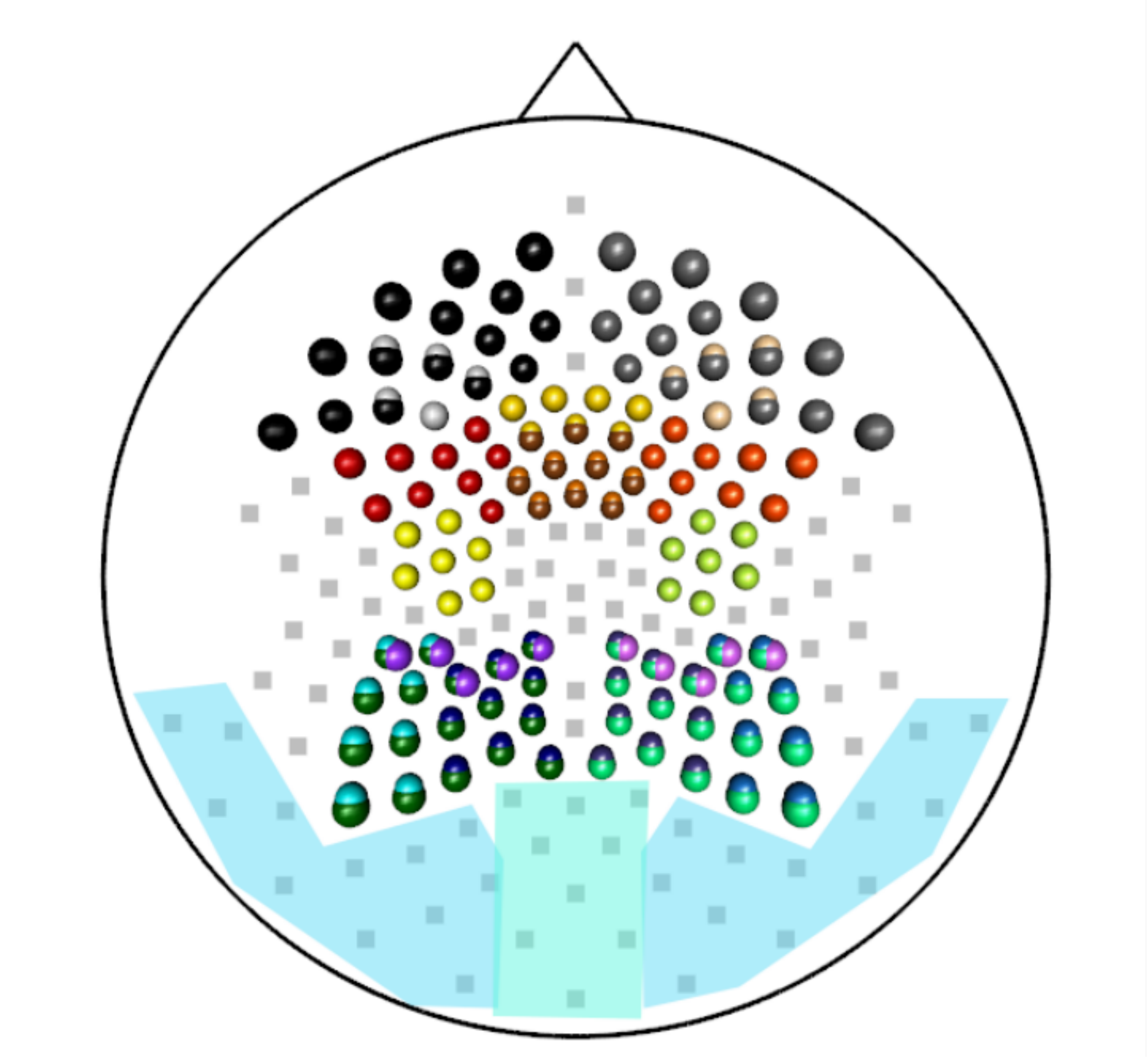}
\includegraphics[scale=.3]{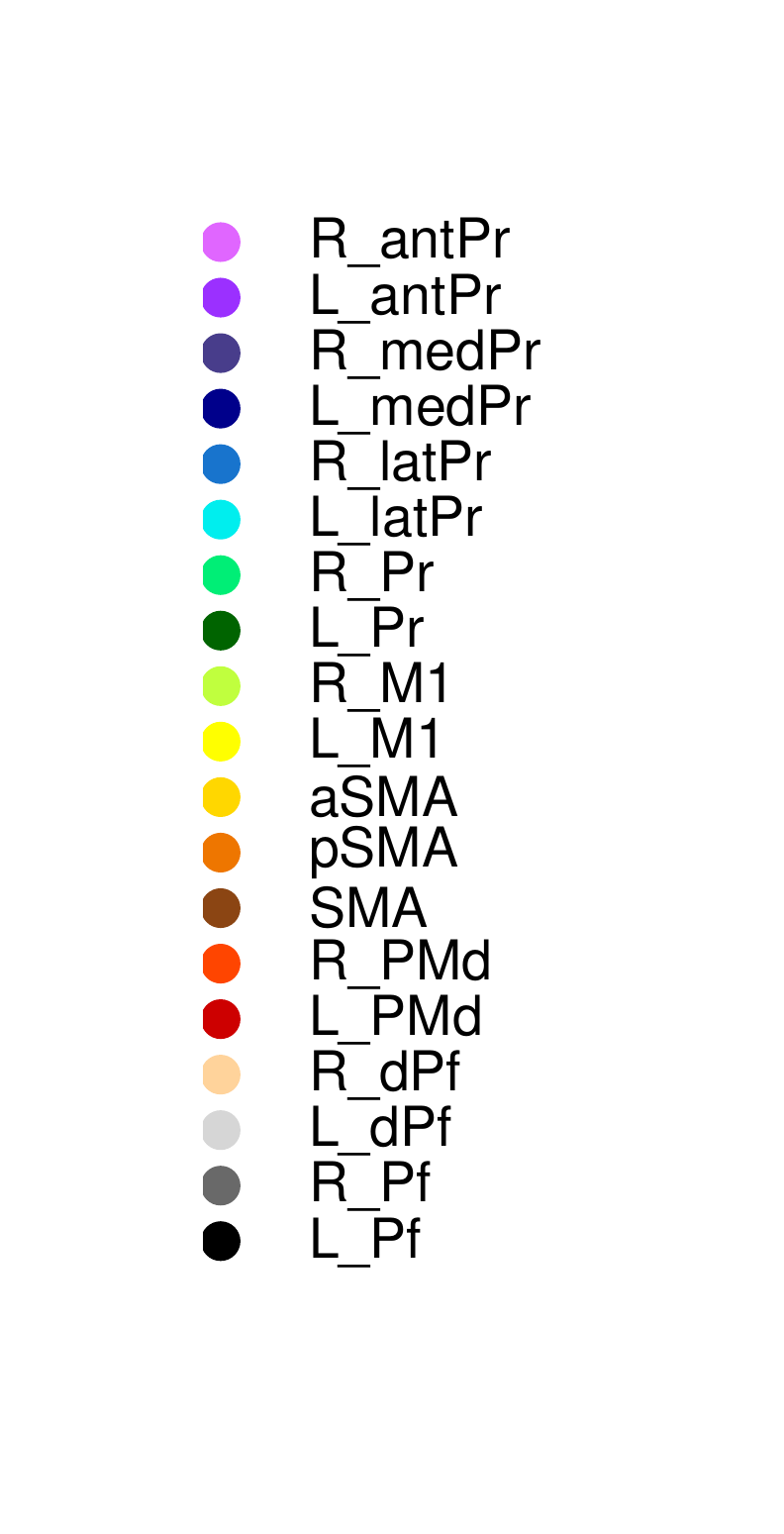}
\caption{Brain regions defined in \cite{Wu14}; Left Prefrontal (L\_Pf), Right Prefrontal(R\_Pr), Left Dorsolateral Prefrontal (L\_dPr), Right Dorsolateral Prefrontal (R\_dPr), Left Pre-motor (L\_PMd), Right Pre-motor (R\_PMd), Supplementary Motor Area (SMA), anterior SMA (aSMA), posterior SMA (pSMA), Left Primary Motor Region (L\_M1), Right Primary Motor Region (R\_M1), Left Parietal (L\_Pr), Right Parietal (R\_Pr), Left Lateral Parietal (L\_latPr), Right Lateral Parietal (R\_latPr), Left Media Parietal (L\_medPr), Right Media Parietal (R\_medPr), Left Anterior Parietal (L\_antPr) and Right Anterior Parietal (R\_antPr). Gray squared channels do not belong to any of these regions; Light blue region corresponds to right and left occipital and light green region corresponds to central occipital.}\label{F1}
\end{figure}

The EEG channels were grouped into $19$ pre-defined regions in the brain as specified in \cite{Wu14}: prefrontal (left-right), dorsolateral prefrontal (left-right), pre-motor (left-right), supplementary motor area (SMA), anterior SMA, posterior SMA, primary motor region (left-right), parietal (left-right), lateral parietal (left-right), media parietal (left-right) and anterior parietal (left-right). Figure \ref{F1} shows the locations of these regions on the cortical surface. The number of channels for the EEG data is $256$.

The data was recorded from a dense array surface using 256-lead Hydrocel net. The complete
data is formed by 17  right-handed individuals who were between 18 and 30 years of age.
During the EEG-Rest period, the participants were asked to hold still with the forearms resting
on the anterior thigh and to direct their gaze at a fixation cross displayed on the computer monitor.
Data were collected at 1000 Hz using a high input impedance Net Amp 300 amplifier (Electrical Geodesics) and Net Station 4.5.3 software (Electrical Geodesics). Data were preprocessed.
The continuous EEG signal was low-pass filtered at 100 Hz, segmented into non-overlapping 1 second epochs, and detrended. The original number of channels 256 had to be reduced to 194 because of the presence of artifacts in channels that could not be corrected (e.g. loose leads).

\subsection{Pre-clustering step of smoothing the periodogram curves}

\begin{figure}
\centering
\includegraphics[scale=.55]{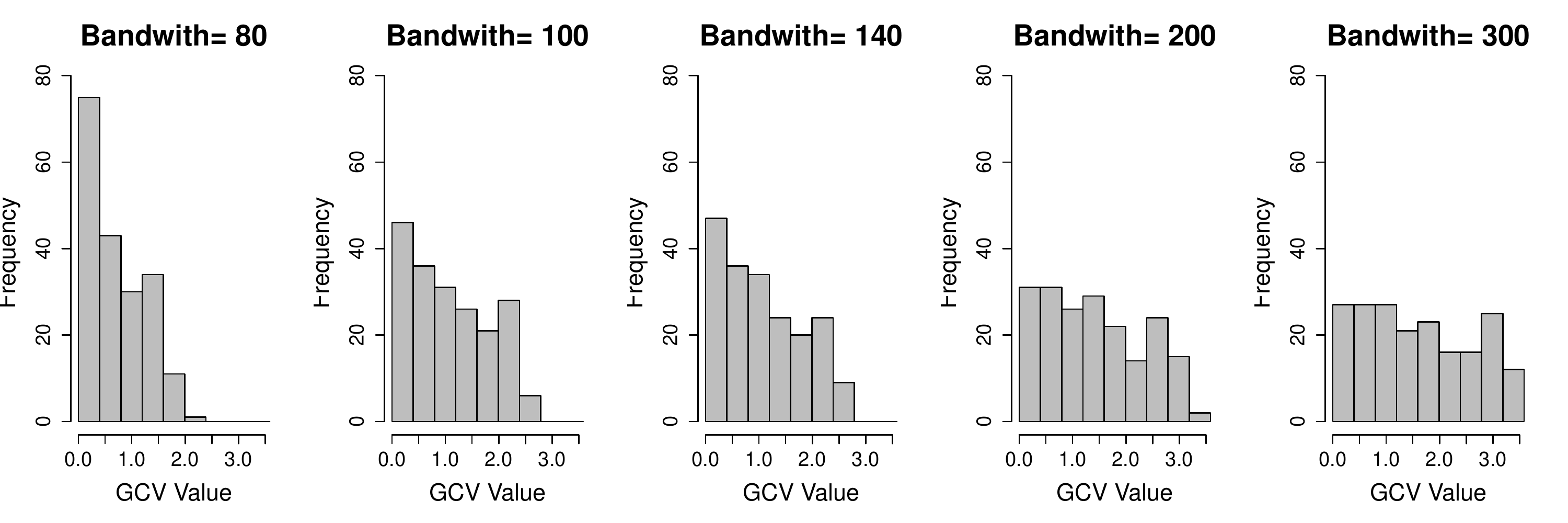}
\caption{Generalised cross validation criteria applied to 194 channels with different bandwidth values. The x axis are scale by a factor of $10^{10}$.}\label{F65}
\end{figure}

To determine a reasonable value for the smoothing bandwidth, we adapted the
Gamma-deviance generalised cross validation (Gamma GCV) criterion in \cite{Ombao2001}
to the multi-channel setting. We applied the Gamma GCV criterion to each channel for all epochs. See Figure \ref{F65} for the histogram of the values of the Gamma GCV criterion
for all channels with different values
of the bandwidth. Trajectories of the Gamma GCV for each channel are very different because this criterion depends on the shape of the estimated spectra. There is not an optimal bandwidth that is common for all channels. From the spectral estimation point of view, one could select $a=80$ over $a=100$. However, in our simulations, choosing the smaller bandwidth results in selecting unnecessarily too many clusters. The choice of a slightly large bandwidth, $a=100$, gave better overall results.

\subsection{Identifying the number of EEG clusters}
Our first data set are EEG recordings from a subject identified as ``BLAK". The entire
resting-state for each subject consisted of $160$ epochs (each is a 1-second recording).
Each epoch has $194$ channels with $T=1000$ time points.
The SMC method (with the spectral merger algorithm) was applied to each epoch.
As a first step, we chose $K=19$ clusters as motivated by the anatomical
parcellation in \cite{Wu14} presented in Figure \ref{F1}.
Our intent here is to check whether or not clustering of channels based on the
anatomical parcellation is consistent with the clustering based on the spectral
features of the EEG as determined by the SMC method. In assessing this, it is
important to observe that the actual cluster {\it label} may change from one
epoch to another even though the actual channel memberships may not change.
For example, \textit{Cluster} $1$
in epoch $30$ is identical to \textit{Cluster} $3$ in epoch $46$ since these
consist of the same channels (even though the cluster labels are different).

\begin{figure}
\centering
\subfigure[\label{F7a}]{\includegraphics[scale=.25]{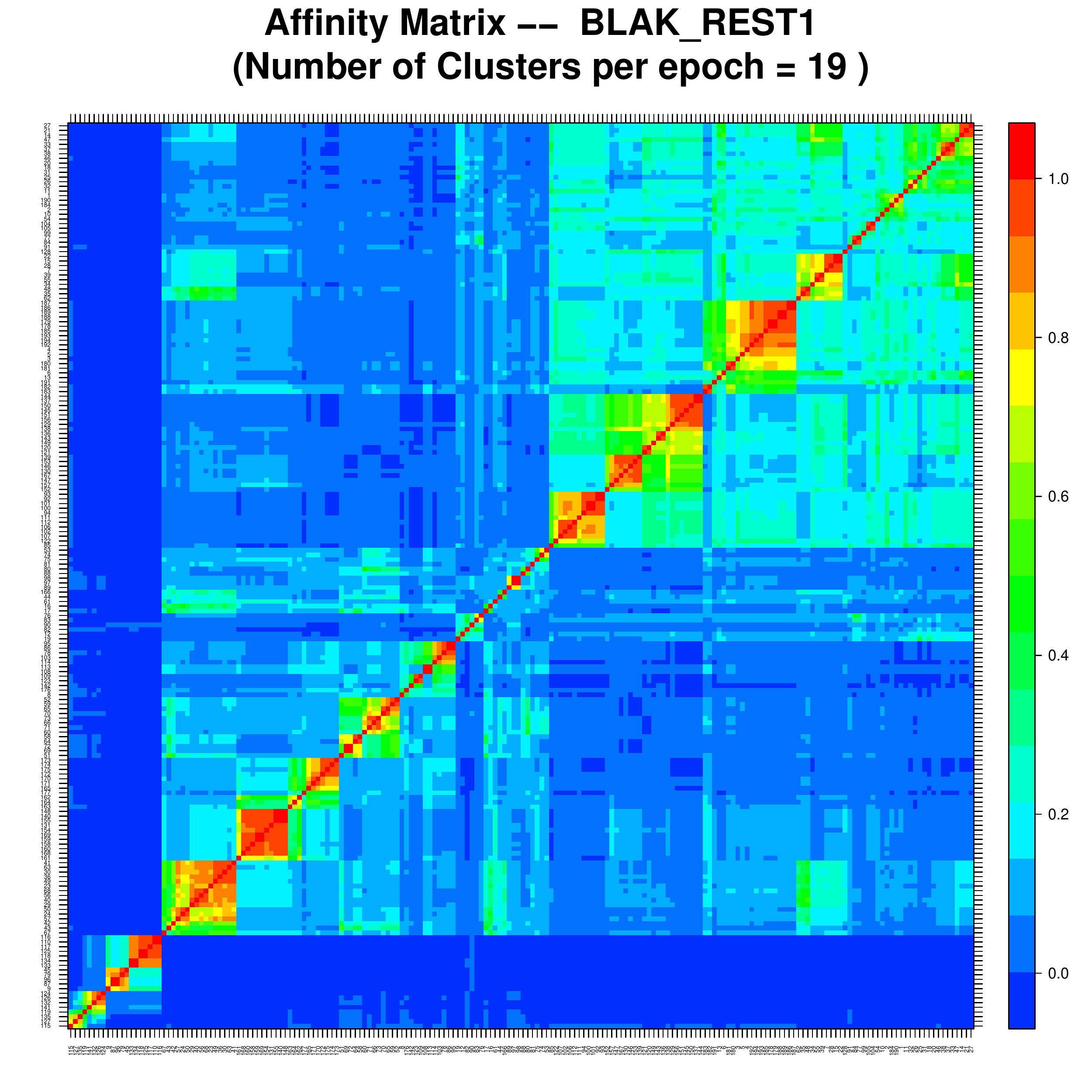}}
\subfigure[\label{F7c}]{\includegraphics[scale=.25]{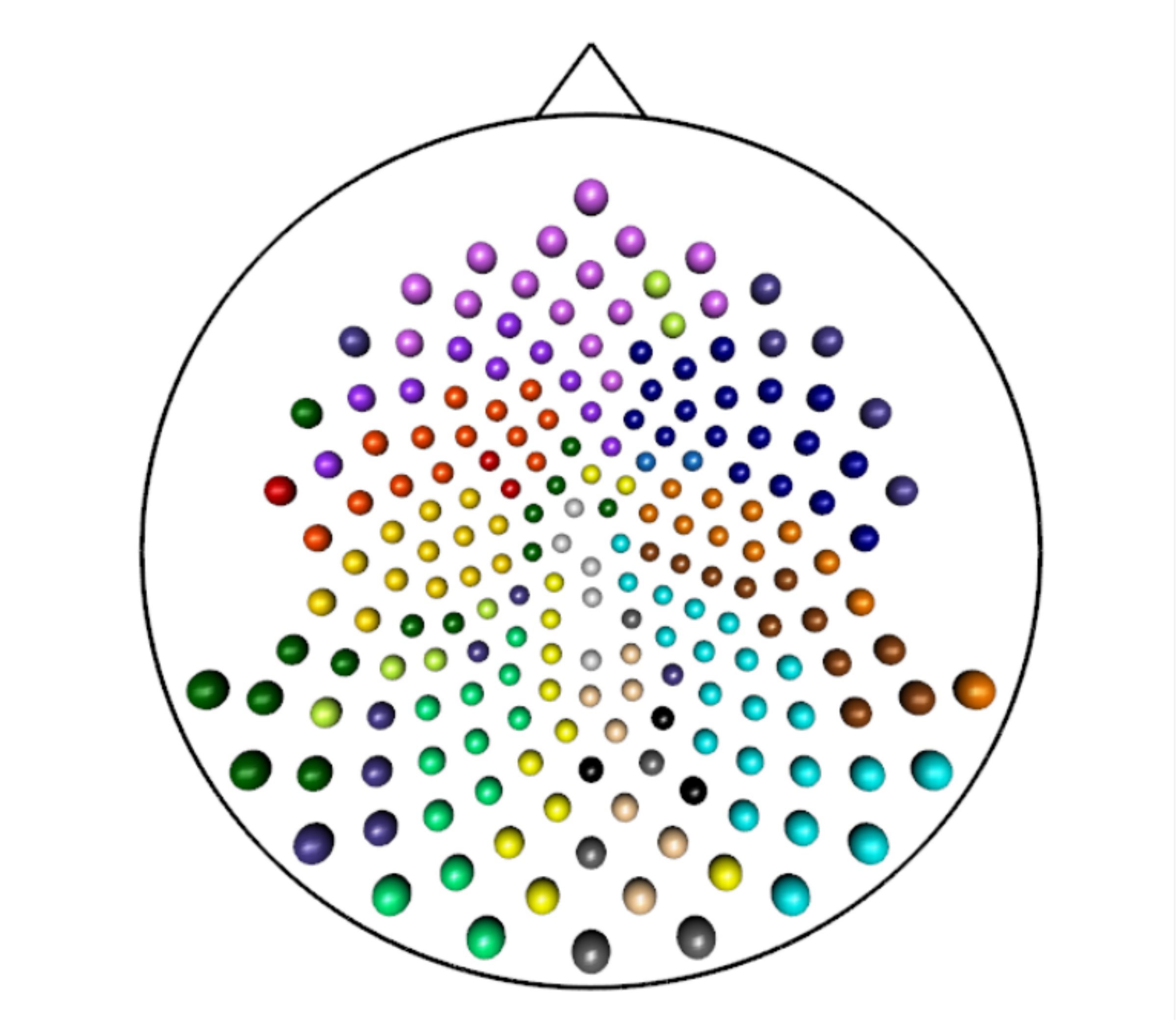}}
\subfigure[\label{F7b}]{\includegraphics[scale=.45]{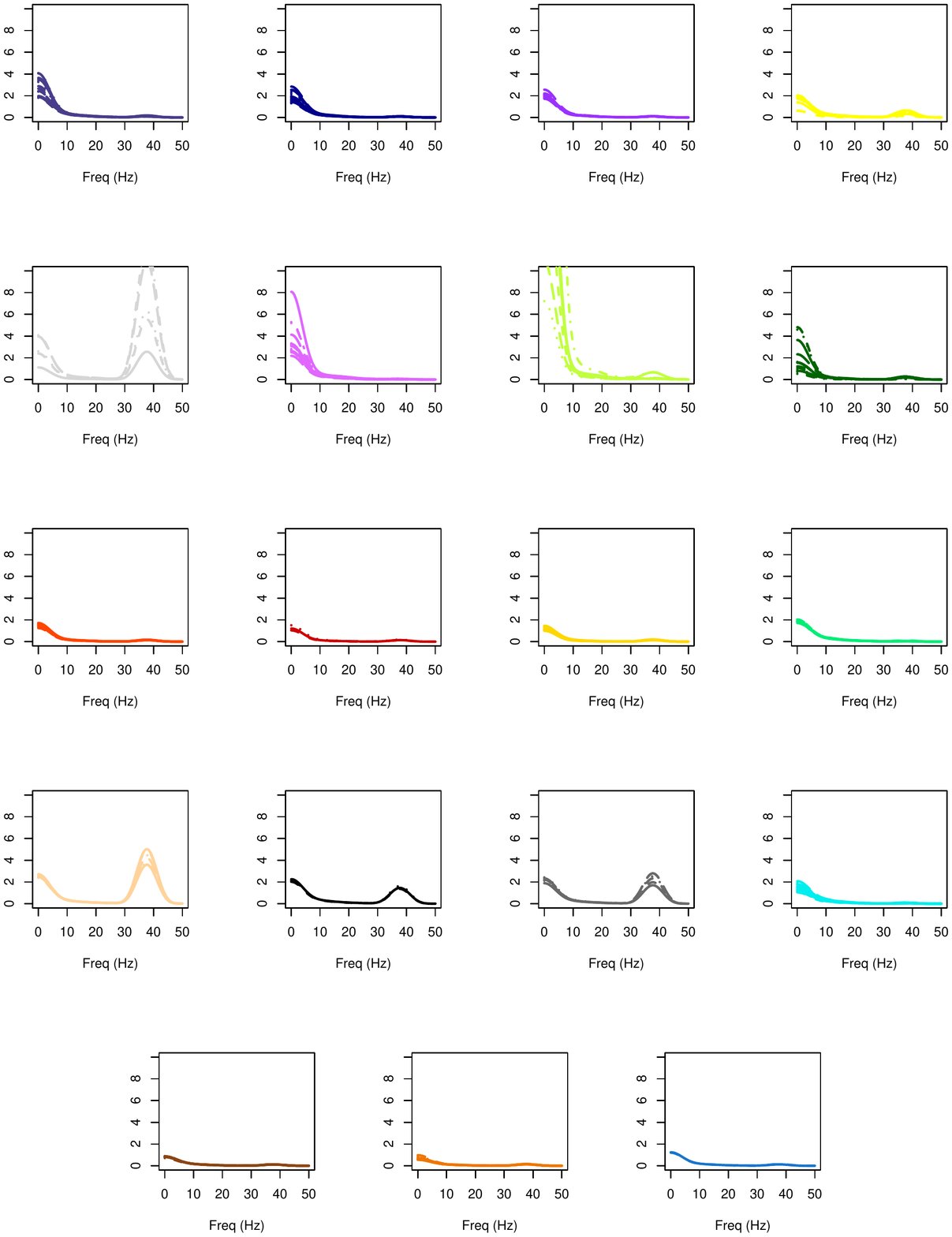}}
\caption{Initial step when there are 19 clusters for the entire resting state. (a) The affinity matrix: proportion (across epochs) when two channels belong to the same cluster. (b) Distribution of clusters across the cortical surface. (c) Mean spectra (across epochs) for each cluster.}\label{F7}
\end{figure}

In Figure \ref{F7a}, we show the ``affinity matrix" which is the proportion of epochs
when a pair of channels belong to the same cluster. The $(i,j)$ element of the
affinity matrix is the proportion of epochs that channels $i$ and $j$ are clustered
together -- regardless of how they cluster with other channels. On the lower left corner
of the affinity matrix, there are a few small red squares that represent channels
that are always clustered together and completely separated from the rest. Here we
describe two of these clusters. The first has channels that belong to the right media
parietal and central occipital regions (as defined in \cite{Wu14}). It turned out that the
spectra at these channels are highly similar (especially in the high power
at the alpha and gamma bands) and thus were joined into a single cluster by the
SMC method. This is not entirely surprising because these two regions are
adjacent to each other (in Figure~\ref{F1}) and thus the channels on the
scalp at these regions could be capturing electrophysiological activity from
the same population of neurons. The second consists of channels in the
primary motor and right parietal regions whose spectra are again similar and hence
were coalesced into one cluster. However, for the remaining clusters, the affinity
matrix suggests a higher level of uncertainty in the clustering structure over all
epochs. That is, joint memberships of pairs of channels tend to vary from one epoch
to another (as reflected by the light-blue or green shades in the affinity matrix).

The variability of the pairwise memberships suggests that $K=19$ is larger than
necessary. In Figure \ref{F7c}, we extracted a ``representative clustering", derived
from the affinity matrix, with $K=19$ over all epochs. Of course, the most accurate
way of presenting the clustering results would be to display the clusters formed
for each epoch. Here, we only show the representative clustering which is composed
of clusters whose channels are joined together for at least $50\%$ of epochs.
As suggested in Figure \ref{F7c}, the clustering determined by the proposed SMC method
(with $K=19$) is consistent with the anatomical parcellation given in \cite{Wu14}.
Next, we estimated the spectra for each channel in each cluster (displayed in
Figure~\ref{F7b}). While some clusters have obviously different spectra,
others show very similar spectra. Thus, if the SMC method was applied with
$K < 19$ then separated clusters with similar spectra would be lumped into
one bigger cluster.


\begin{figure}
\centering
\subfigure[\label{F8a}]{\includegraphics[scale=.6]{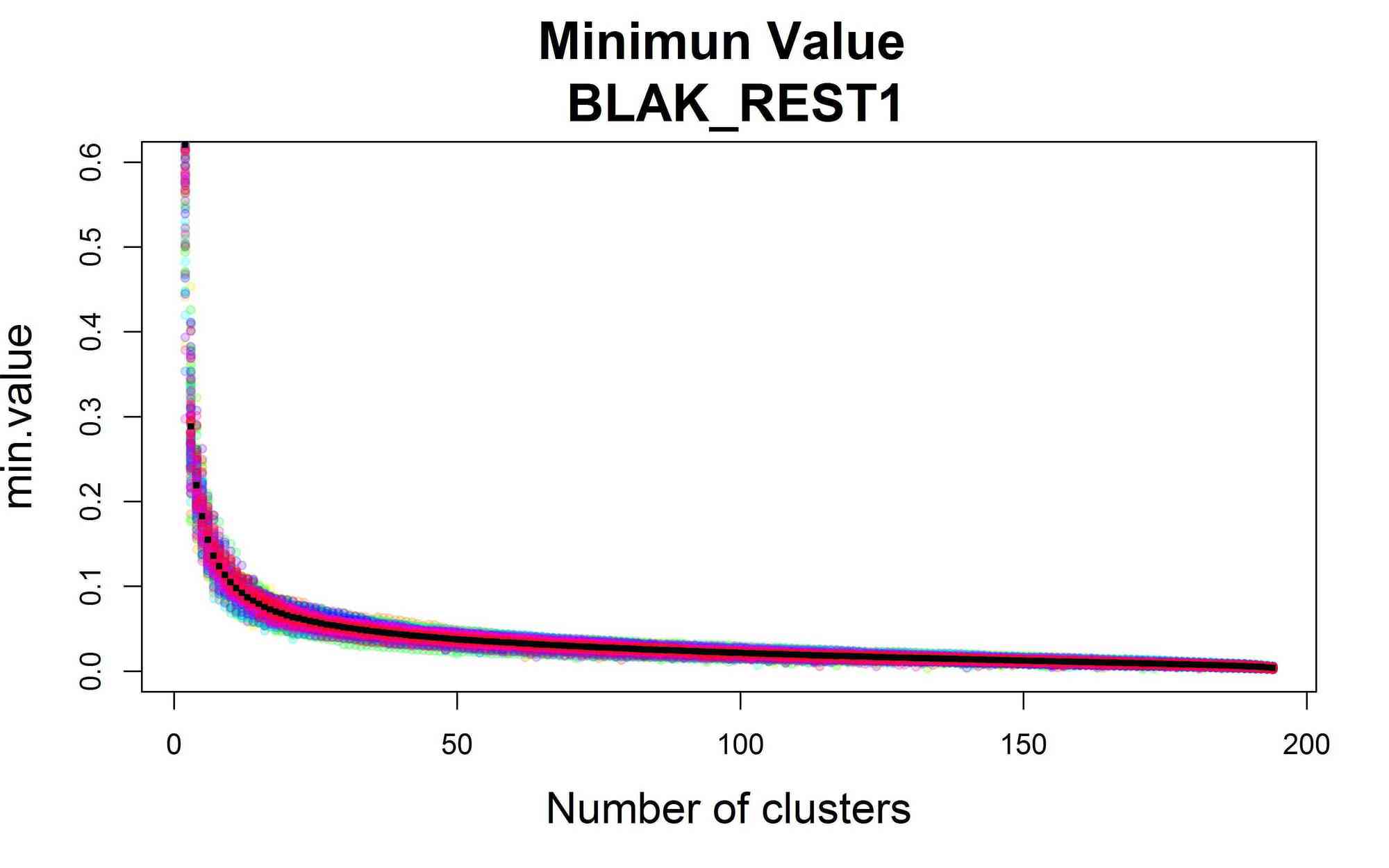}}
\subfigure[\label{F8b}]{\includegraphics[scale=.6]{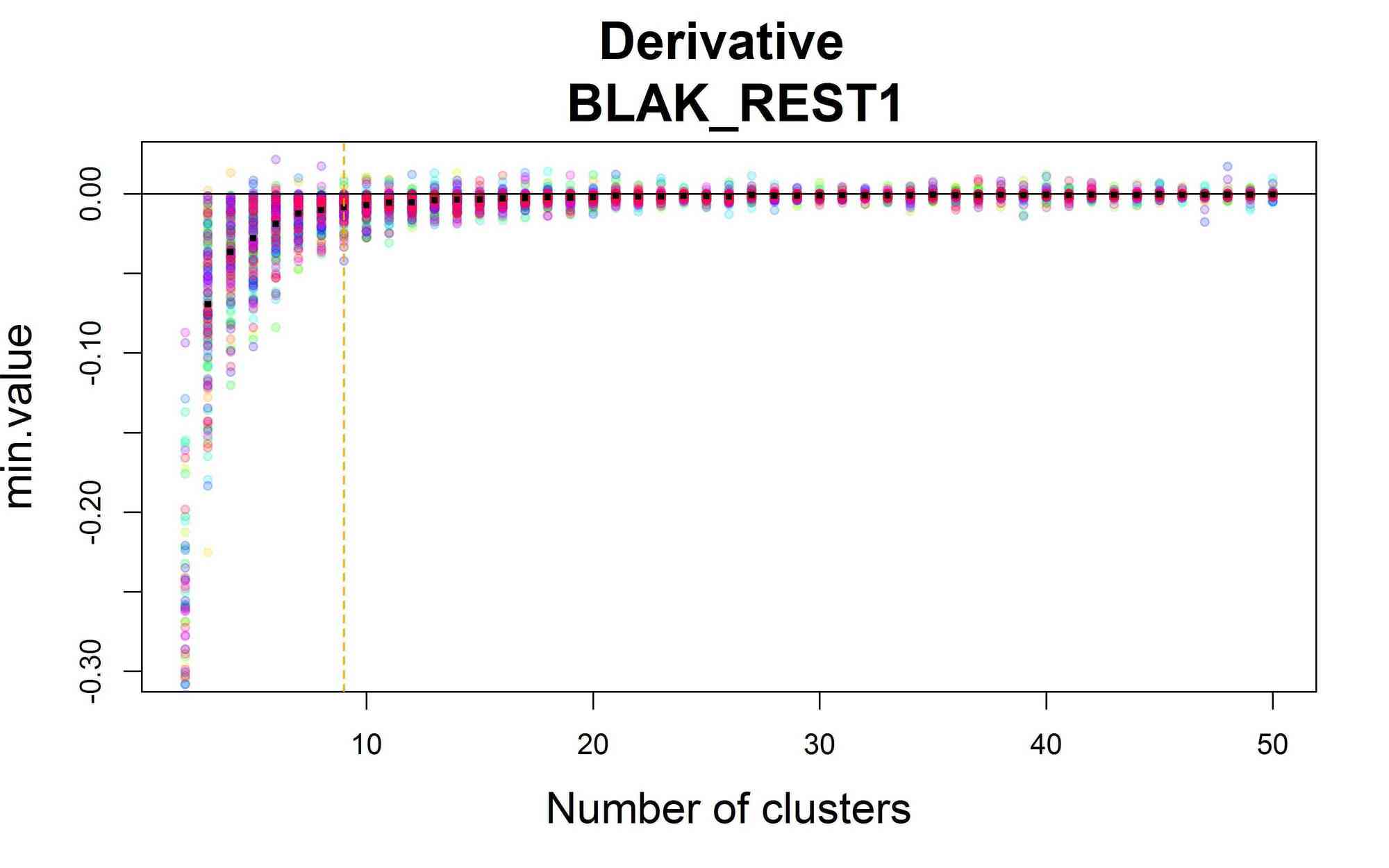}}
\caption{(a) Trajectories of the minimum TVD between clusters for the 160 epochs (subject name BLAK). (b) Numerical derivative of the minimum distance.}\label{F8}
\end{figure}

Now we want to determine a reasonable number of clusters for this particular data
by checking the analogue of  the elbow of the scree plot which in this case is
the trajectory of the minimum value of the TVD. Figure \ref{F8a} shows the value
of the minimum TVD between any two clusters at each step of the clustering algorithm,
as a function of the number of clusters. The curves for the 160 epochs are plotted along
with the mean curve (black dotted curve). As can be seen there is not a lot of variability in
the curves across epochs. Figure \ref{F8b} presents the numerical derivative of the
curves which is an effective visual tool for selecting the number of clusters by identifying
the first value of $K$ where the numerical derivative was below a small threshold
(here, we used $0.01$, based on empirical evidence from previous simulations).
According to the first derivative tool, $K=9$ was selected as the optimal number of clusters.

\begin{figure}
\centering
\subfigure[\label{F10a}]{\includegraphics[scale=.5]{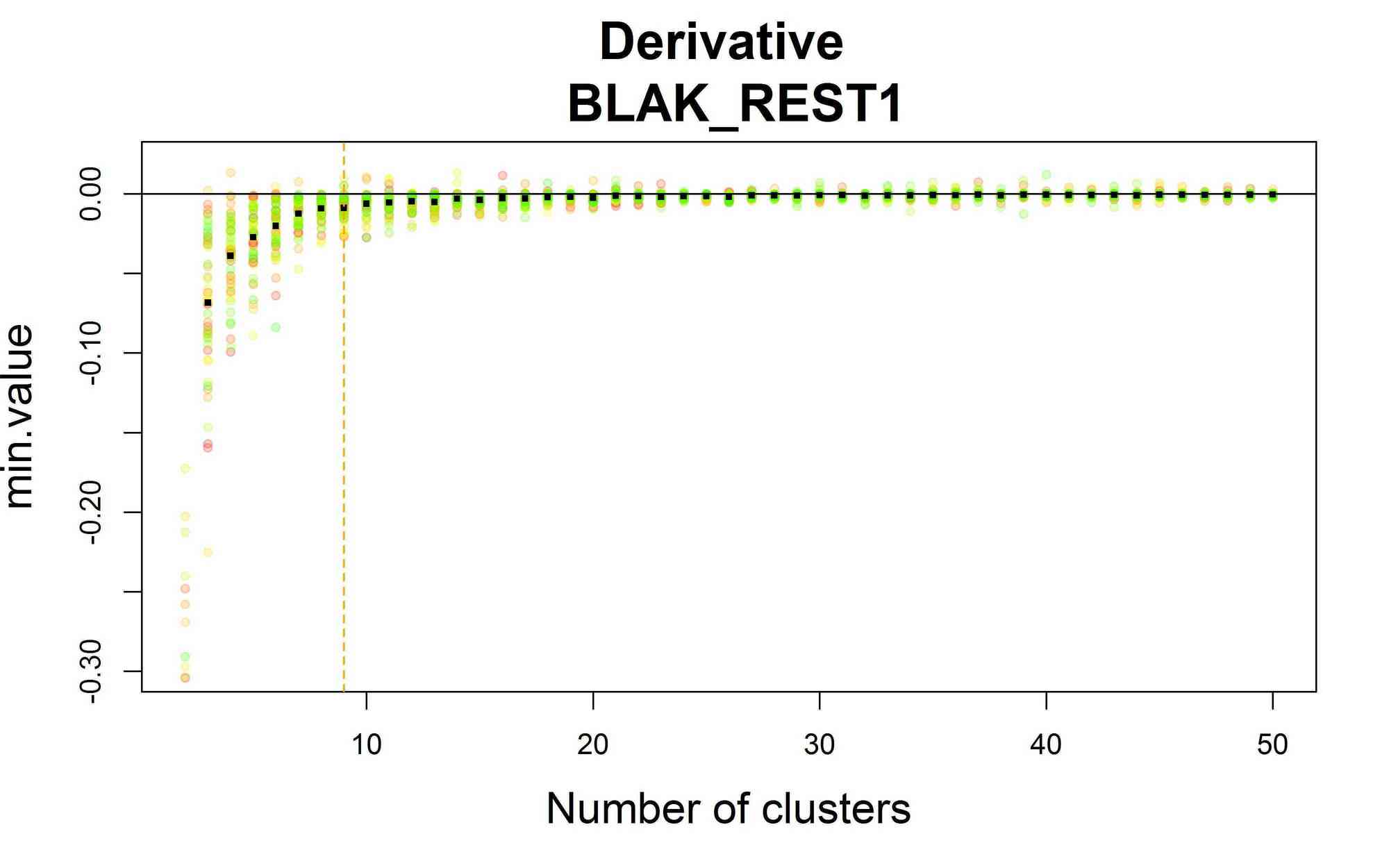}\hspace{.2cm}\includegraphics[scale=.5]{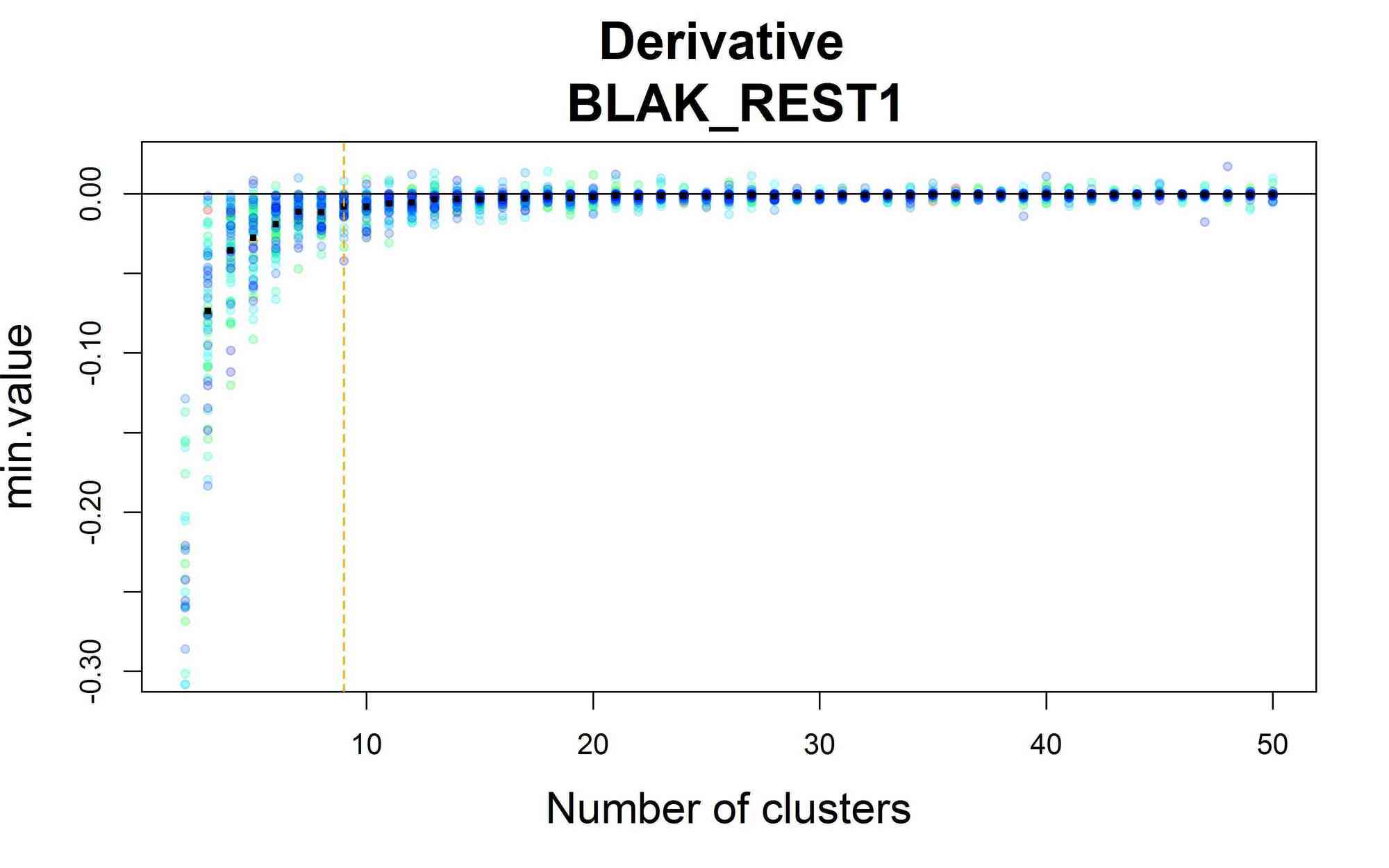}
             \hspace{.2cm}\includegraphics[scale=.5]{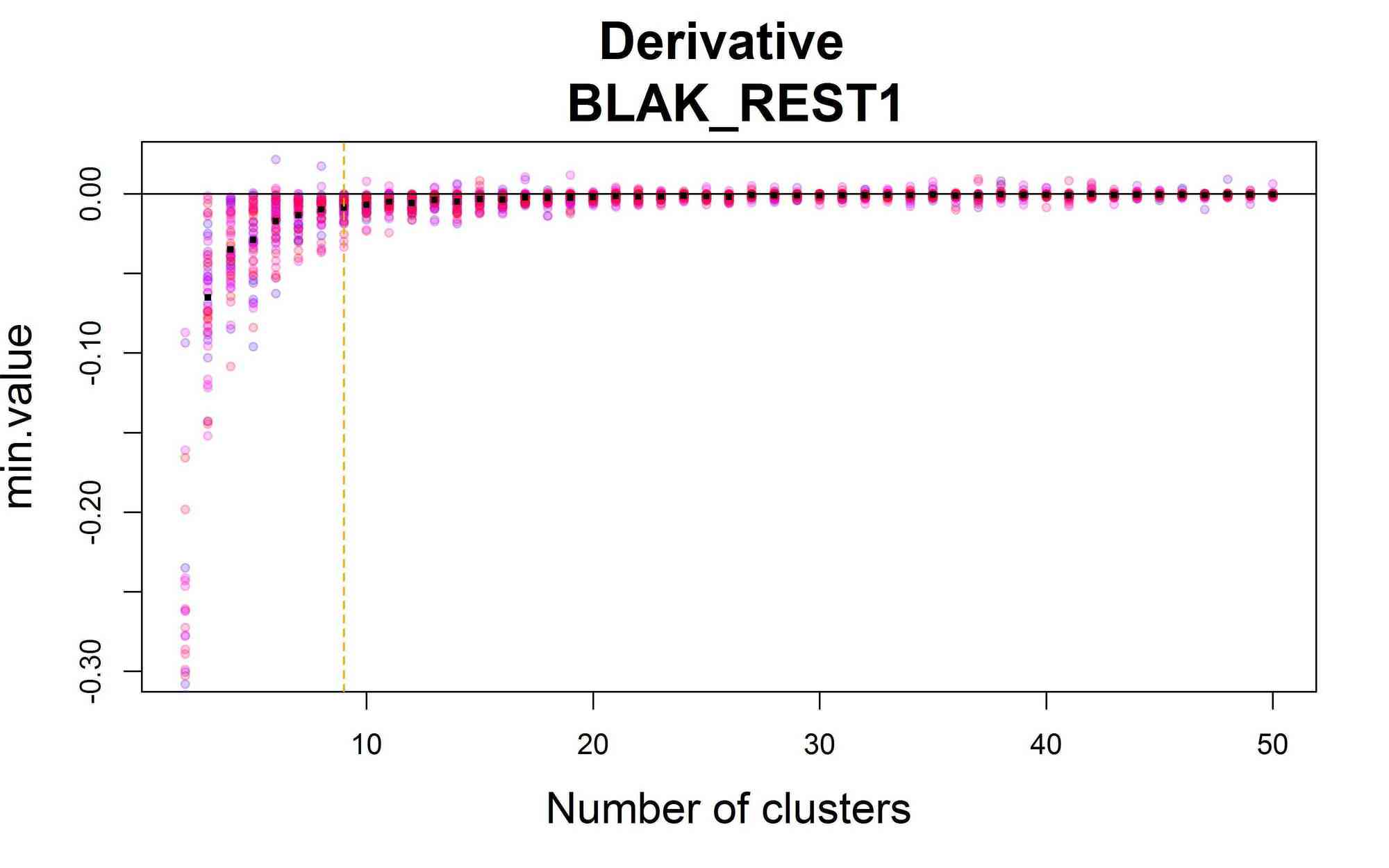}}
\subfigure[\label{F10b}]{\includegraphics[scale=.21]{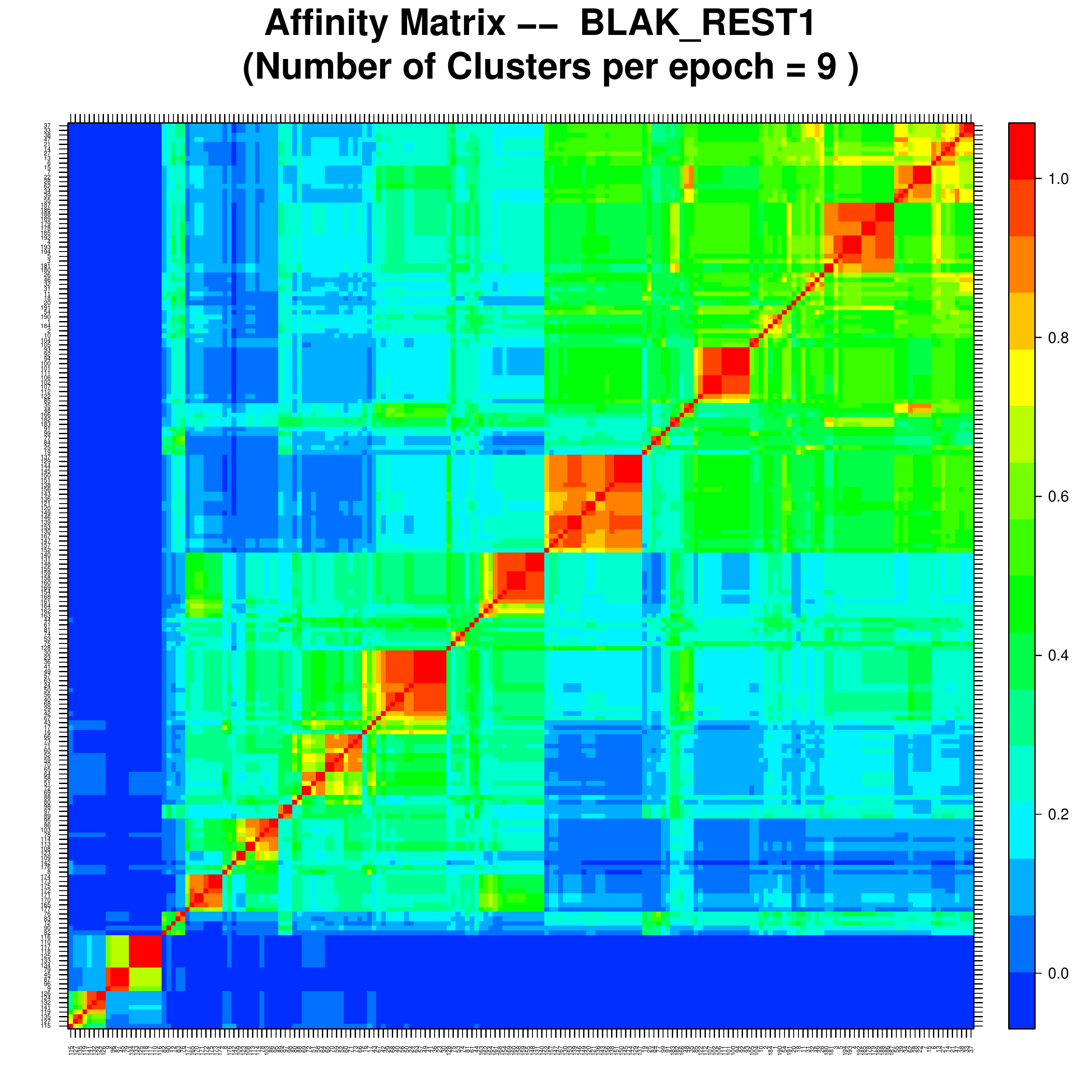}\hspace{.3cm}\includegraphics[scale=.21]{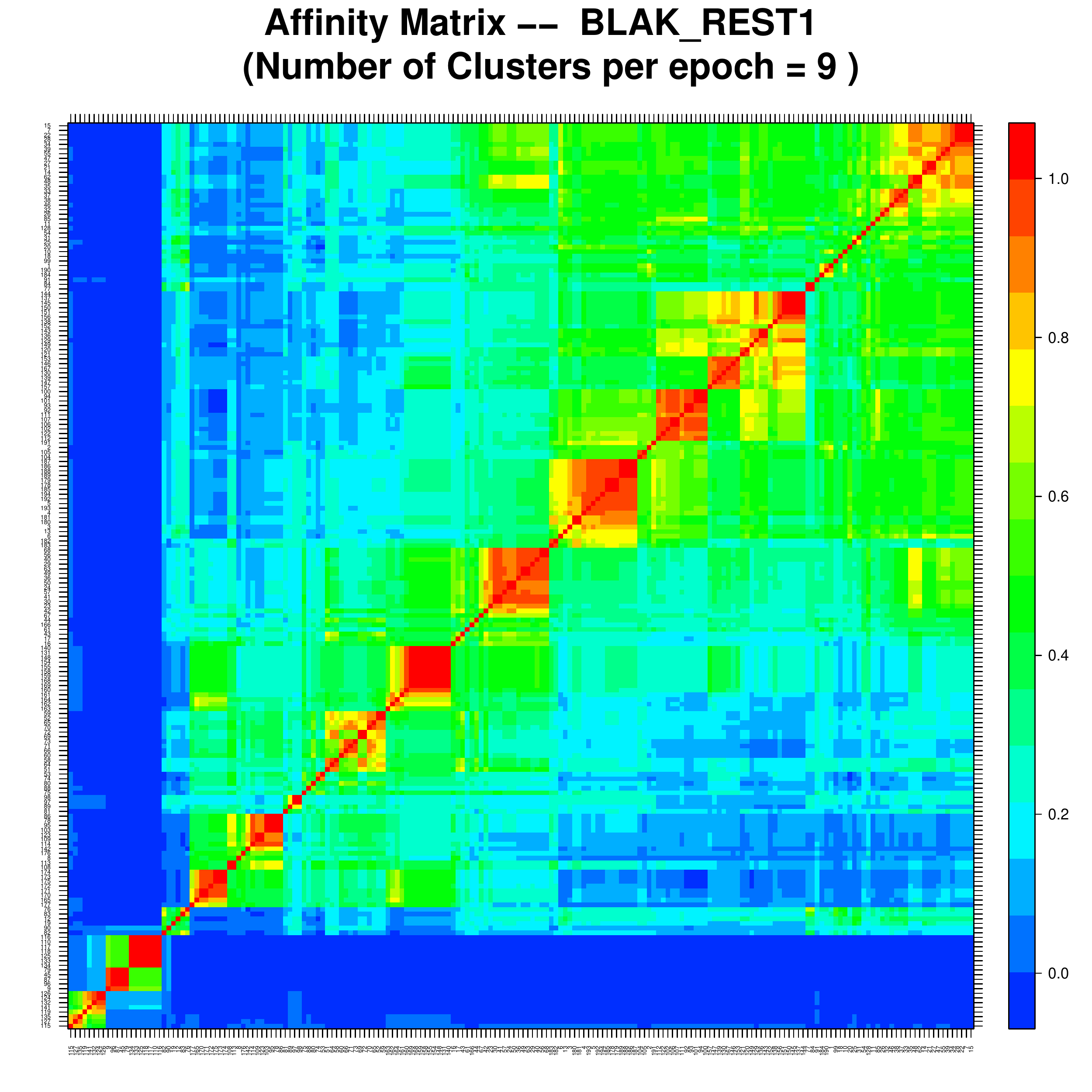}\hspace{.3cm}\includegraphics[scale=.21]{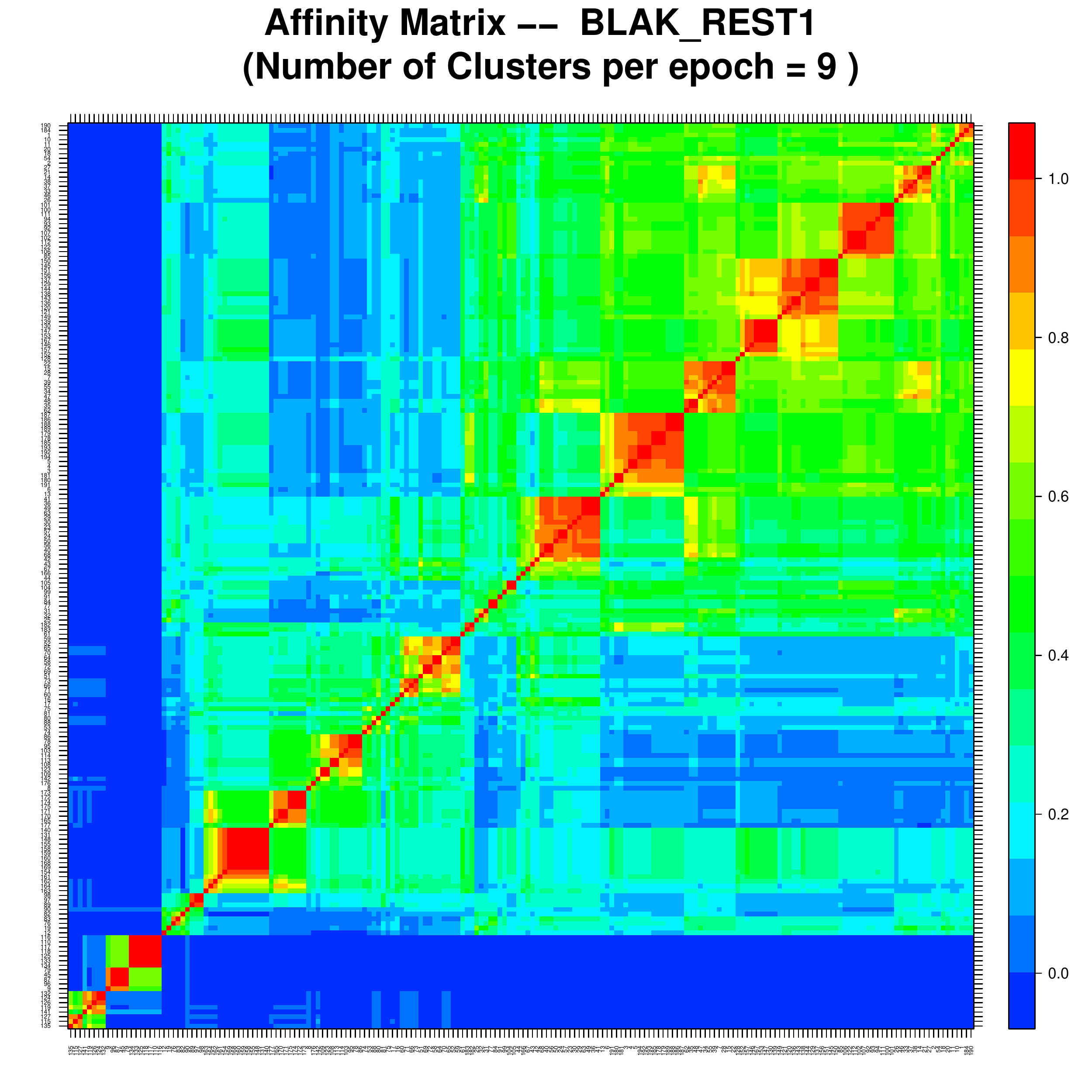}}
\caption{a) Minimum value trajectories by segments 1-50, 51-110 and 111-160. b) The affinity matrix: proportion of epochs where channel $i$ and $j$ belong to the same cluster, with $9$ clusters. }\label{F10}
\end{figure}

Even though the number of clusters remain constant across epochs, the cluster formation
(i.e., location, spatial distribution, specific channel memberships) of the 9 clusters
may vary across epochs. In this EEG analysis, the total number of epochs was divided into
three different phases of the resting state: early (epoch numbers 1 to 50), middle
(epochs 51-110) and late (epochs 111-160). Figure \ref{F10a} shows the derivative curves
for the minimum value trajectories. Again it appears reasonable to conclude that the number
of clusters across epochs remains constant. Figure \ref{F10b} shows that the affinity matrix
when $K=9$ shows more red squares compared to $K=19$. This indicates more consistent
clustering (across epochs) which is expected because the proportion of a pair of channels
belonging to the same cluster is higher when $K=9$ compared to $K=19$.

It is evident that clustering evolved across the three phases. The affinity matrix for the
early and late phases shows darker red colors which has a wider spread than that during the
middle phase. This indicates that (a.) the proportion of times that a pair of channels
belongs to the same cluster is higher in the early and late phases than during the middle
phase; and (b.) there are more pairs with higher proportions of joint membership during
the early and late phases than during the middle phase. Some of the well defined types of
clusters are located at the left pre-motor and right anterior parietal regions for the
early phase and located at the central parietal and left pre-motor regions for the late phase.
The interpretation here is that
the clustering formation during the early and late phases have a higher degree of certainty
and that there is clearer delineation of the boundaries between clusters. This waning of
cluster formation (weakening of the ties between channels within a cluster) during the
middle phase is not easily explainable for a resting state. It would have been easier
if the subject had to attend to stimuli presentations and in that case the mind drifts and
wanders while in the middle of a long experiment.

\subsection{Descriptive comparison of cluster formation for two individuals}

In this analysis, we examined two representative subjects: one identified as BLAK
whose motor skill performance showed very little improvement during the experiment
(whose percentage improvement score of $9.78$ was low relative to this cohort) and MOHK
the highest performer with percentage improvement of $25.68$.  The neurologist
collaborator was keenly interested in investigating differences between spectral
synchronicity between these two individuals who, very roughly, represent the
``low" and ``high" performers.

\begin{figure}
\centering
\subfigure[\label{F11a}]{\includegraphics[scale=.6]{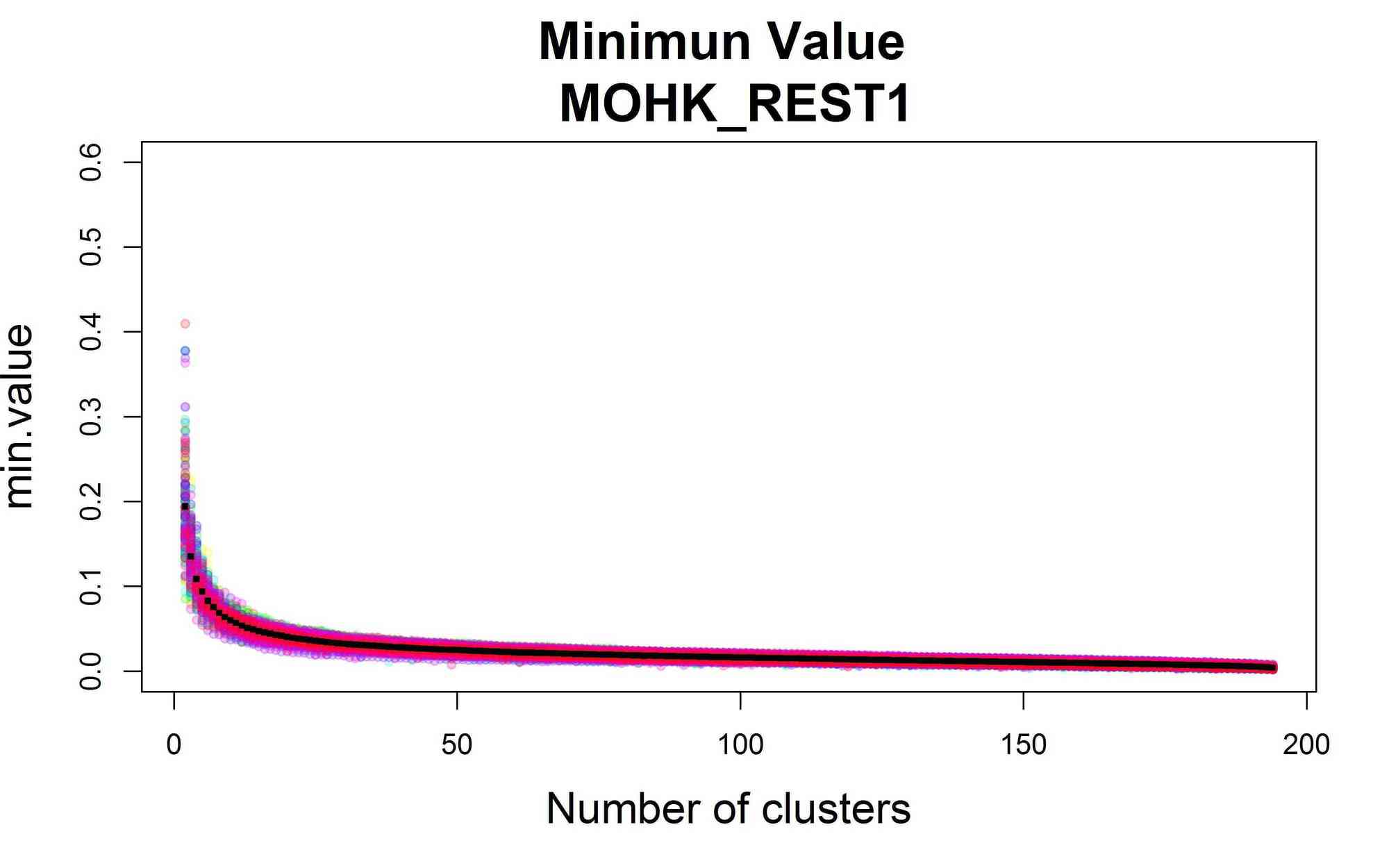}}
\subfigure[\label{F11b}]{\includegraphics[scale=.6]{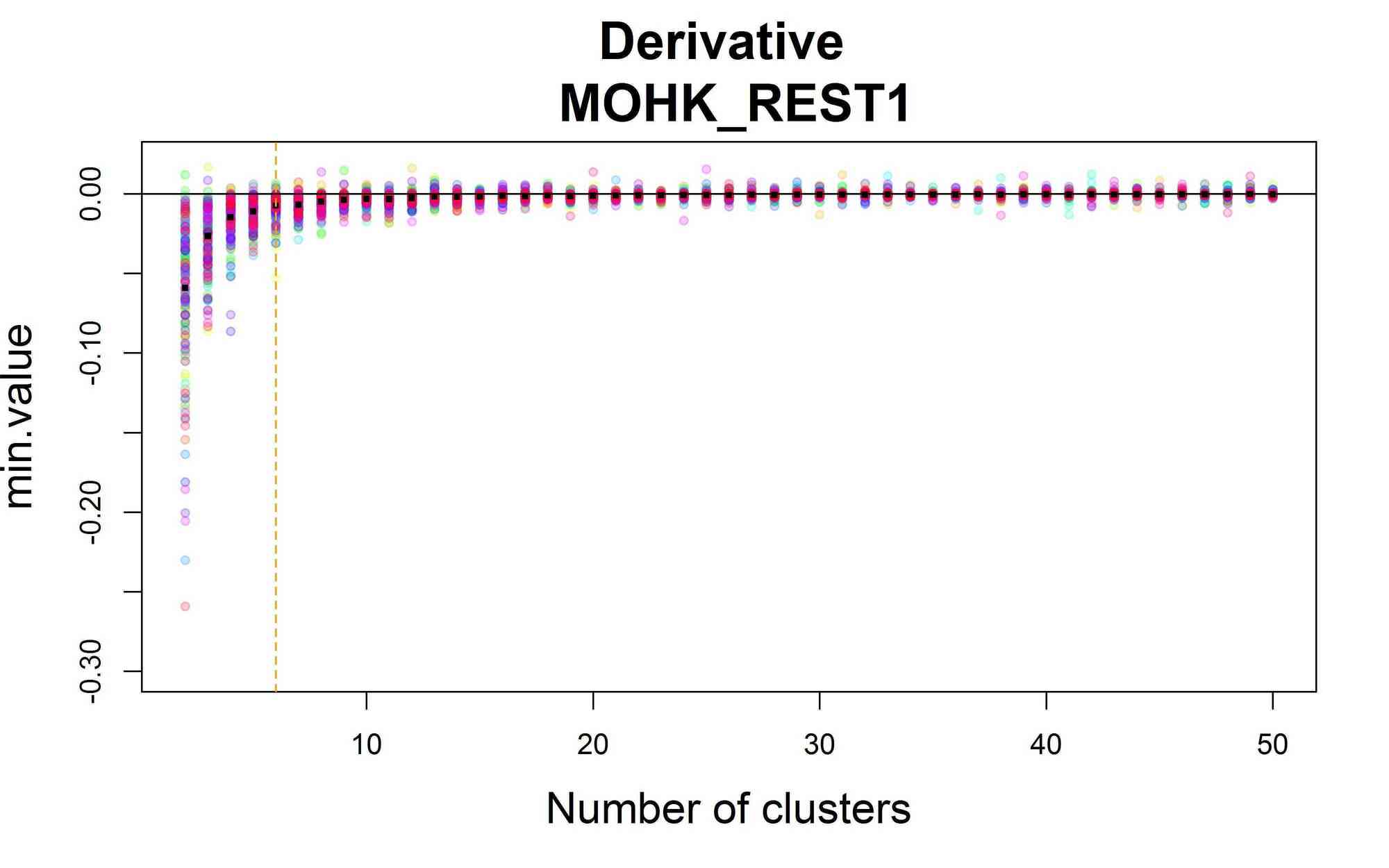}}
\caption{Trajectories of the minimum value stored by the algorithm for all the epochs (160, subject name MOHK).}\label{F11}
\end{figure}

\begin{figure}
\centering
\subfigure[\label{F12a}]{\includegraphics[scale=.5]{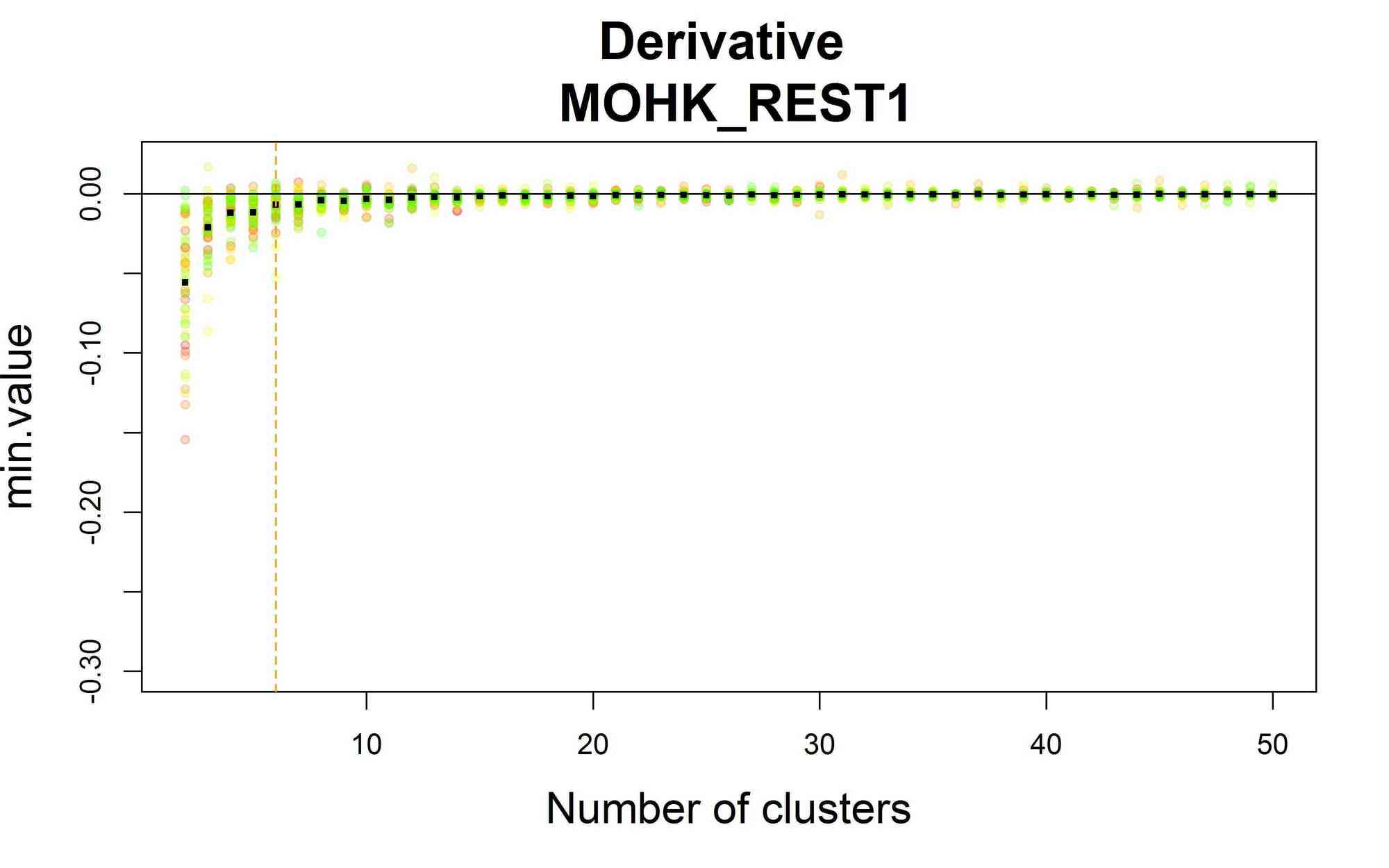}\hspace{.2cm}\includegraphics[scale=.5]{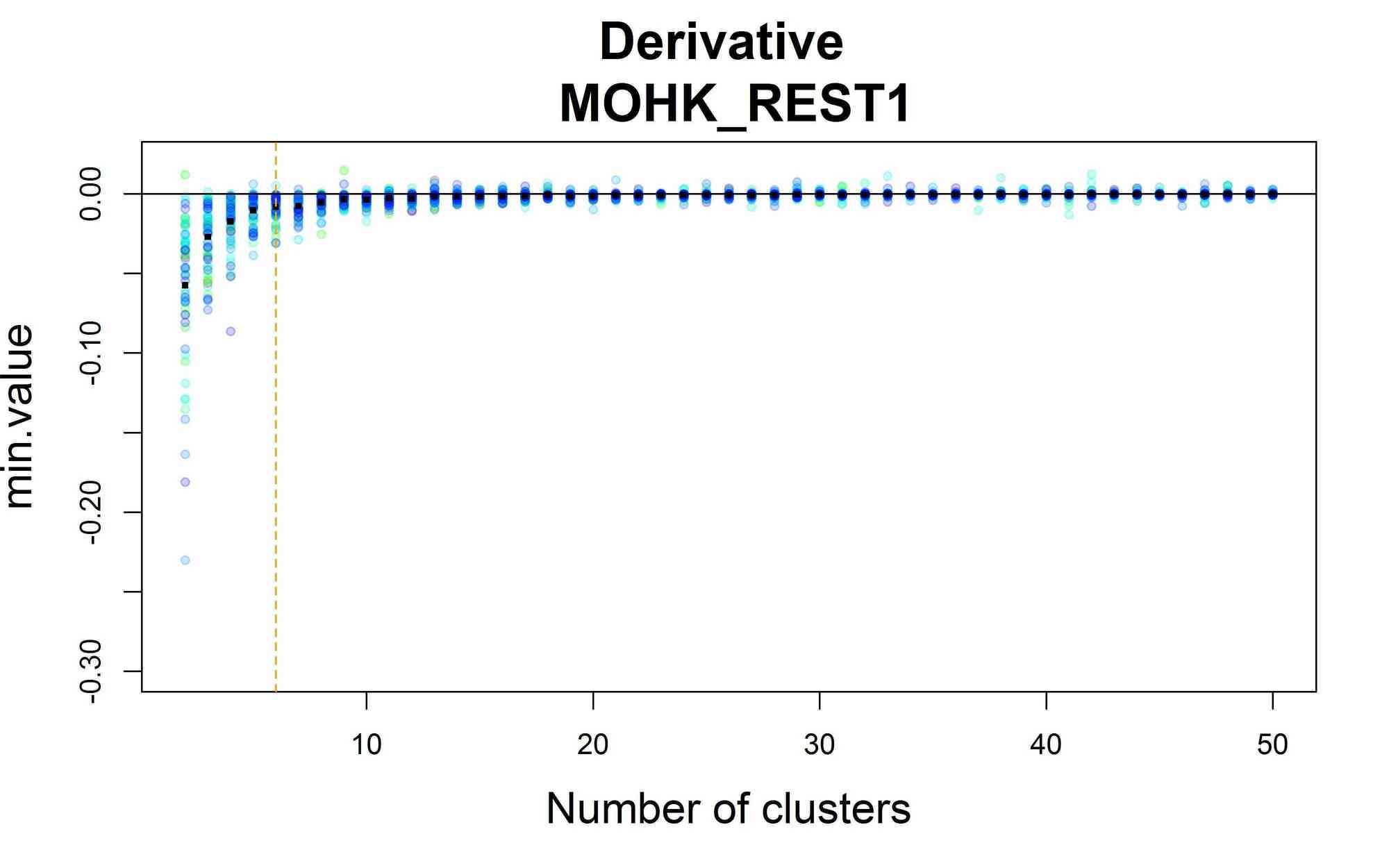}\hspace{.2cm}\includegraphics[scale=.5]{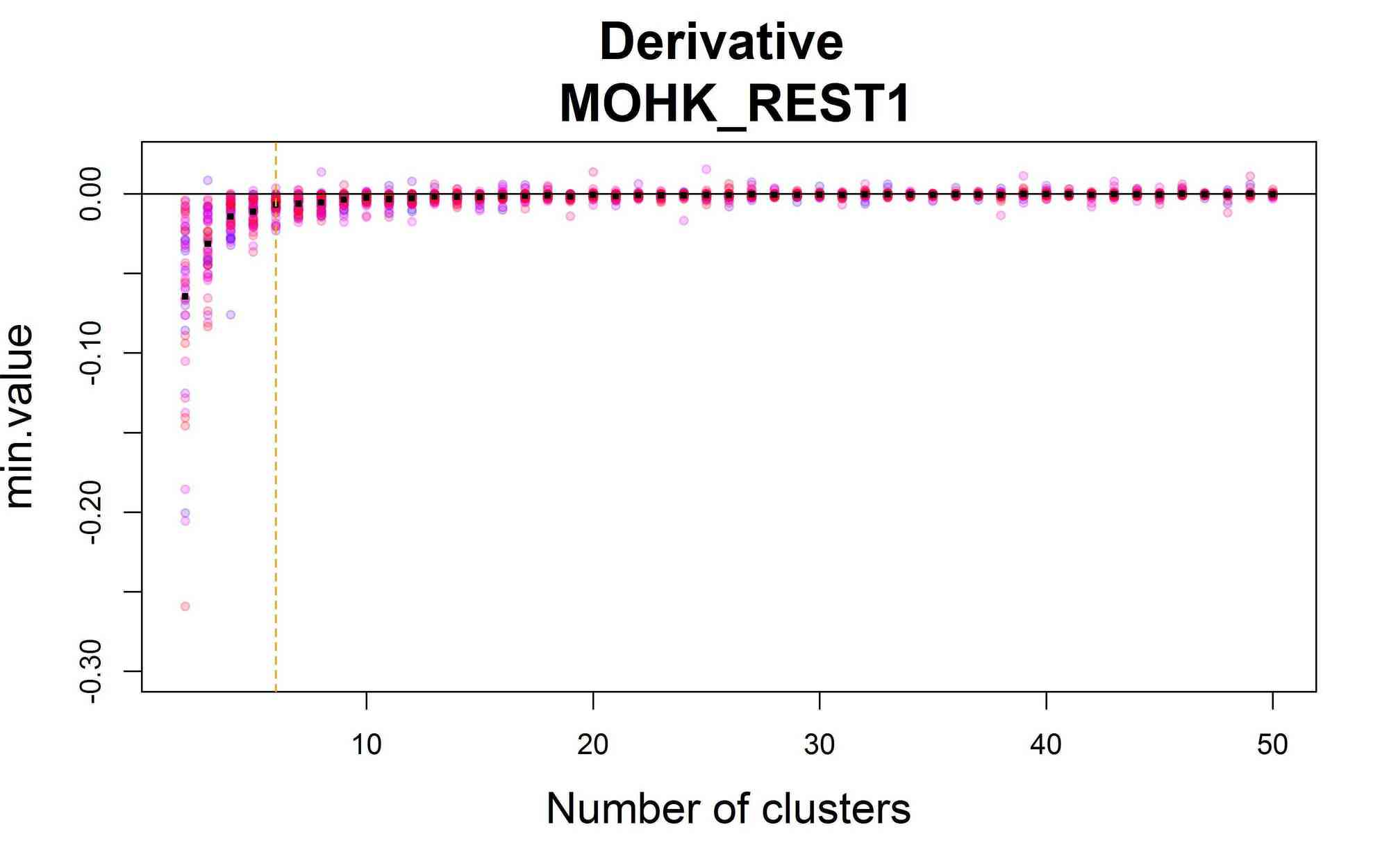}}
\subfigure[\label{F12b}]{\includegraphics[scale=.21]{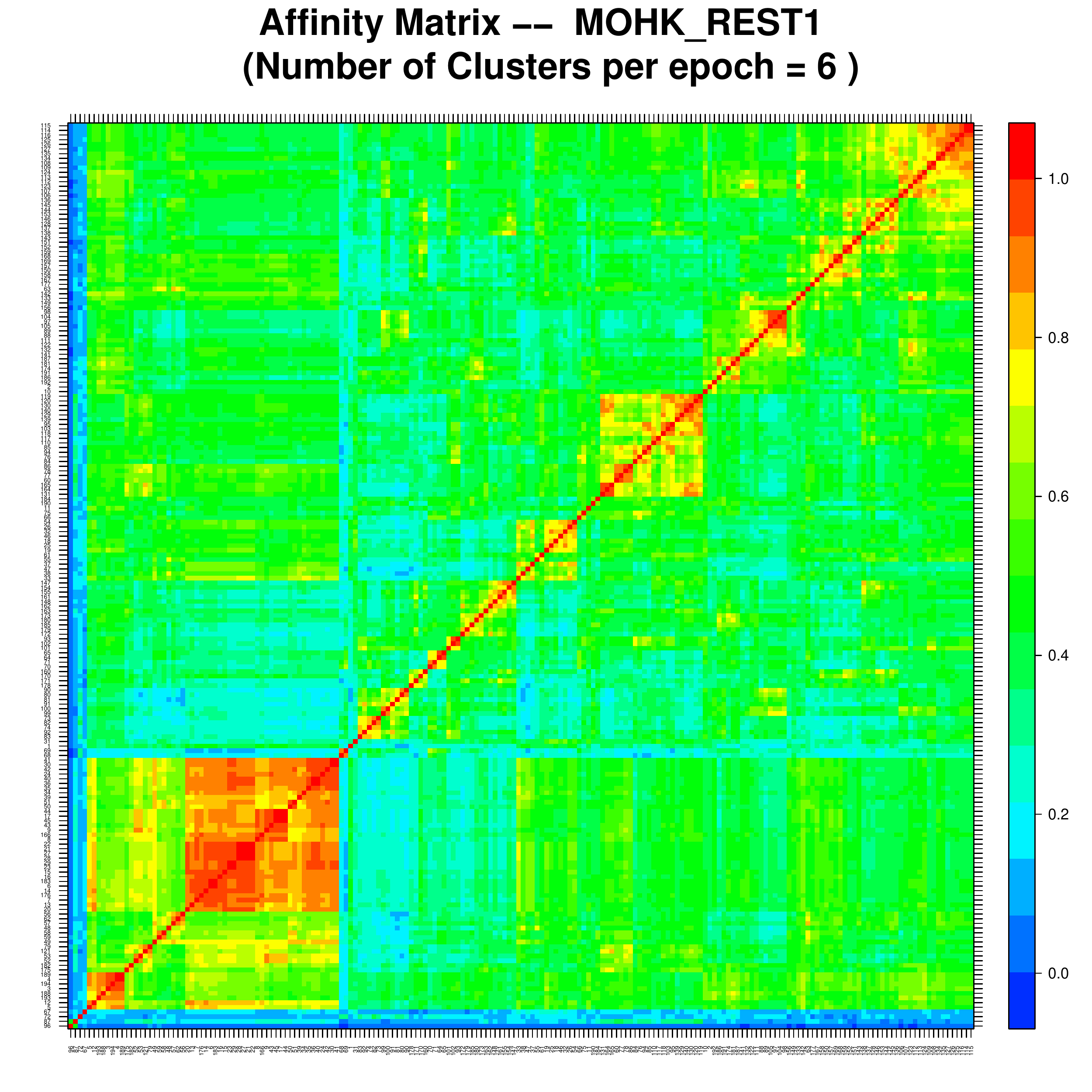}\hspace{.3cm}\includegraphics[scale=.21]{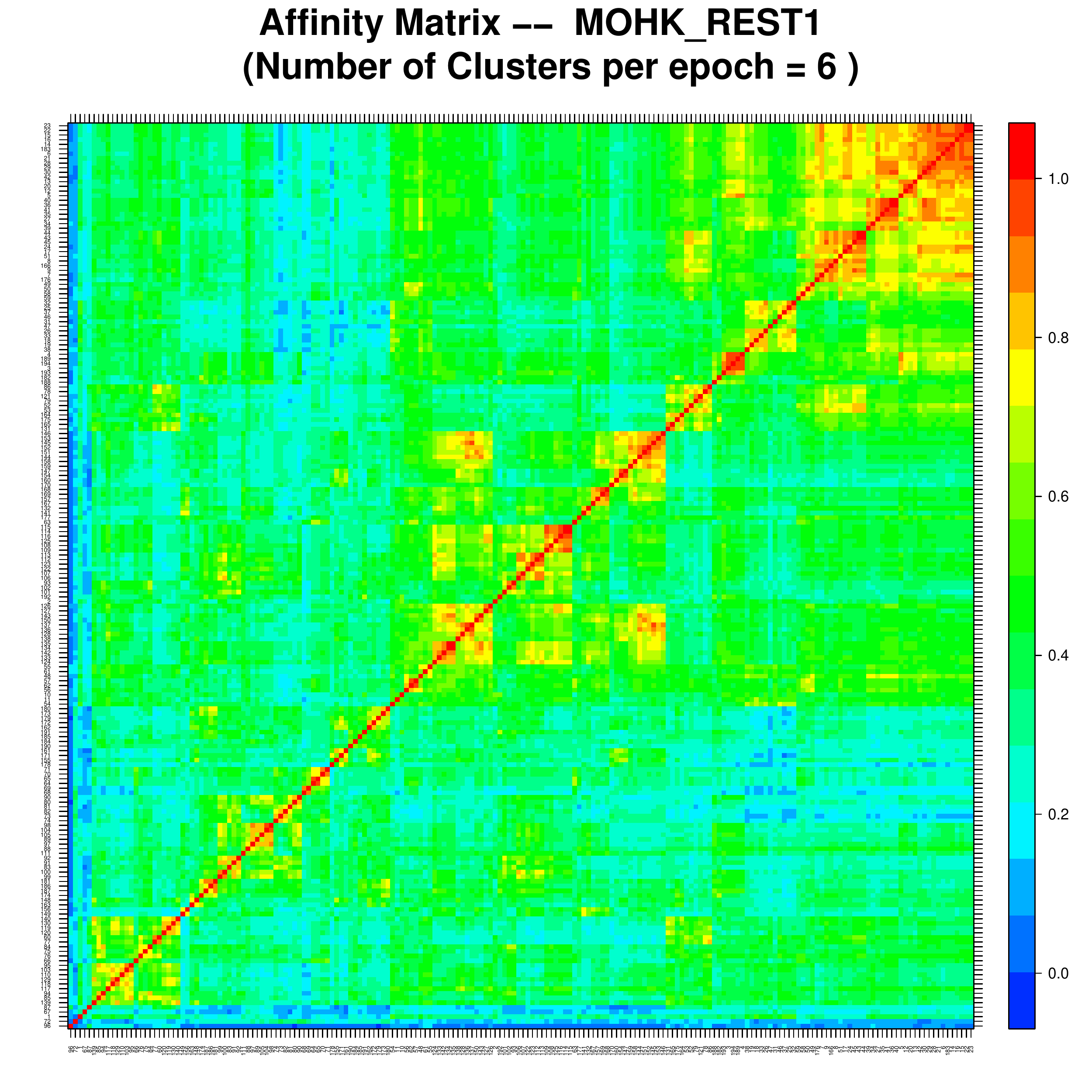}\hspace{.3cm}\includegraphics[scale=.21]{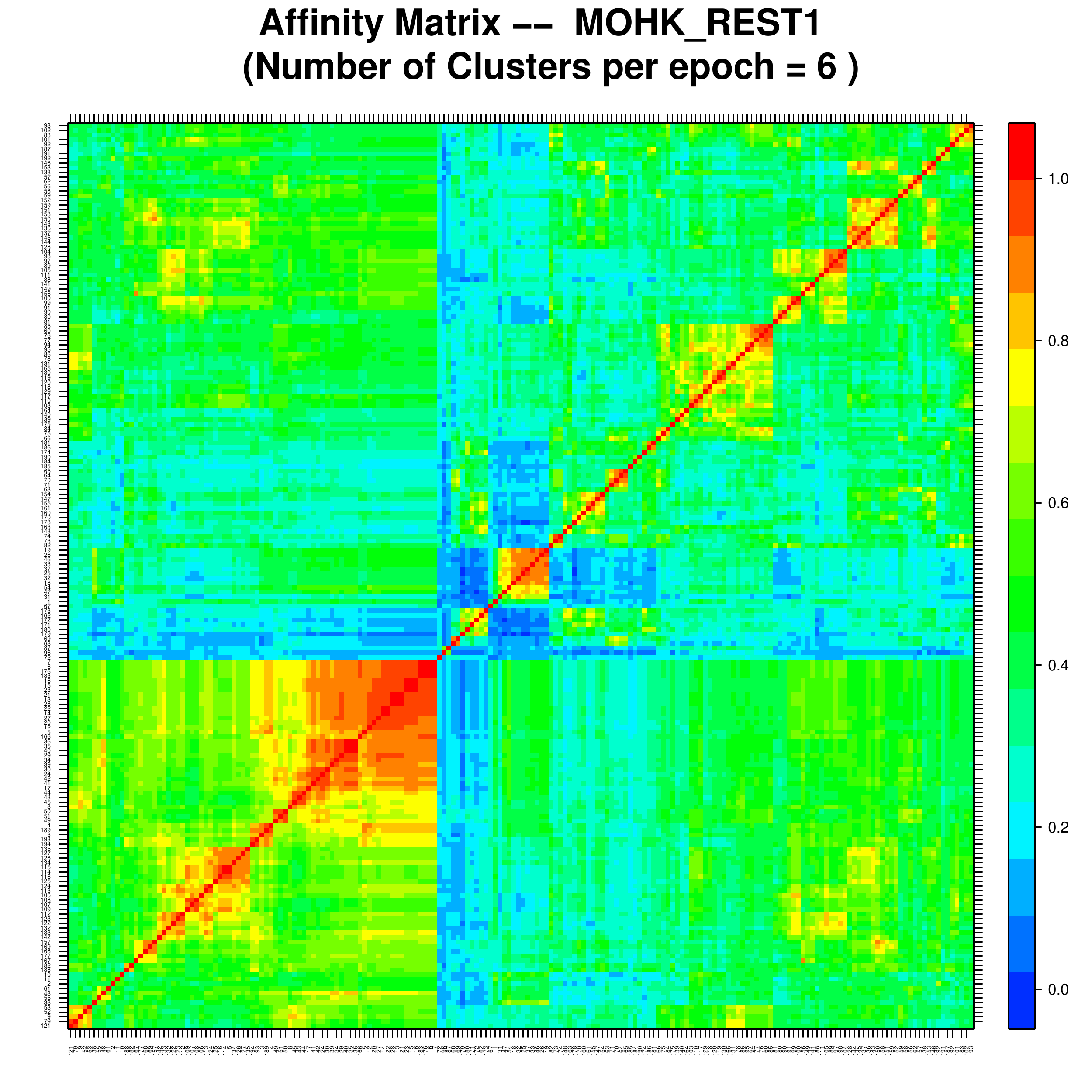}}
\caption{a) Minimum value trajectories by segments 1-50, 51-110 and 111-160. b) The affinity matrix: proportion of epochs where channel $i$ and $j$ belongs to the same cluster, with $7$ clusters and the total number of epochs divided. }\label{F12}
\end{figure}

The results for both subjects are clearly different. Figure \ref{F11} shows the trajectories
for the minimum value of the TVD. These trajectories do not appear to vary a lot across
epochs. The optimal number of clusters for subject MOHK was $K=6$ (as opposed to $K=9$
for BLAK). Figure \ref{F12} shows the derivative for each stage and, as noted, the number
of clusters seem to be constant across epochs. However, in this case there is more uncertainty regarding the clustering structure in each stage. The middle phase also appears to be the one
with more variability (less red squares and more green regions). At the early phase, there is
only one group of channels that is tightly clustered together which are located at the left
pre-motor area and supplementary motor area. At the late phase this cluster is located at predominantly at the left prefrontal area. In this sense, one big difference between these
individuals is the location of the clusters that have less changes across epochs on the different phases. The subject BLAK had more at the parietal regions while MOHK had more at the pre-motor area and prefrontal.

To summarize the results at this point.
The number of clusters that were created based on the similarity of the spectra
is smaller than the number of regions obtained from anatomical parcellation.
This suggests that some regions, though anatomically distinct, share similarities
in terms of the actual underlying local brain process that gave rise to the observed
EEG signals during resting state.

\subsection{Comparing clusters across the different phases}\label{S5.3}

The next step in our analysis is to compare the clustering results across the different
phases of resting state. The goal here is to address the question about whether and how
brain organization, as expressed by cluster formation, evolves during resting state.
As noted, even when the number of clusters remains constant across the entire resting
state, the spectra of the underlying processes could change which will then be reflected
and also on the formation of clusters at the cortical surface. Since there are at least
50 epochs per phase, in order to present a summary of the clustering results for each phase,
we focus only on the ``representative'' clustering. Using the affinity matrices (in Figures \ref{F10} and \ref{F12}), we applied a hierarchical cluster analysis with the complete
linkage function to obtain the representative clustering.

\begin{figure}
\centering
\subfigure[\label{F9b}]{\includegraphics[scale=.23]{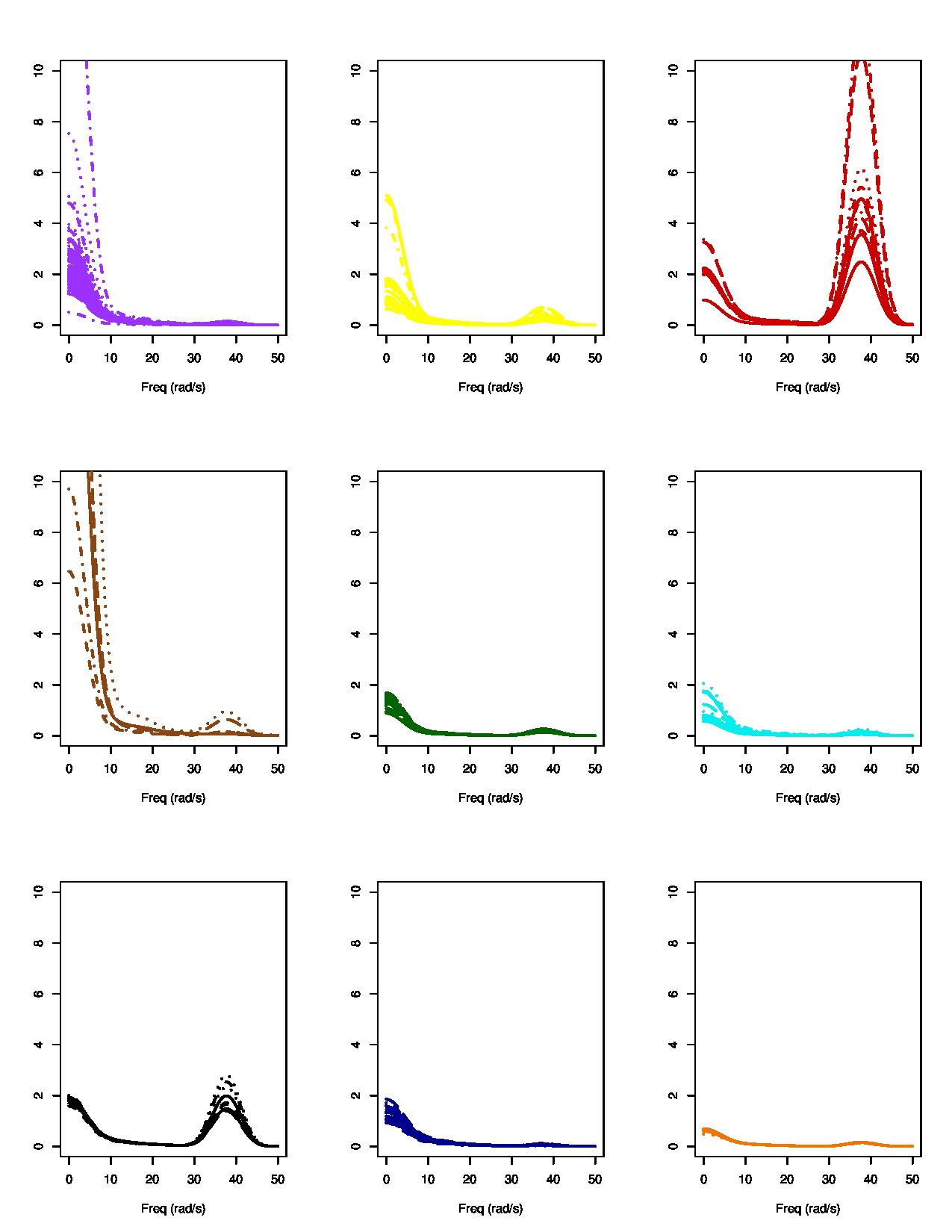}\hspace{.3cm}\includegraphics[scale=.23]{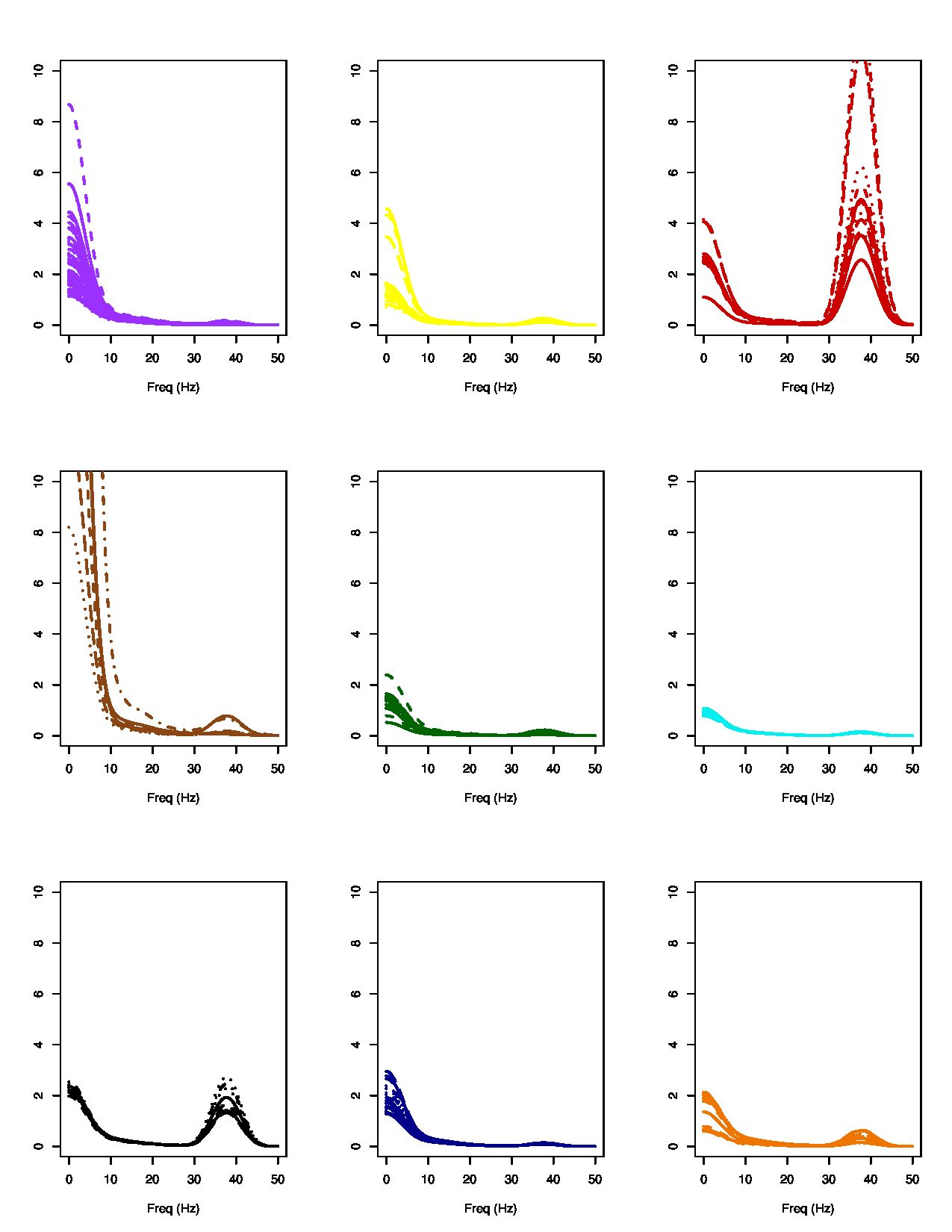}\hspace{.3cm}\includegraphics[scale=.23]{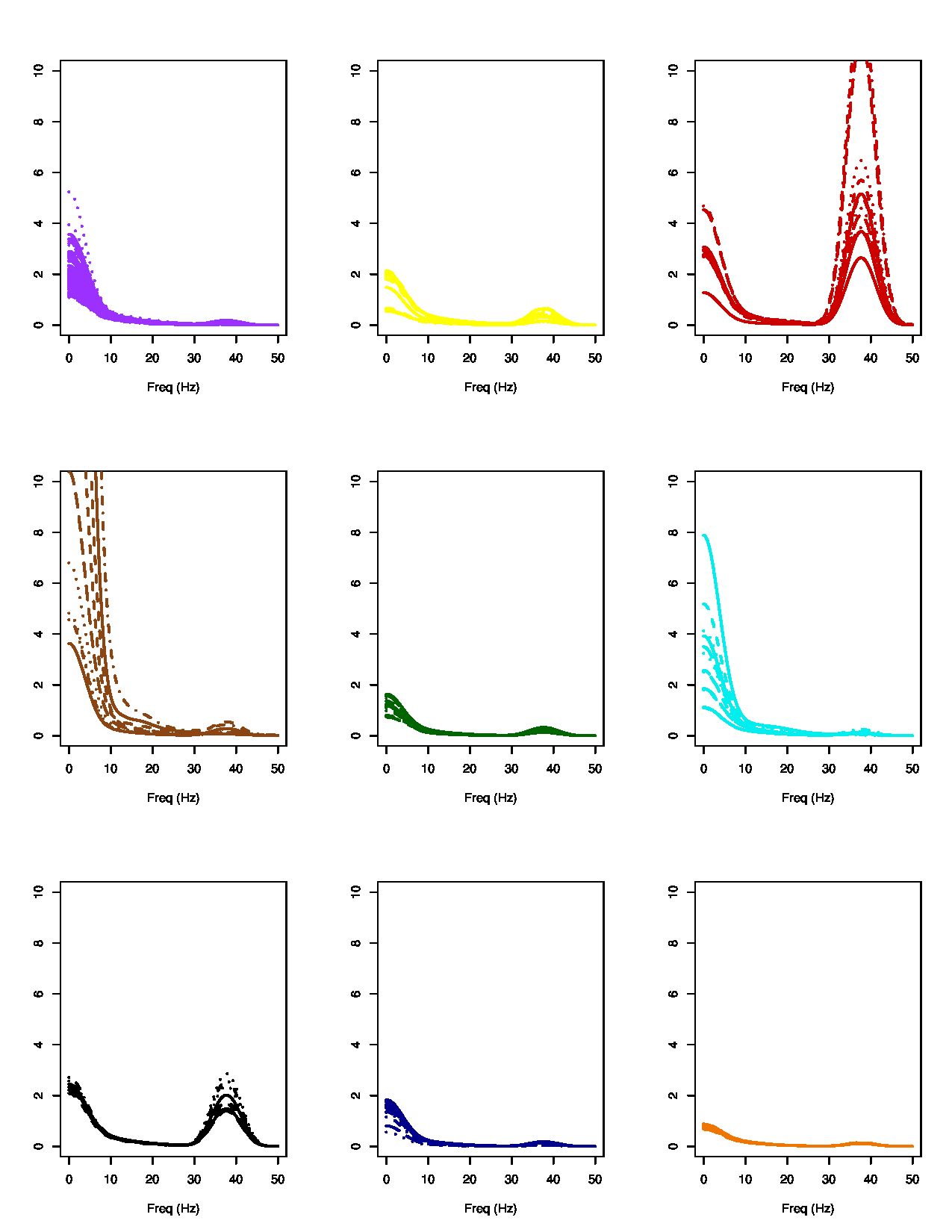}}
\subfigure[\label{F9c}]{\includegraphics[scale=.22]{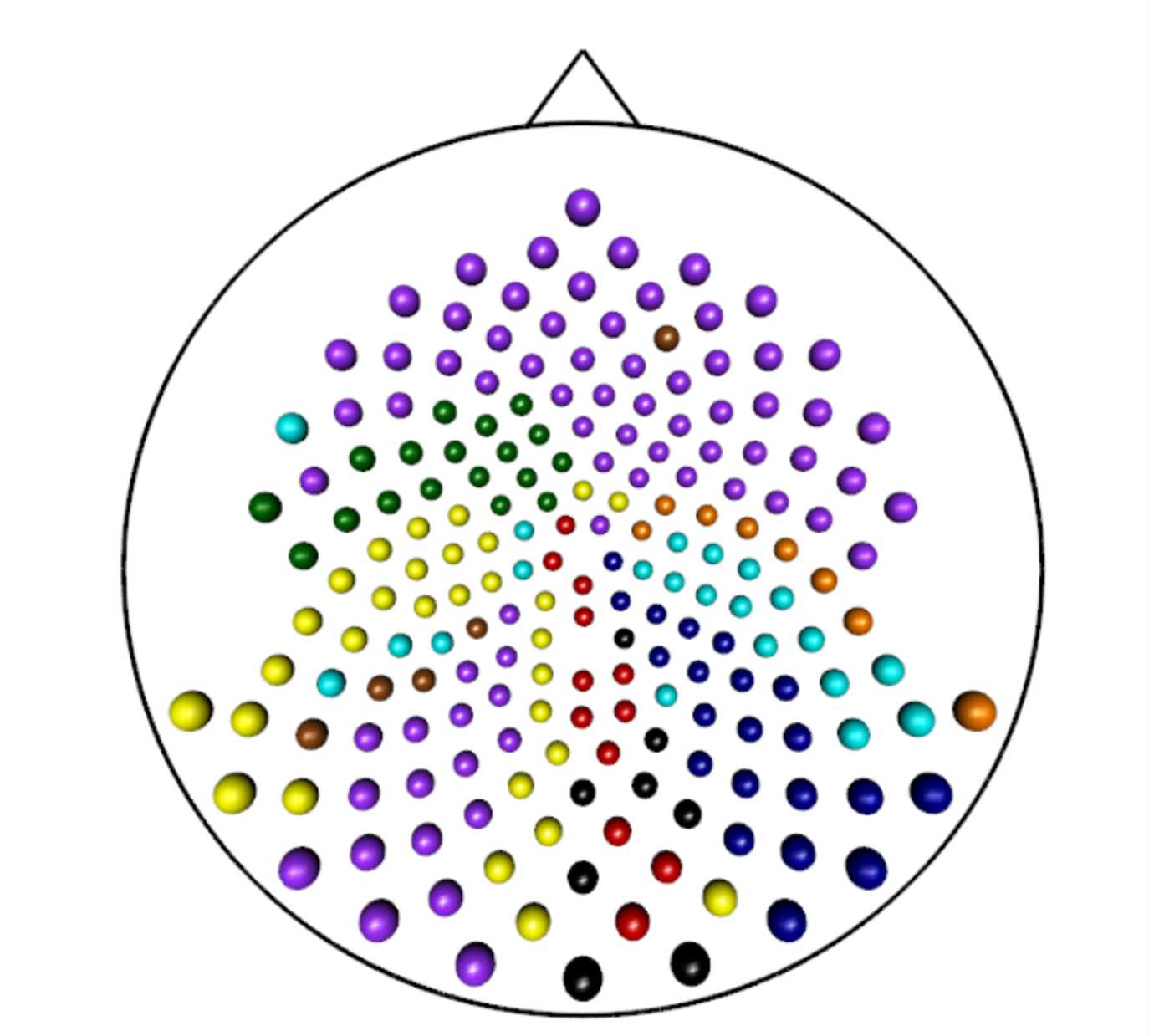}\hspace{.2cm}\includegraphics[scale=.22]{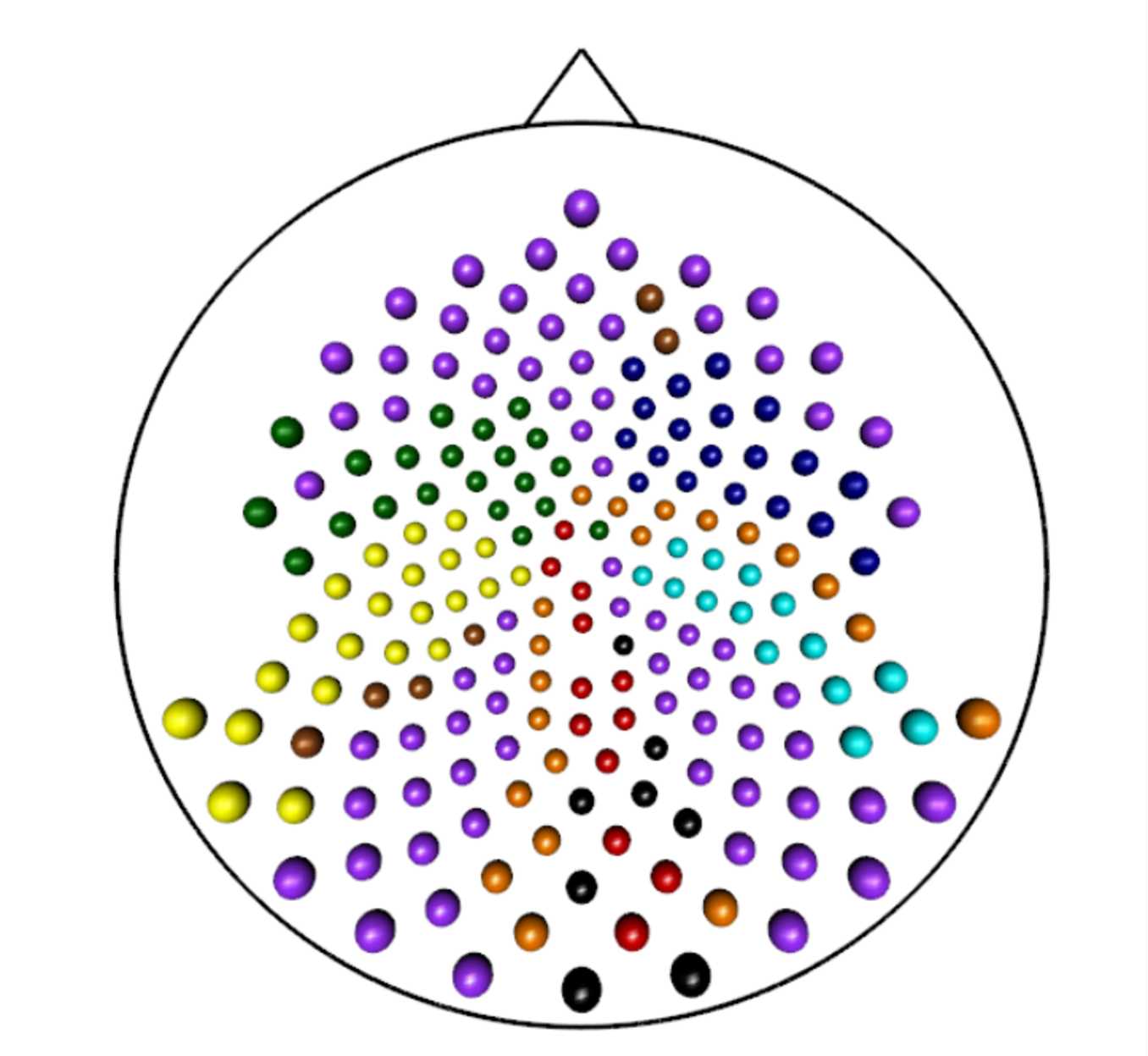}
             \hspace{.2cm}\includegraphics[scale=.22]{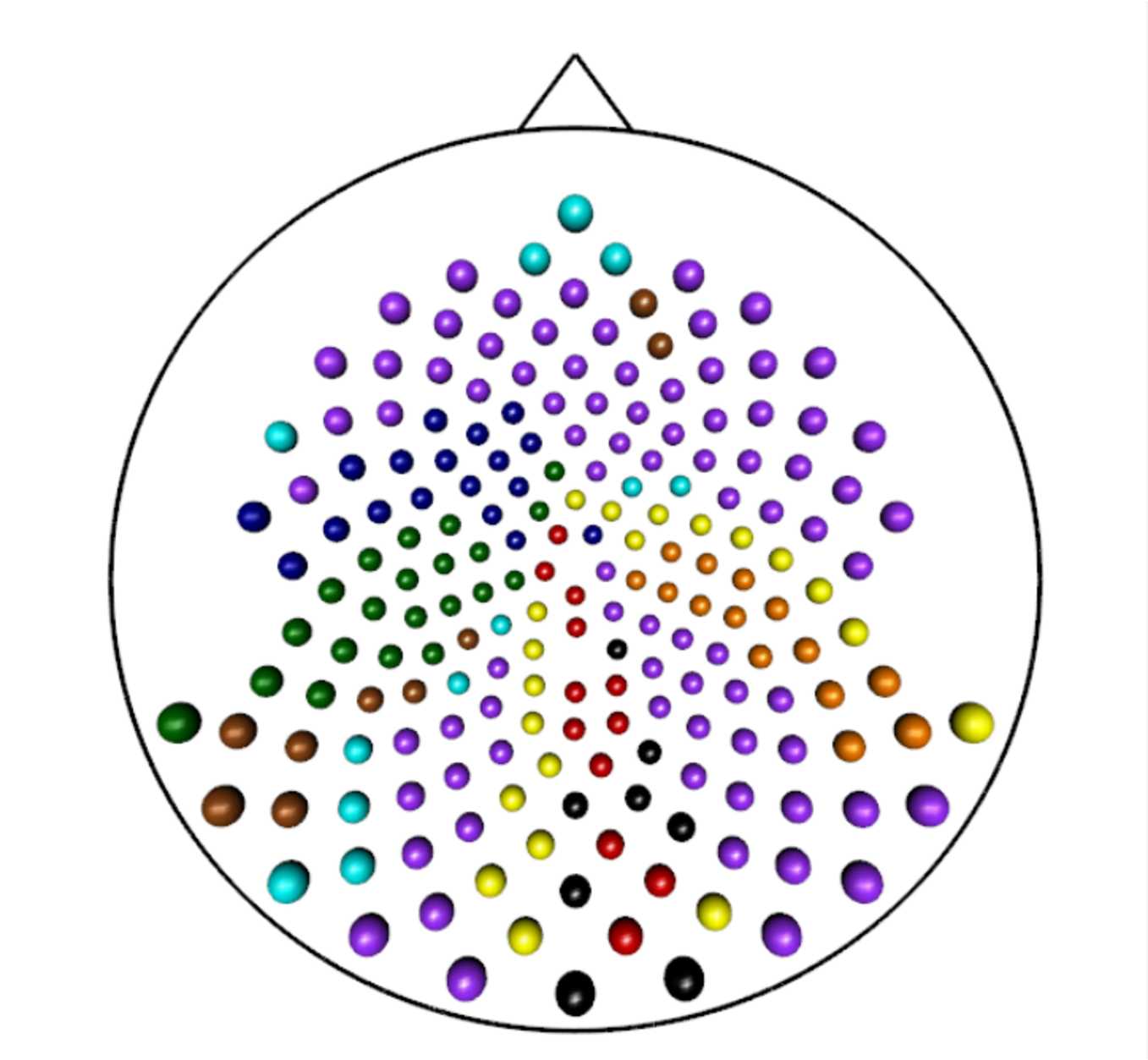}}
\caption{Clustering results for BLAK's resting state during different phases: early resting state (epochs 1-50), middle resting state (51-110) and late resting state (111-160). a) Mean spectral estimates across epochs by cluster and b) Distribution of clusters across the cortical surface}\label{F9}
\end{figure}

Figure \ref{F9} shows formation of the clusters (location, spatial distribution and
specific channel membership) and the shape of the corresponding spectral densities,
coded in different colors, for BLAK in each phase. Comparing the early and middle phases of
resting state, we noted that the formation of clusters during the early and middle phases
were heavily influenced by specific bands: seven (out of the nine) clusters were dominated
by the alpha and beta bands; while the formation of the remaining two clusters were driven by
the gamma band. During the late phase, the influence of the alpha band was reduced in some
of the clusters but the influence of the beta and gamma bands were increased. The increased
power in the beta and gamma band is interesting. This suggests that these individuals
were engaged in some cognitive task (which could not have been a response to an
experimental stimulus but something that is self-induced). Moreover, \cite{Barry}
report that higher beta and gamma power during resting state for healthy controls
compared to children diagnosed with attention-deficit hyperactivity disorder (ADHD).

The formation of the clusters at the cortical surface varies across the three phases
during resting state. In the early phase, channels at the left pre-motor region
belong to one cluster (green) and most of the channels at prefrontal and right
pre-motor region belong to another cluster (purple). However, the clustering structure
at these regions changes during the middle phase where the channels in the pre-motor
(which were originally clustered with the other non pre-motor channels) are assigned
back with the rest of the pre-motor channels (dark blue cluster). As we transition
from the middle to the late phase, channels that were assigned to the right pre-motor
reverted back to the channels at the prefrontal region. These changes in cluster assignment was not entirely unexpected since many of these channels lie at the boundaries between the
two anatomical regions.

Also, some channels which belong to the yellow cluster during the early phase
switched to the orange cluster during the middle phase of resting-state. In this
switch, the gamma and alpha bands played the key roles. The late phase of the resting
state shows more changes. For example, three channels located at the right occipital
region switched from the yellow to the brown cluster, due to an increase of power in
the alpha band and a decrease at the gamma band.
Another interesting change appears on the prefontal region. There we observe
three of the purple-colored channels switched to light blue cluster and a
new cluster was formed. This fact lets the dark blue channels at the middle
state go back to the purple ones and the underlying process that characterized
this cluster (dark blue) changes completely its location.


While some channels displayed dynamic behavior across phases, there some clusters,
such as the red and black, which showed consistent membership. The red cluster is
characterized by the presence of the delta, theta and gamma bands while the black
cluster was dominated by the theta, alpha and gamma bands. These
channels are located at the central occipital and right occipital regions. It is true
that the term ``resting state" is misleading since the brain does not really shut off even
during the so-called rest. That term refers more to the fact that no external stimulus was
applied as a part of the experiment. Indeed, during ``rest", the brain performs a
number of functions (e.g., basal) which are known to be associated with beta and gamma band
activity.

\begin{figure}
\centering
\subfigure[\label{F13b}]{\includegraphics[scale=.23]{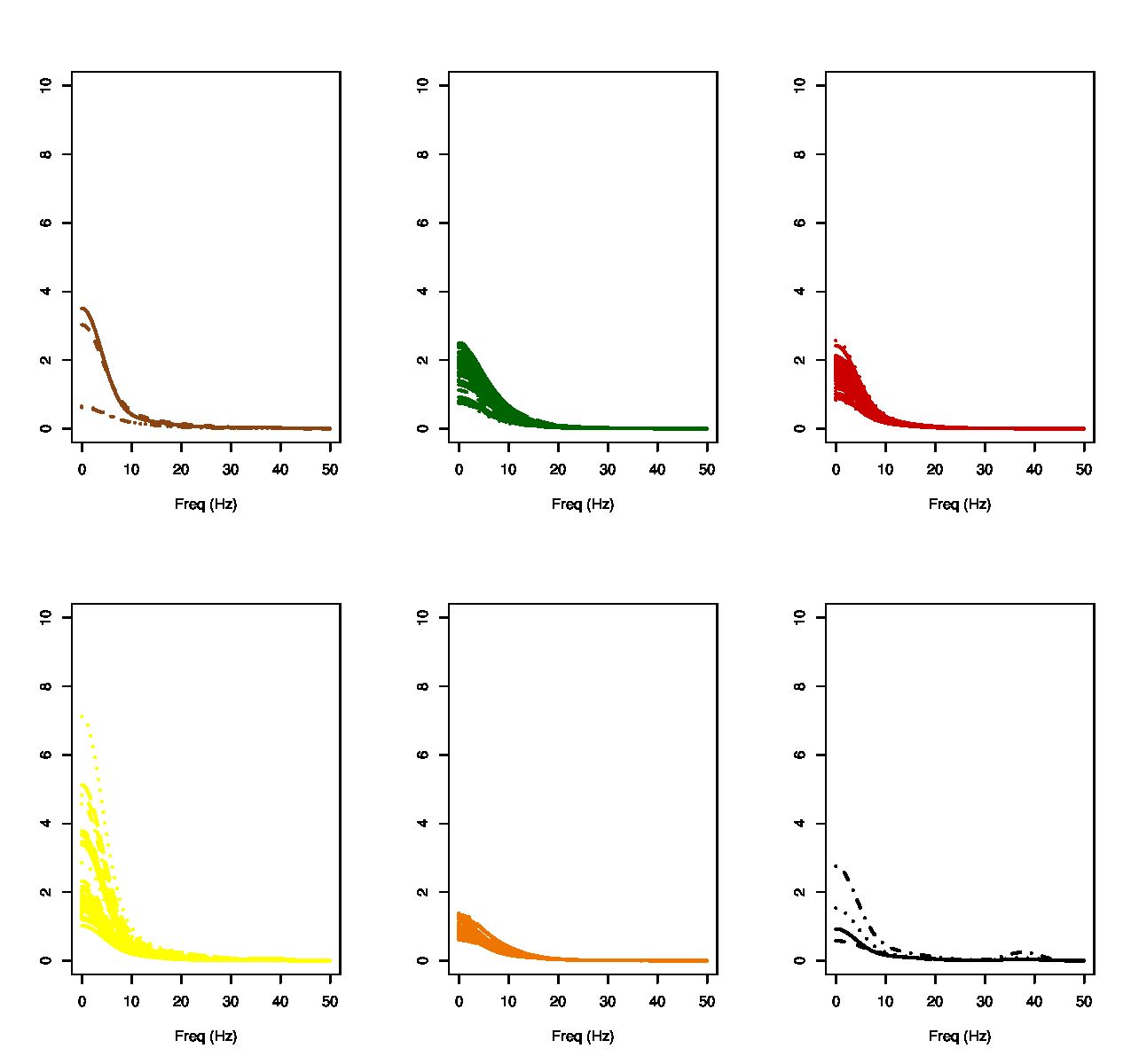}\hspace{.3cm}\includegraphics[scale=.23]{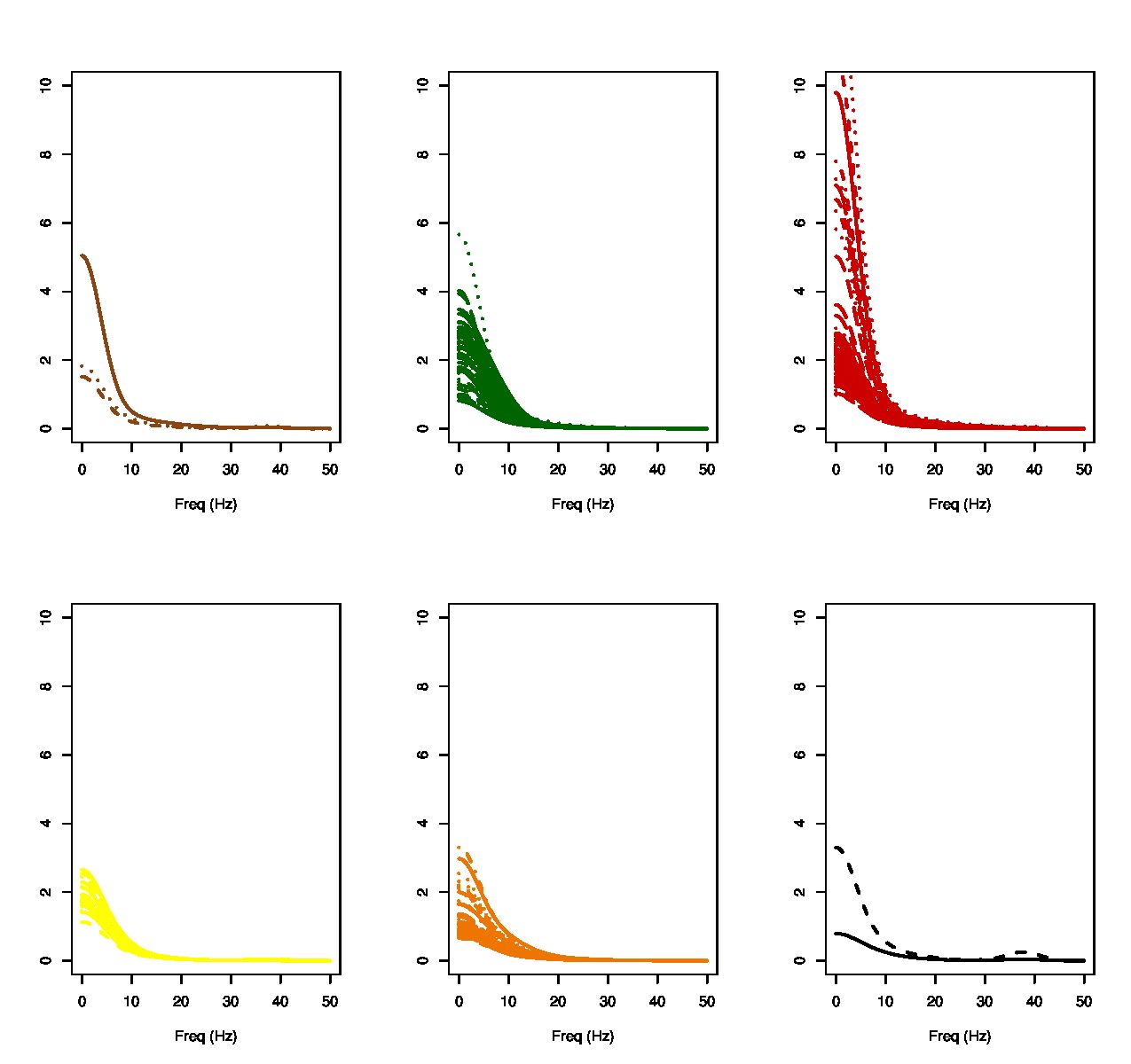}\hspace{.3cm}\includegraphics[scale=.23]{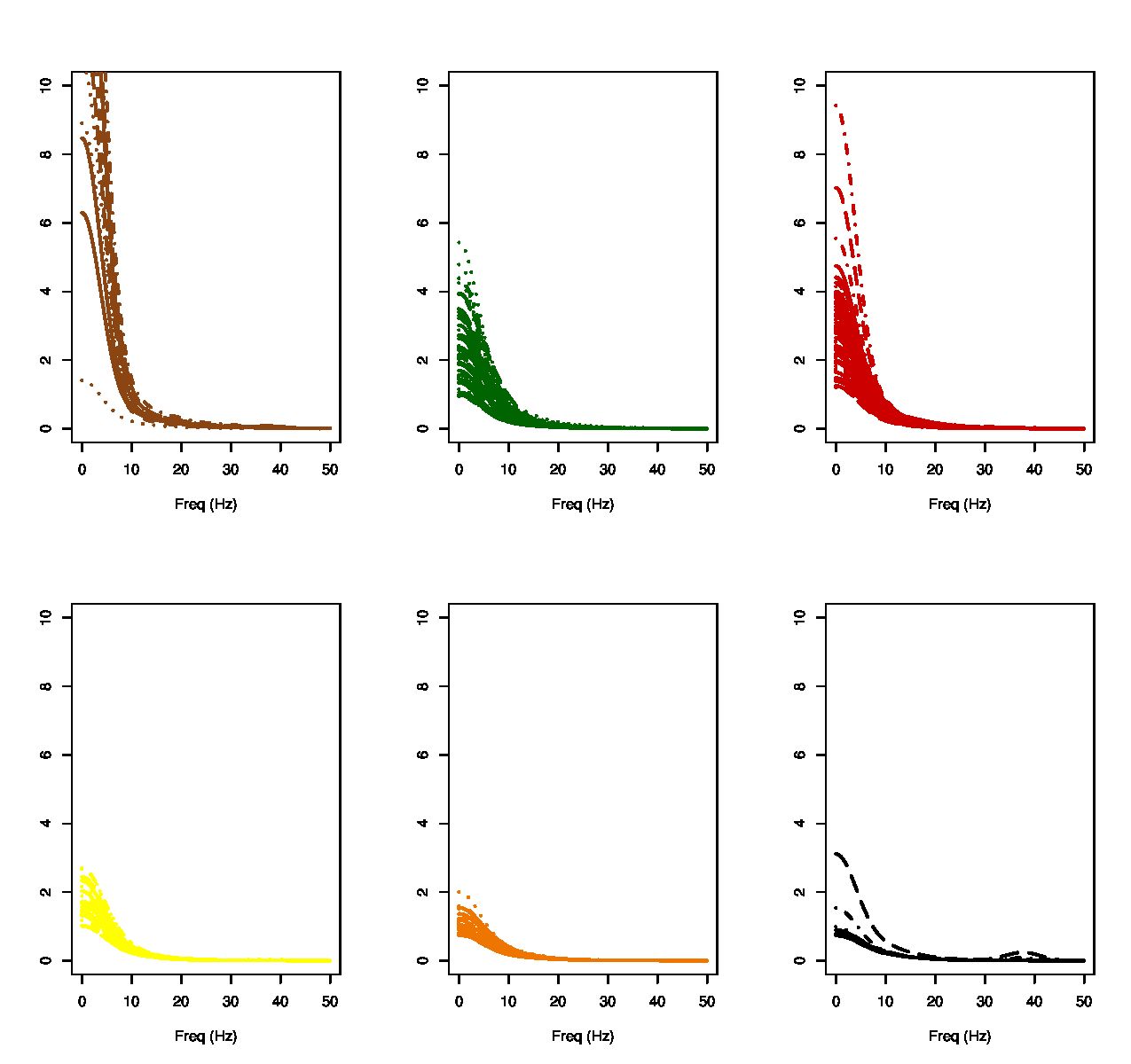}}
\subfigure[\label{F13c}]{\includegraphics[scale=.22]{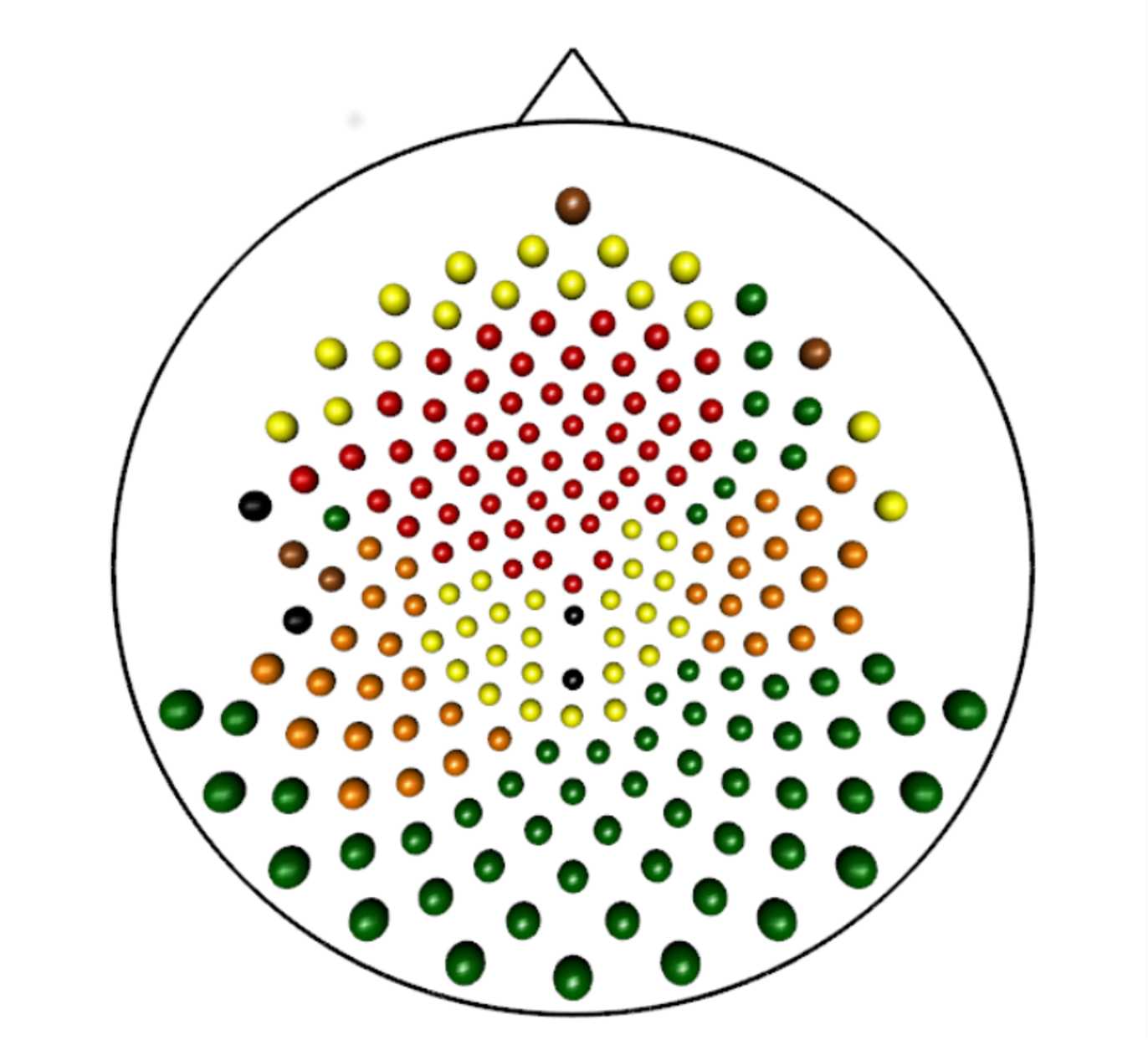}\hspace{.2cm}\includegraphics[scale=.22]{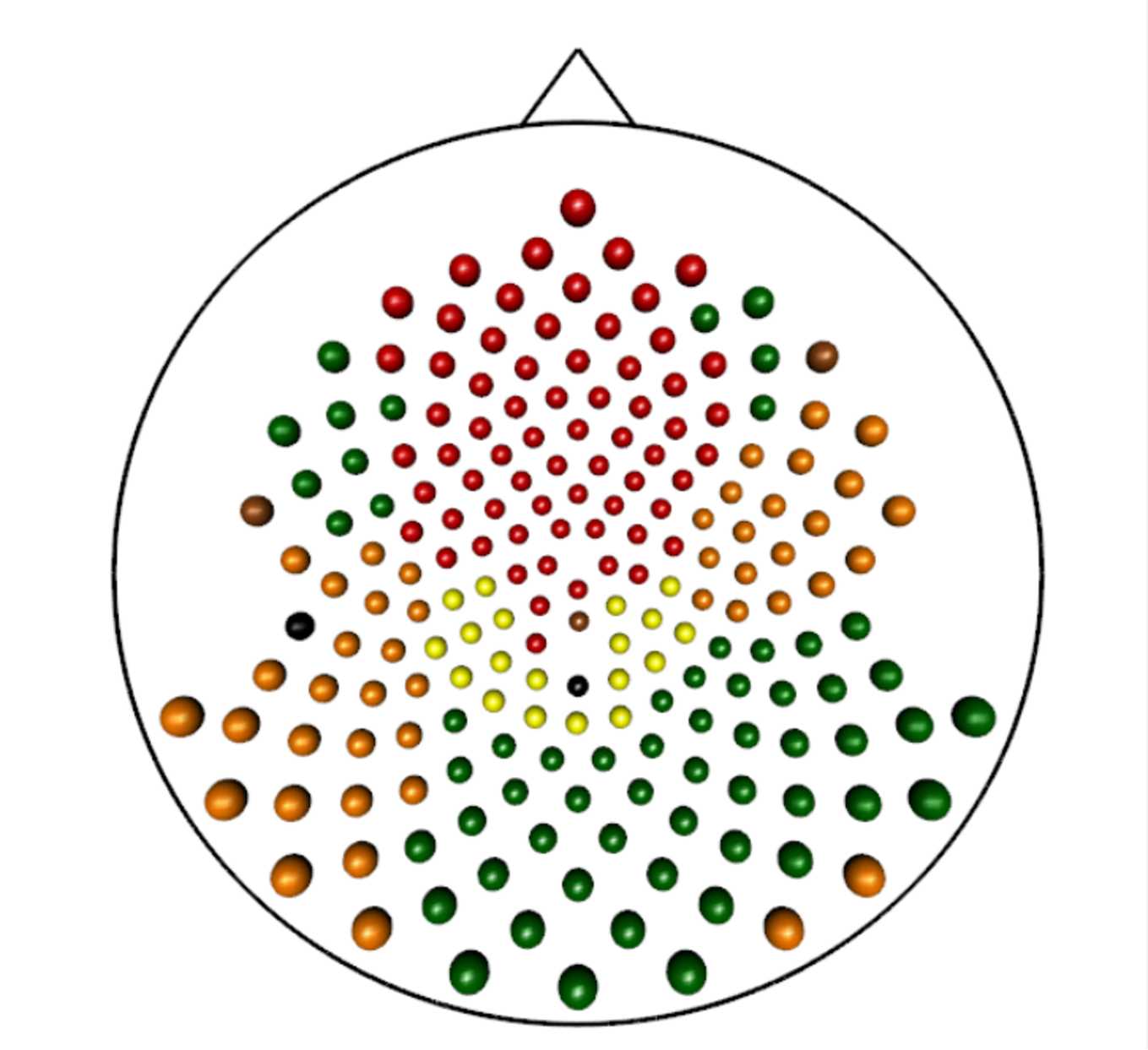}
             \hspace{.2cm}\includegraphics[scale=.22]{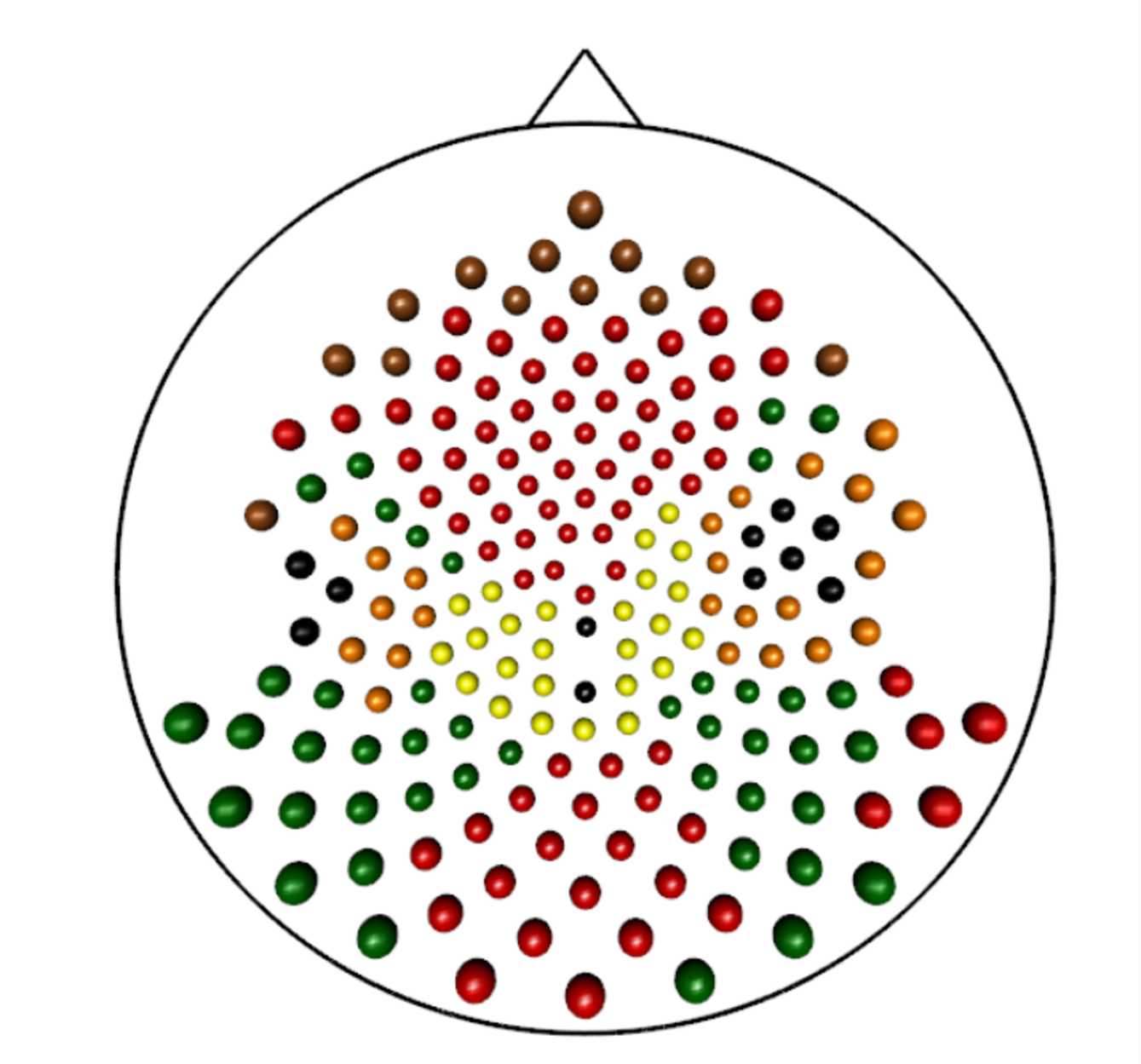}}
\caption{Clustering results for MOHK's resting state during different phases: early resting state (epochs 1-50), middle resting state (epochs 51-110) and late resting state (epochs 111-160). (a.) Mean spectral estimates across epochs by cluster and (b.) Distribution of
clusters across the cortical surface}\label{F13}
\end{figure}

In Figure \ref{F13}, we display the spectra of each cluster and the corresponding
location of the clusters on the cortical surface for MOHK. There does not seem to be much
activity during the early resting stage and spectra belonging to each cluster
have only small variations in the delta and theta bands. This fact could explain
the presence of so much variability in the clustering structure. During the middle
and late resting phases there are some underlying signals that change and
produce a different structure. Some channels at the prefontal region in the early
stage are yellow, in the middle are red and finally at the late stage are brown;
these changes are produced by variations on the delta band. Another interesting
structure is the black cluster, since these are the only channels that have activity
on the gamma frequency band, and by the late resting stage some channels in the
pre-motor area belongs to this cluster.

The SMC method captures the dynamic behavior of the EEG during the resting state as the cluster
memberships change across phases. In addition, the SMC method also identified the frequency bands
that were the primarily drive the formation of the clusters. It is interesting to note that the
dominant frequency bands (i.e., those that dictate how the channels are clustered together)
vary for the two individuals. It is not possible to say if this has implications for the
change on perceptual improvement since a causal analysis was not performed, but there seems to
be an important difference between the underlying processes for both individuals during the
resting state. The subject with lower improvement, BLAK, has more clusters
which implied that the underlying brain process is more complex in a sense that it requires a
greater number of distinct components. In contrast, the subject with higher improvement, MOHK,
had fewer clusters which could be interpreted as having an
underlying process with a simpler structure since it is characterized by a fewer number of
distinct processes.
This behavior has an analogy in the study of ocean waves. In quieter sea states, the effect of
small perturbations of the sea surface (such as those produced by local winds or the presence of swell) are more noticeable. These leads to less precise clustering than during hurricanes
when there is usually a dominant frequency throughout the build-up stage, that makes the
clustering structure more stable. For both MOHK and BLAK, the clusters produced are
consistent for the most part with the anatomical-based parcellation of the cortical surface
and thus cluster formation based on the spectra of the EEGs can be used to recover the
spatial structure of the underlying brain process.


\section{Discussion and Future Work}
We developed the SMC method -- a new clustering method based on the spectra of
EEGs and the total variation distance. The SMC method was applied to a resting-state
EEG data to produce cluster formations, i.e., identify channels that are spectrally
synchronized. Motivated by the fact that the underlying brain process is not
necessarily stationary across resting-sate, we divided the resting-state period
into three phases: early, middle and late resting-state phases to determine how
cluster formation evolves across the entire resting state. This analysis is the
first, to the best of our knowledge, to demonstrate non-stationarity in EEGs
during resting state. The SMC method was able to identify rough spatial boundaries
(i.e., clusters) based on the similarity of the spectra and also how these
boundaries change across resting state as channels, especially at or near the
borders, change memberships.

An important step towards understanding the brain process -- while at rest -- is to
identify the number of distinct networks (or clusters) that are present during
resting state. The SMC method identifies the number of clusters based on
analogue of the elbow-scree plot. Moreover, the results of the analysis
suggests that the SMC method could be used to identify patterns, such as the ranges
of frequencies delta, theta, alpha, beta or gamma that play the most important
role in grouping the channels or identifying spatial boundaries between clusters.
Moreover, we provided a qualitative report about how the location of the spectrally
synchronized channel could be associated with the participant's ability to learn a
new motor skill.

The SMC method for the most part provided a data-based confirmation of the
parcellation of the cortical surface based on the anatomy. However, the method
also gave additional interesting results. The activity in the gamma band was an
unexpected behavior since basal activity is often seen to occur only in lower
frequencies. Nevertheless there is a small accumulation of evidence that
high frequency activity could be present during resting state but this has
not been replicated. This high gamma activity could be used as an indicator of
a state outside from the expected resting behavior. It is also interesting that
we did not find activity of the gamma band in the clusters related with the
motor region during resting state which is a question of interest for motor studies.

The new method has limitations and will need to be further developed in different
fronts. The criteria to determine the number of clusters need to be tightened. At
this point it merely parallels the scree plot that is often used in principal
components analysis. Future work will focus on deriving the approximate distribution
of the total variation distance using asymptotic theory or resampling (bootstrap)
methods. This methodology will give us some measures of uncertainty of our conclusions.
Also, we need a more rigorous model to present the clustering results for several
epochs, a possible option being the use of probabilistic graphical models. Here
it would be ideal to model cluster membership as a function of the epoch $r$. One
way to characterize cluster membership is through the affinity matrix which, for
each epoch, contains only $0$'s and $1$'s. This matrix will have components
$(p,q,r)$ which takes a value of $1$ if channels $p$ and $q$ belong to the same
cluster during the $r$-th epoch.

Finally, we conclude that, though emphasis is on the application to the
spectra of multi-channel EEG, our method is general with a wide range of applications
such as analysis of time-varying amplitude of ocean waves.


\section*{Acknowledgements}
\hide{
This work was partially supported by 1) CONACYT, M\'exico, scholarship AS visiting research student, This work was partially supported by 1) CONACYT, M\'exico, scholarship AS visiting research student, 2) CONACYT, M\'exico, Proyecto An\'alisis Estad\'istico de Olas Marinas, Fase II, and 3) Center of Research in Mathematics (CIMAT), A.C. Eu\'an and Ortega wish to thank Prof Pedro C. Alvarez Esteban for several fruitful conversations on the methodology proposed on this paper. Eu\'anwish to thank to UC Irvine Space Time Modeling Group for the invitation to collaborate as a visiting scholar in their research group.}

\hide{The research conducted at the UC Irvine Space-Time Modeling
group (PI: Ombao) is supported in part by the National Science
Foundation Division of Mathematical Sciences and the Division
of Social and Economic Sciences. The authors thank Dr. Steven
C. Cramer of the UC Irvine Department of Neurology for sharing
the EEG data that was used in this paper.}

\hide{This work was done while J.O. was visiting, on sabbatical leave from CIMAT and with
support from CONACYT, M\'exico, the Departamento de Estadística e I.O., Universidad de
Valladolid. Their hospitality and support is gratefully acknowledged.}


\end{document}